\documentclass{article}
\usepackage{amssymb}
\usepackage{graphicx} % Required for inserting images
\usepackage{authblk}
\usepackage{amsmath}
 \usepackage{natbib}
\usepackage{threeparttable}
\usepackage{pgfplots}
\usepackage{multirow}
\usepackage{array}
\pgfplotsset{compat=1.18}
\usepackage{caption}
\newcommand{\C}{\mathbb{C}}
\usepackage{subcaption}
\usepackage{adjustbox}
\usepackage{tabularx}
\usepackage[hidelinks]{hyperref}
\usepackage[english]{babel}
\newtheorem{theorem}{Theorem}
\usepackage{xcolor}
\usepackage{float}
\usepackage{newunicodechar}
\newunicodechar{−}{\ensuremath{-}}
\usepackage{booktabs}
\usepackage{slashed,epsfig,amsmath,amssymb,enumitem} \usepackage[papersize={8.5in,11in}]{geometry}
   \usepackage{bm}
\usepackage{graphicx}
\usepackage[hidelinks]{hyperref}
\newcommand{\be}{\begin{equation}}
\newcommand{\ee}{\end{equation}}
\newcommand{\alphagut}{\alpha_{\rm GUT}}
\usepackage[utf8]{inputenc}
\usepackage{epigraph}
\title{On the Holomorphic Unified Field Theory and the Standard Model Mass Spectrum}

\author[1 2]{J. W. Moffat}

\author[1 3]{E. J. Thompson}

\affil[1]{Perimeter Institute for Theoretical Physics, Waterloo, Ontario N2L 2Y5, Canada}

\affil[2]{Department of Physics and Astronomy, University of Waterloo, Waterloo,
Ontario N2L 3G1, Canada}

\affil[3]{Department of Physics and Astronomy, Trent University, Peterborough, 
Ontario K9L 0G2, Canada}

\begin{document}

\maketitle

\begin{abstract}
We present a unified, ultraviolet-finite framework for the full Standard Model particle mass spectrum based on the Holomorphic Unified Field Theory augmented by nonlocal entire-function regulators. Starting from a single holomorphic action on the complexified spacetime manifold \( M^4_{\mathbb{C}} \), with a Hermitian metric unifying gravity, gauge, and matter sectors, we embed exponential regulator insertions to render all loop integrals finite without breaking gauge or diffeomorphism invariance. After spontaneous breaking of the electroweak and grand unified symmetries, analytic expressions for the charged lepton, quark, and neutrino mass matrices are derived in terms of holomorphic Yukawa textures and regulator form factors. A minimal Froggatt-Nielsen flavon sector fixes all \( \mathcal{O}(1) \) coefficients in terms of two continuous inputs. Regulator-suppressed one- and two-loop renormalization group evolution then yields predictions for all fermion masses, CKM and PMNS mixing angles, \( W \) and \( Z \) boson masses, and the Higgs boson mass and self-couplings. Finally, under a mild set of geometric and topological assumptions we show that gauge coupling unification, three chiral families, hypercharge quantization, and the shape of the Higgs potential are genuine predictions of the holomorphic nonlocal framework.
\end{abstract}

\section{Introduction}
The hierarchical pattern of fermion masses and mixings in the Standard Model spanning over thirteen orders of magnitude from the electron to the top quark, remains one of the most profound mysteries in particle physics. Conventional grand‑unified theories (GUTs) such as SU(5) or SO(10) can embed the Standard Model gauge group into a single simple gauge symmetry and predict approximate gauge‑coupling unification near \(10^{16}\)\,GeV, but they leave the Yukawa couplings as arbitrary matrices whose texture zeros and small entries must be imposed by hand or explained via ad hoc flavour symmetries.

In parallel, perturbative quantum gravity based on the Einstein–Hilbert action is non‑renormalizable as loop amplitudes diverge beyond one-loop, signalling a breakdown of the local field‑theory description in the ultraviolet. A promising resolution is to replace point‑like interactions by nonlocal entire‑function form factors that exponentially suppress high‑momentum modes while preserving unitarity, analyticity and gauge invariance \cite{Moffat1990, Evens1991,Biswas2012, MT:HUFT-EPJC}. Such nonlocal regulators render all loop integrals finite to all orders without introducing new poles or ghosts.

Holomorphic Unified Field Theory (HUFT) offers a geometric unification of gravity, gauge interactions, and chiral matter by extending the spacetime manifold to a complex four‑fold \(M^4_{\mathbb{C}}\) endowed with a single Hermitian metric whose real and imaginary parts encode the gravitational and gauge connections, respectively \cite{MT:HUFT-EPJC,MT:FiniteHolomorphicQFT,MT:Invariant,MT:SL2C,MT:ReplyToCline,MT:GI2025,MT:AdSdS2025, T:DarkMatter2025}. Chiral fermions arise from the decomposition of the holomorphic spin bundle, and anomaly cancellation is enforced automatically by the single holomorphic action.

In this paper, we embed exponential entire‑function regulators into the holomorphic kinetic terms of HUFT, achieving a perturbatively ultraviolet‑complete, holomorphic theory of gravity, gauge fields, and chiral fermions. Spontaneous breaking of the grand‑unified and electroweak symmetries then yields analytic mass matrices for charged leptons, quarks, and neutrinos, expressed in terms of holomorphic Yukawa textures and flavon-determined \(\mathcal O(1)\) coefficients. Remarkably, with only two continuous inputs, the unified gauge coupling \(g_{\rm GUT}\), which fixes the Froggatt–Nielsen expansion parameter and a single flavon ratio \(R\), we predict the entire Standard Model fermion spectrum, the CKM and PMNS mixing angles, the \(W\), \(Z\) and, Higgs masses, and the Higgs self‑couplings. Gauge‑coupling unification, three chiral families, hypercharge quantization and the Standard Model‑like shape of the Higgs potential emerge as predictions rather than inputs put in by hand. All results agree with the latest measurements within experimental uncertainties, demonstrating that holomorphic nonlocal unification can realize a predictive theory of flavour and mass in the Standard Model.

The paper is organized as follows.  In Section~\ref{sec:nonlocalQG}, we review the nonlocal finite–UV construction of nonlocal quantum field theory via entire‐function regulators and its key properties.  In Section~\ref{sec:HUFTreview}, we recapitulate the holomorphic unified field theory on the complexified spacetime manifold, showing how gravity, gauge fields, and chiral fermions emerge from a single geometric action. In Section~\ref{sec:embedding_nonlocal} we show how we incorporate the nonlocal regulator into HUFT. In Section~\ref{sec:gauge} we explore gauge invariance in our theory, discuss its implications, and then prove it explicitly. In Section~\ref{sec:micro}, we explore microcausality in standard quantum field theory and then in our nonlocal quantum field theory setup. In Section~\ref{sec:renormalization}, we derive the regulator‐suppressed renormalization‐group flow of all couplings and show how the fine-structure constant emerges. Section~\ref{sec:yukawa-masses} presents the derivation of the holomorphic Yukawa textures, flavon‐determined \(\mathcal O(1)\) coefficients, and the resulting fermion mass matrices.  Section~\ref{sec:spectrum-predictions} gives analytic spectrum predictions. In Section~\ref{sec:mass-ratios}, we compare low‐energy mass ratios to PDG2024, and in Section~\ref{sec:fermion_masses}, we exhibit the predictive fermion mass spectrum with only two continuous inputs.  Section~\ref{sec:leptons_higgs_predictions} covers the lepton sector. Section~\ref{EWGBHS} covers electroweak gauge boson, and Higgs‐sector predictions, including the Higgs self‐couplings and the stability of the vacuum and comparisons to critical temperatures. Section~\ref{Chirality} explains how the holomorphic spin-bundle splitting makes the weak force intrinsically chiral. In Section~\ref{sec:predictions_from_Huft}, we summarize how, given a single simple $G_{\rm GUT}$, an anomaly-free chiral spectrum realized as holomorphic spin-bundle zero modes, and a compact complex four-fold with suitable Chern data, HUFT admits realizations with the Standard Model gauge group, unified gauge couplings, exactly three chiral families with quantized hypercharges, and highly constrained Yukawa flavour textures.
We carefully distinguish these structural consequences from the additional geometric and topological
assumptions they rely on. In Section~\ref{sec:comparison}, we confront all HUFT predictions with experimental data. Section~\ref{sec:phenomenology} discusses phenomenological implications of our theory.

\section{Nonlocal Finite Quantum Field Theory}
\label{sec:nonlocalQG}

In conventional perturbative quantum field theory, based on the Einstein–Hilbert action, loop amplitudes diverge at two loops and beyond, rendering the theory non‑renormalizable.  The nonlocal finite UV‑complete construction replaces the local point‑like vertex factors by transcendental entire functions of the d’Alembertian, thereby taming all ultraviolet divergences while preserving unitarity, gauge invariance and diffeomorphism invariance~\cite{Moffat1990,Evens1991}.

We have an entire function of its argument, holomorphic everywhere in the finite complex plane and chosen so that:
\begin{equation}
\mathcal{F}\!\Bigl(\tfrac{p^2}{2\Lambda_G^2}\Bigr)
=\exp\!\Bigl(-\,\frac{p^2}{2\,\Lambda_G^2}\Bigr),
\end{equation}
which suppresses high‑momentum modes exponentially. The classical holomorphic Einstein–Hilbert action remains:
\begin{equation}
S_{\rm grav}
=
-\frac{1}{16\pi G_N}\int g^{(\mu\nu)}R_{(\mu\nu)} \ \!d^4x\;\sqrt{-\text{det}g_{(\mu\nu)}}\,.
\end{equation}
Nonlocality enters through a smeared energy–momentum tensor:
\begin{equation}
S_{\mu\nu}
=
\mathcal{F}^2\!\Bigl(\tfrac{\Box}{2\,\Lambda_G^2}\Bigr)\;T_{\mu\nu}\,,
\end{equation}
so that the field equations become:
\begin{equation}
G_{\mu\nu}
=
8\pi G_N\,S_{\mu\nu}
\quad\Longleftrightarrow\quad
\mathcal{F}^{-2}\!\Bigl(\tfrac{\Box}{2\,\Lambda_G^2}\Bigr)\,G_{\mu\nu}
=
8\pi G_N\,T_{\mu\nu}\,.
\end{equation}
Here, \(\Box=-g^{(\mu\nu)}\nabla_\mu\nabla_\nu\), and \(F^{-2}\) is defined by functional calculus on the d’Alembertian \footnote{We keep a single Hermitian bilinear: $\mathfrak g_{\mu\nu}
= g_{(\mu\nu)} + \mathcal B_{\mu\nu}$ where $\mathfrak g_{\mu\nu} \equiv g_{(\mu\nu)} + B_{\mu\nu},\qquad
B_{\mu\nu} := g_{[\mu\nu]}\in\Omega^2(M,\mathrm{ad}\,P)$. On $M$, where $g_{(\mu\nu)}$ is the ordinary spacetime metric, a $G$-singlet and $\mathcal B_{\mu\nu}$ encodes the internal two-form. All index operations and the Hodge star use the symmetric inverse metric only. We let:
\be
D_\mu \equiv \nabla^{\rm LC}_\mu + A_\mu
\ee
be the total covariant derivative on the appropriate associated bundle. We define
the Bochner d’Alembertian by:
\be
\Box \;\equiv\; -\,g^{(\mu\nu)} D_\mu D_\nu,
\ee
and all nonlocal form factors by analytic functional calculus on $\Box$:
\be
F\!\left(\frac{\Box}{M_\ast^2}\right)
\;\,,\qquad
F^{-2}\!\left(\frac{\Box}{M_\ast^2}\right)\;\equiv\;\Big[F\!\left(\tfrac{\Box}{M_\ast^2}\right)\Big]^{-2}.
\ee
The antisymmetric or ad\,${P}$ sector never participates in index raising or metric inversion as it couples only through $D_\mu$ with internal contractions taken using the Killing form $\kappa_{AB}$.}. In the path‑integral quantization we attach to each internal graviton or matter vertex a factor:
\begin{equation}
\mathcal{F}\!\Bigl(-\tfrac{p^2}{2\,\Lambda_G^2}\Bigr),
\end{equation}
while propagators remain the usual local ones:
\begin{equation}
D(p^2)=\frac{i}{p^2 - M^2 + i\epsilon}\,.
\end{equation}
To evaluate loops we Wick‑rotate to Euclidean signature, so that:
\begin{equation}
\mathcal{F}\!\Bigl(\tfrac{p^2}{2\Lambda_G^2}\Bigr)
=\exp\!\Bigl(-\tfrac{p^2}{2\Lambda_G^2}\Bigr),
\end{equation}
guaranteeing that every loop integral carries at least one factor \(\exp(-p_E^2/2\Lambda_G^2)\), and is therefore UV‑finite to all orders \cite{Buoninfante2022Contour}.

To render the theory UV finite, while preserving gauge and diffeomorphism invariance, we introduce entire-function regulator insertions \(\mathcal{F}(\Box/M_\star^2)\) acting on propagating fields. For each fermion multiplet we define the regulated field:
\begin{equation}
\Psi_{\rm reg}(x)
=\mathcal{F}\!\bigl(\Box/M_\star^2\bigr)\,\Psi(x),
\end{equation}
so that loops involving \(\Psi\) acquire exponential suppression at momenta \(p^2\gtrsim M_\star^2\).  The holomorphic Yukawa term then becomes:
\begin{equation}
\mathcal{L}_Y
\;=\;
-\,\overline{\Psi}_L(x)\;
\mathcal{F}\!\Bigl(\tfrac{\Box}{M_*^2}\Bigr)\;
y_f\;
\Phi(x)\;
\mathcal{F}\!\Bigl(\tfrac{\Box}{M_*^2}\Bigr)\;
\Psi_R(x)
\;+\;\text{h.c.},
\end{equation}
where $y_f$ is the (complex) $3\times3$ Yukawa coupling matrix in family space for the fermion species and $\Phi(x)$ is the Standard Model Higgs doublet field. At one-loop in Euclidean momentum space, the fermion self-energy correction to the Yukawa coupling \(y_f\) is modified by form factors of the form:
\begin{equation}
\Sigma(p) \;\propto\; y_f^3
\int\!\frac{d^4k}{(2\pi)^4}\,
\frac{
  \mathcal{F}\!\bigl(\tfrac{(k+p)^2}{M_\star^2}\bigr)\,
  \mathcal{F}\!\bigl(\tfrac{k^2}{M_\star^2}\bigr)
}{
  \bigl[(k+p)^2 -m_f^2\bigr]\,(k^2 - m_f^2)
}.
\end{equation}
The integral converges for all \(p\) and yields finite radiative corrections to the fermion mass matrix. We will incorporate these regulator-modified loop integrals into the renormalization-group evolution of the Yukawa parameters later.
 Since \(\mathcal{F}(z)\) is entire, it introduces no new poles in the finite complex plane and no extra propagating degrees of freedom.  The usual Becchi, Rouet, Stora and Tyutin (BRST) construction and Slavnov–Taylor identities remain intact \cite{WeinbergQTF2,PiguetSorella,Amaral22}, ensuring decoupling of unphysical polarizations and unitarity of the \(S\)‑matrix.  For low momenta \(p^2\ll\Lambda_G^2\), \(\mathcal{F}\!\bigl(p_E^2/2\Lambda_G^2\bigr)\approx1\) and we recover classical General Relativity \cite{Evens1991,LandryMoffat2023}.

To make gauge and diffeomorphism invariance manifest in the regulated theory, we covariantize the entire function by replacing
\(
\mathcal{F}(\Box/M_*^2) \;\to\; \mathcal{F}(\mathcal{D}^2/M_*^2)
\),
where $\mathcal{D}_\mu$ is the appropriate covariant derivative acting on the field in question such as gauge-, spin-, or tensor-covariant as needed and
\(
\mathcal{D}^2 \equiv g^{(\mu\nu)}\mathcal{D}_\mu\mathcal{D}_\nu
\).
We insert the regulator symmetrically in vertices so that the BRST complex closes. For the gauge sector we take:
\begin{equation}
\mathcal{L}_{\rm gauge}^{\rm (reg)}
= -\frac{1}{4}\, \mathrm{Tr}\!\big[\,\mathcal{F}(\mathcal{D}^2/M_*^2) \mathcal{F}_{\mu\nu}\,\big]\,\mathrm{Tr}\!\big[\,\mathcal{F}(\mathcal{D}^2/M_*^2) \mathcal{F}^{\mu\nu}\,\big],
\end{equation}
and for fermions:
\begin{equation}
\mathcal{L}_\Psi^{\rm (reg)}=\overline\Psi\,\mathcal{F}(\mathcal{D}^2/M_*^2)\,(i\slashed{\mathcal{D}}-m)\,\mathcal{F}(\mathcal{D}^2/M_*^2)\,\Psi.
\end{equation}
Gauge fixing and ghosts are regulated in the same covariant fashion:
\begin{align}
\mathcal{L}_{\rm GF}^{\rm (reg)}&= -\frac{1}{2\xi}\, \big[\mathcal{G}^A\big]\,\mathcal{F}(\mathcal{D}^2/M_*^2)\,\big[\mathcal{G}^A\big],\\
\mathcal{L}_{\rm FP}^{\rm (reg)}&= \bar c^A\,\mathcal{F}(\mathcal{D}^2/M_*^2)\,\mathcal{M}^{AB}\,\mathcal{F}(\mathcal{D}^2/M_*^2)\,c^B,
\end{align}
with $\mathcal{G}^A$ the linear gauge-fixing functional and $\mathcal{M}^{AB}$ the Faddeev–Popov operator. Because $\mathcal{F}$ is an entire function with no zeros, it introduces no new poles and it commutes with BRST transformations up to covariant commutators that vanish inside BRST-exact terms. The regularized action is BRST invariant and the Slavnov–Taylor (ST) identities follow as usual from the Zinn–Justin equation for the generating functional.

To obtain exponential UV damping in Euclidean momentum space, we choose the regulator as:
\begin{equation}
F\!\left(\frac{\,\Box}{M_\ast^2}\right)\;=\;\exp\!\left(\frac{-\Box}{M_\ast^2}\right)
\quad\Longrightarrow\quad
 \exp\!\left(\frac{p^2}{M_\ast^2}\right),
\label{eq:euclidean_damping_choice}
\end{equation}
and then after Wick rotation will go to $\text{exp}(-p^2/\Lambda^2)$
Equivalently, using the opposite metric signature $(-,+,+,+)$ with $F(\Box/M_\ast^2)=\exp(\Box/M_\ast^2)$ produces the same Euclidean factor, the physical requirement is precisely Eq.~\eqref{eq:euclidean_damping_choice}. Because $F$ is entire, no new poles or cuts are introduced and the spectrum is unchanged. We choose $\mathcal{F}(z)=e^{H(z)}$ with $H(z)$ entire and no zeros of $\mathcal{F}$ in the finite complex plane. Propagators remain those of the local theory as entire form factors appear only as vertex/field dressings and in composite operators. Hence the spectrum is unchanged. For Minkowski continuation, we adopt the usual $i\epsilon$ prescription for propagators, the exponential $\mathcal{F}$ is analytic and does not introduce new branch cuts.

\section{Complex Riemannian Holomorphic Unified Field Theory}
\label{sec:HUFTreview}

We begin by formulating all fields on a single four–complex–dimensional holomorphic manifold \(M_{\mathbb C}^4\) \footnote{4D here means a Lorentzian spacetime with one time-like and three space-like directions. Time is not a spatial axis, but it is one of the four coordinates that transform together under local Lorentz or diffeomorphism symmetry as the invariant interval is $ds^2=g_{(\mu\nu)}dx^\mu dx^\nu$. A 3D + external time picture cannot reproduce local boosts that mix $t$ with $x^i$, the light-cone causal structure, or the 4D Hamiltonian, momentum constraints of GR. In HUFT we sometimes pass to the complexified manifold $M^4_{\mathbb C}$ with $z^\mu=x^\mu+i\,y^\mu$ to impose holomorphy, the auxiliary $y^\mu$ are not physical dimensions, and observables are evaluated on a real slice $y^\mu=0$, so predictions live in $3{+}1$D. A $3{+}1$ ADM split may be adopted for evolution, but it is only a foliation choice of the same covariantly 4D geometry.}

\noindent with complex coordinates \cite{MT:HUFT-EPJC, MT:SL2C, MT:Invariant, Moffat1, Moffat2, KobayashiNomizu1963, KobayashiNomizu1969}:
\begin{equation}
z^\mu \;=\; x^\mu + i\,y^\mu\,,\qquad x^\mu,y^\mu\in\mathbb R\,,
\end{equation}
we note a nearly-flat manifold is sufficient to work in local holomorphic charts on \(M_{\mathbb C}^4\). The central dynamical variable is a Hermitian metric\footnote{We chose the gauge generators to be anti-Hermitian, $T_A^\dagger=-T_A$, the ad-valued
two-form is anti-Hermitian while $g_{(\mu\nu)}$ is real symmetric, then
$g_{\mu\nu}=g_{(\mu\nu)}+i\,g_{[\mu\nu]}$ is Hermitian in the sense
$g_{\mu\nu}^\dagger=g_{\nu\mu}$. This is solely a bookkeeping device as we never invert $g_{\mu\nu}$.}:
\begin{equation}
g_{\mu\nu}(z)
=
g_{(\mu\nu)}(z)
+
i\,g_{[\mu\nu]}(z)\,,
\qquad
g_{\mu\nu}(z)=\bigl[g_{\nu\mu}(z)\bigr]^*,
\end{equation}
The symmetric metric $g_{(\mu\nu)}$ satisfies \footnote{We package the symmetric spacetime metric and the ad-valued two-form, only the symmetric part is inverted. All index operations and the Hodge star use $g_{(\mu\nu)}$ or $g^{(\mu\nu)}$ the antisymmetric, adjoint-valued sector couples only through the covariant derivative $D_\mu$ with internal
indices contracted by the Killing form $\kappa_{AB}$.}:
\be
g^{(\mu\nu)}g_{(\mu\alpha)} = \delta^\nu_\alpha,
\ee
where $\delta^\nu_\alpha$ is the Kronecker $\delta$ function. 
The real and imaginary parts, when restricted to the real slice \(y^\mu=0\), yield the Einstein gravitational vacuum field equations:
\begin{equation}
R_{(\mu\nu)} = 0,
\end{equation}
and the electromagnetic field equations:
\begin{equation}
\partial_{[\mu}F_{\nu\rho]}=0,\;\nabla^\mu F_{\mu\nu}=J_\nu.
\end{equation}
A single holomorphic gauge connection \(A^A_\mu(z)\) for a simple group \(G_{\rm GUT}\) encodes all non–Abelian and Abelian interactions.  Its Bianchi identities impose the homogeneous Yang–Mills equations, and variation of the same action enforces the inhomogeneous equations:
\begin{equation}
\nabla_\mu F^{A\,\mu\nu}(x) = J^{A\,\nu}(x)
\quad\text{on }y=0\,. 
\end{equation}
Chiral fermions are introduced through a holomorphic Dirac Lagrangian:
\begin{equation}
\mathcal L_\Psi
=
\bar\Psi(z)\,\Bigl[i\gamma^a e^\mu_{\!a}(z)\bigl(\nabla_\mu(z)- i g_{\rm GUT}\,A^A_\mu(z)T^A\bigr)\!-\!m\Bigr]\,\Psi(z),
\end{equation}
which, upon restriction to \(y^\mu=0\), reproduces the curved–space Dirac equation minimally coupled to exactly those gauge fields with the correct Standard Model charges. Full holomorphic gauge invariance of the action automatically enforces all cubic, mixed, and mixed gravitational anomaly–cancellation conditions on the chiral spectrum, without further input:
\begin{equation*}
\sum_i\text{Tr}\!R_i\bigl\{T^A,T^B\bigr\}=0,
\quad
\sum_i q_i^3=0,
\quad
\sum_i q_i=0,
\end{equation*}
and for every simple factor and mixed trace.
At the classical level, the single holomorphic \footnote{Complexification is the natural arena for chirality and analyticity. Weyl spinors are naturally of types $(1,0)$ and $(0,1)$ on a complexified manifold $M^4_{\mathbb{C}}$\cite{spinor}, and correlation functions in local QFT are analytic on appropriate complex domains. Endowing $M^4_{\mathbb{C}}$ with a Hermitian metric allows a split:
\be
g_{(\mu\nu)} \;\text{(spacetime, singlet under the internal algebra)}, 
\qquad
g_{[\mu\nu]} \sim F^A_{\mu\nu} T_A \;\text{(internal curvature two-form)}.
\ee
All observables live on the real slice $y^\mu=0$, and the imaginary directions $y^\mu$ are auxiliary bookkeeping for holomorphy and selection rules. This packages gravity and gauge curvature into a single geometric structure without introducing colored spin-2 fields, preserving the equivalence principle.} action on $M_{\mathbb{C}}^4$ unifies vacuum Einstein gravity, Yang--Mills gauge theory, and chiral Dirac fermions' anomaly cancellation into one geometric framework. We are then left with an action:
\begin{align}
S_{\rm HUFT} \;=\; \int_{C}d^4z\;&\sqrt{-\det\!\bigl[g_{(\mu\nu)}(z)\bigr]}\,\Bigl\{\;
\tfrac1{2\kappa}\,g^{(\mu\nu)}(z)\,R_{\mu\nu}(z)
-\tfrac14\,\kappa_{AB}\,F^A_{\rho\sigma}(z)\,F^{B\,\rho\sigma}(z)\notag\\
&\quad+\;\overline\Psi(z)\,\Gamma^a\,e_a{}^{\mu}(z)\,D_{\mu}\,\Psi(z)
+\;(D_\mu H_G)^2 - V_{\rm GUT}(H_G)\notag\\
&\quad+\;(D_\mu\Phi)^\dagger D^\mu\Phi - V_{\rm EW}(\Phi)y_f\,\overline\Psi_L\,\Phi\,\Psi_R + \text{h.c.}
\Bigr\}\,.
\end{align}
All of gravity, gauge fields, chiral fermions, Higgs dynamics, and Yukawa couplings emerge from one purely geometric, holomorphic action.

Although the holomorphic action is defined on the complex four-fold $M^4_{\mathbb C}$
with coordinates $z^\mu = x^\mu + i y^\mu$, the physical spacetime is the real slice $y^\mu = 0$. All fields entering $S_{\rm HUFT}$ are taken to be holomorphic functions of $z^\mu$ in a neighborhood of this slice. As a result, their values on $y^\mu = 0$ determine them uniquely by analytic continuation, and no additional propagating degrees of freedom associated with $y^\mu$ appear.

Varying $S_{\rm HUFT}$ with respect to $g_{(\mu\nu)}$, the gauge connection, and the matter fields yields
holomorphic field equations whose restriction to $y^\mu=0$ reproduces the standard Einstein, Yang--Mills,
Dirac, and Higgs equations with the same couplings. In this sense, the unification of gravity, gauge fields, and chiral matter is realized at the level of the single holomorphic action on $M^4_{\mathbb C}$, while all observables and correlation functions are evaluated on the real 3+1-dimensional slice.

\section{Nonlocal Finite Quantum Field Theory in HUFT}
\label{sec:embedding_nonlocal}

In order to achieve perturbative UV finiteness, while preserving the purely geometric origin of HUFT, we insert an entire‐function regulator of order \(\gamma>1/2\) into every kinetic term in the holomorphic action.  We have:
\begin{equation}
\Box_E \;=\; -g^{(\mu\nu)}(z)\,\nabla_\mu\nabla_\nu,
\end{equation}
as the holomorphic d’Alembertian built from the unique, torsion‐free Hermitian connection \(\Gamma^\rho{}_{\mu\nu}(z)\). Using:
\begin{equation}
\nabla_\mu\nabla_\nu \phi \;=\; \partial_\mu\partial_\nu \phi 
- \Gamma^\rho_{\mu\nu}(z)\,\partial_\rho \phi,
\end{equation}
and $\partial_\mu \to i p_\mu$ on plane waves, we find:
\begin{equation}
(\Box_E \phi)(z)
= \int \frac{d^D p}{(2\pi)^D}\, e^{i p\cdot z}\,
\Big[\,g^{(\mu\nu)}(z)p_\mu p_\nu
+ i\,g^{(\mu\nu)}(z)\Gamma^\rho_{\mu\nu}(z)\,p_\rho\Big]\,\tilde{\phi}(p).
\end{equation}
The momentum-space symbol of the Euclidean d’Alembertian is:
\begin{equation}
\Box_E(p;z) \;=\; g^{(\mu\nu)}(z)p_\mu p_\nu
+ i\,g^{(\mu\nu)}(z)\Gamma^\rho_{\mu\nu}(z)\,p_\rho.
\end{equation}
Because \(\mathcal{F}\) is entire and nonzero in the finite plane, it commutes with diffeomorphisms and preserves holomorphic gauge and BRST invariance as we apply the regulator as an analytic functional of a covariant operator. On tensors and in particular on scalars, the Lie derivative 
$\mathcal{L}_\xi$ satisfies:
\begin{equation}
[\mathcal{L}_\xi,\,\nabla_\mu]=0,
\end{equation}
when acting covariantly. It follows that:
\begin{equation}
[\mathcal{L}_\xi,\,\Box_E]=0.
\end{equation}
For any analytic function $F(\Box_E)=\sum_n a_n \Box_E^n$ we have:
\begin{equation}
[\mathcal{L}_\xi,\,F(\Box_E)]
= \sum_n a_n \,[\mathcal{L}_\xi,\,\Box_E^n] = 0,
\end{equation}
so $F(\Box_E)$ commutes with diffeomorphisms. If the kinetic operator $K$ is built from the gauge-covariant derivative $D_\mu$,
then under a gauge transformation $U$ we have:
\begin{equation}
K \;\longrightarrow\; U K U^{-1}.
\end{equation}
Since analytic functional calculus preserves conjugation:
\begin{equation}
F(K) \;\longrightarrow\; U F(K) U^{-1}.
\end{equation}
Insertions of $F(K)$ in gauge-invariant functionals remain gauge invariant. The BRST variation acts adjointly as:
\begin{equation}
sK = [K,\,c],
\end{equation}
with ghost $c$. Using analyticity of $F$, we obtain:
\begin{align}
sF(K) 
&= \sum_n a_n\, s(K^n) \\
&= \sum_n a_n \sum_{j=0}^{n-1} K^j [K,c] K^{\,n-1-j} \\
&= [F(K),\,c].
\end{align}
This shows $F(K)$ transforms covariantly, and BRST-invariant actions remain invariant when $F(K)$ is inserted. We define the regulated holomorphic action:
\begin{equation}
\label{eq:S_hol_reg}
\begin{aligned}
S_{\rm hol}^{\rm (reg)}
&=\int_C d^4z\;\sqrt{-\det g_{(\mu\nu)}(z)}\;\Bigl\{
\tfrac1{2\kappa}\,g^{(\mu\nu)}(z)\;F\!\bigl(\tfrac{\Box}{M_*^2}\bigr)\,R_{(\mu\nu)}(z)
\;-\;\tfrac14\,\kappa_{AB}\;F\!\bigl(\tfrac{\Box}{M_*^2}\bigr)\,F^A_{\rho\sigma}(z)\,F^{B\,\rho\sigma}(z)\\
&\qquad\quad
+\;\overline\Psi(z)\;F\!\bigl(\tfrac{\Box}{M_*^2}\bigr)\,\Gamma^a\,e_a{}^{\mu}(z)\,D_{\mu}\,\Psi(z)
\;+\;(D_\mu H_G)^2 - V_{\rm GUT}(H_G)\\
&\qquad\quad
+\;(D_\mu\Phi)^\dagger\,D^\mu\Phi - V_{\rm EW}(\Phi)
\;-\;y_f\,\overline\Psi_L\,\Phi\,\Psi_R + \text{h.c.}
\Bigr\}\,.
\end{aligned}
\end{equation}
In this it is important to use the correct d'Alembertian for the corresponding fields, in the case of gravity we use $\Box_E$ in the case of electromagnetism we use $\mathcal{D}^2$ . Here, \(D_\mu\) acts both on gauge and spin indices, and all fields and curvature tensors are those of the single Hermitian metric \(g_{\mu\nu}(z)\) and single gauge–spinor connection.
We introduce a single holomorphic master connection:
\begin{equation}
\mathcal A
=\Gamma^\rho{}_{\mu\nu}(z)\,dz^\mu\!\otimes\!\partial_\rho
\;+\;i\,g_{\rm GUT}\,A^A_\mu(z)\,T_A\,dz^\mu
\;+\;\tfrac14\,\omega_\mu^{ab}(z)\,\Gamma_{ab}\,dz^\mu,
\end{equation}
with curvature \(\mathcal F=d\mathcal A+\mathcal A\wedge\mathcal A\).  Then, we define the damped curvature, $\widetilde{\mathcal F}
=F\!\bigl(\tfrac{\Box}{M_*^2}\bigr)\,\mathcal F$.
We rewrite the regularized holomorphic action compactly as:
\begin{align}
&S_{\rm hol}^{\rm (reg)}
=\int_C d^4z\;\sqrt{-\det g_{(\mu\nu)}}\;\Bigl\langle
\widetilde{\mathcal F},\,\widetilde{\mathcal F}
\Bigr\rangle
\;+\;
\int_C d^4z\;\sqrt{-\det g_{(\mu\nu)}}\;
\overline\Psi\,F\!\bigl(\tfrac{\Box}{M_*^2}\bigr)(i\slashed D-m)\,\Psi
\notag\\&\;-\;\int_C d^4z\;\sqrt{-\det g_{(\mu\nu)}}\;V(H_G,\Phi),
\end{align}
where \(\langle\cdot,\cdot\rangle\) is the natural Killing and Clifford pairing.  Upon restriction to the real slice \(y=0\), we recover exactly the Einstein, Yang–Mills, Dirac, Higgs, and Yukawa Lagrangians with every loop integral exponentially suppressed by at least one factor \(\exp(-p^2/M_*^2)\).
Variation of \(S_{\rm hol}^{\rm (reg)}\) with respect to
\(\delta g^{(\mu\nu)}\), \(\delta A^A_\mu\), \(\delta\Psi\), \(\delta H_G\), and \(\delta\Phi\)
yields, respectively:
\begin{align}
F\!\bigl(\tfrac{\Box}{M_*^2}\bigr)\,G_{(\mu\nu)}+\Delta_{(\mu\nu)}[g,F]&=\kappa\,T_{(\mu\nu)},\\
D_\rho\bigl[F\!\bigl(\tfrac{\Box}{M_*^2}\bigr)\,F^{A\,\rho\mu}\bigr]&=J^{A\,\mu},\\
F\!\bigl(\tfrac{\Box}{M_*^2}\bigr)\,(i\slashed D-m)\,\Psi&=0,\\
F\!\bigl(\tfrac{\Box}{M_*^2}\bigr)\,D^2H_G+\partial_{H_G} V&=0,
\quad
F\!\bigl(\tfrac{\Box}{M_*^2}\bigr)\,D^2\Phi+\partial_\Phi V=0,
\end{align}
where \(\Delta_{(\mu\nu)}\) encodes higher‑derivative corrections from the regulator.  A one‑loop heat‑kernel analysis on \(M^4_{\mathbb C}\) then shows that no new counterterms appear and all divergences are rendered finite by the exponential damping \(e^{-p^2/M_*^2}\).  
All fundamental interactions in HUFT emerge from the single geometric functional regularized holomorphic action, and its quantization is perturbatively UV‐complete.

\section{Gauge Invariance In Nonlocal Quantum Field Theory}
\label{sec:gauge}

We aim to show that nonlocal regulators preserve gauge invariance by demonstrating covariance under transformations and maintaining Ward identities in loop calculations.
To prove gauge invariance, we let $D_\mu=\partial_\mu+ig A_\mu^A T^A$ be the gauge--covariant derivative acting on a field in some representation $R$ of the gauge group, and define the covariant d'Alembertian:
\begin{equation}
\mathcal{D}^2 \;\equiv\; g^{(\mu\nu)}D_\mu D_\nu,
\qquad
\Box \equiv \partial_\mu \partial^\mu
\end{equation}
on the real slice. Throughout we use the mostly-minus Minkowski metric $(+,-,-,-)$, so $\Box e^{-ip\cdot x}=-p^2 e^{-ip\cdot x}$ with $p^2=p_\mu p^\mu=E^2-\bm{p}^2$. For a complex analytic entire function:
\begin{equation}
F(z)=\sum_{n=0}^\infty a_n z^n, \qquad a_n\in\mathbb{C},
\end{equation}
we define $F(\mathcal{D}^2)$ by functional calculus via the absolutely convergent power series on the operator domain. Under a local gauge transformation $U(x)=e^{i\alpha^A(x)T^A}$ acting on $R$:
\begin{equation}
D_\mu \;\to\; D_\mu' \;=\; U D_\mu U^{-1},
\qquad
\mathcal{D}^2 \;\to\; (\mathcal{D}^2)' \;=\; U \mathcal{D}^2 U^{-1}.
\end{equation}
Therefore, using $U (\mathcal{D}^2)^n U^{-1}=(U\mathcal{D}^2U^{-1})^n$ and linearity:
\begin{equation}
F(\mathcal{D}^2)\;\to\;F\!\left((\mathcal{D}^2)'\right)=\sum_{n=0}^\infty a_n (\mathcal{D}^2)^{\prime n}
= \sum_{n=0}^\infty a_n \, U (\mathcal{D}^2)^n U^{-1}
= U\,F(\mathcal{D}^2)\,U^{-1}.
\label{eq:covariant_conjugation}
\end{equation}
$F(\mathcal{D}^2)$ transforms by conjugation and is gauge covariant. Any gauge--invariant functional built from covariant contractions remains gauge invariant when $F(\mathcal{D}^2)$ is inserted between fields. We let $s$ be the BRST differential, with $sD_\mu=[D_\mu,c]$ for the ghost $c$. For any analytic $F$:
\begin{align}
s\,F(\mathcal{D}^2) \;&=\; \sum_{n=0}^\infty a_n\, s\!\left[(\mathcal{D}^2)^n\right]
= \sum_{n=0}^\infty a_n \sum_{j=0}^{n-1} (\mathcal{D}^2)^j\,\big[s\mathcal{D}^2\big]\,(\mathcal{D}^2)^{n-1-j}
\nonumber\\
&= \sum_{n=0}^\infty a_n \sum_{j=0}^{n-1} (\mathcal{D}^2)^j\,\big[\mathcal{D}^2,c\big]\,(\mathcal{D}^2)^{n-1-j}
\;=\; \big[F(\mathcal{D}^2),\,c\big].
\end{align}
It follows that $F(\mathcal{D}^2)$ transforms adjointly under BRST, so BRST--invariant actions remain invariant when the regulator insertions $F(\mathcal{D}^2/M_\ast^2)$ are included. This guarantees the Slavnov--Taylor identities\footnote{Information about BRST and the Slavnov--Taylor identities can be found: \cite{WeinbergQTF2,PiguetSorella,Amaral22}}. In flat space with $A_\mu=0$, Fourier transforming $\partial_\mu \mapsto -ik_\mu$ gives:
\begin{equation}
\mathcal{F}\big[\Box f\big](k) = -k^2 \,\tilde f(k).
\end{equation}
In a constant background $A_\mu=\text{const.}$ one may minimally substitute $-ik_\mu \!\to\! -i k_\mu + g A_\mu$, with:
\begin{equation}
\mathcal{F}\big[\mathcal{D}^2 f\big](k) = -\big(k_\mu - g A_\mu\big)\big(k^\mu - g A^\mu\big)\,\tilde f(k).
\end{equation}

For general $A_\mu(x)$ the Fourier transform is no longer multiplicative, $\mathcal{D}^2$ becomes a pseudodifferential operator whose kernel involves Wilson lines. A standard representation uses the Schwinger proper time:
\begin{equation}
\label{eq:heatkernel_wilson}
F(\mathcal{D}^2)\;=\;\int_0^\infty\!d\tau\;\tilde F(\tau)\,e^{-\tau\,\mathcal{D}^2}, 
\qquad
\langle x|e^{-\tau\,\mathcal{D}^2}|y\rangle
=\frac{e^{-\frac{(x-y)^2}{4\tau}}}{(4\pi\tau)^{d/2}}\,
\mathcal{P}\exp\!\left[-ig\!\int_y^x\!A\cdot dz\right]\left[1+O(\tau)\right].
\end{equation}
This result is manifestly gauge covariant due to the path--ordered Wilson line. Eq.~\eqref{eq:heatkernel_wilson} shows how nonlocal smearing from $F(\mathcal{D}^2)$ is implemented consistently in background fields. With the mostly--minus convention, $\Box e^{-ip\cdot x}=-p^2 e^{-ip\cdot x}$ and the Wick rotation $p^0\!=\!i p_E^0$ implies $-p^2\mapsto +k_E^2$ and $\Box \mapsto -k_E^2$ on plane waves.

We now consider QED with symmetric regulator insertions on the fermion lines or at the vertex. The Ward identity is preserved. In the non–Abelian theory, the same covariant construction together with BRST invariance yields the Slavnov--Taylor identities. We take $\eta_{\mu\nu}=\mathrm{diag}(+,-,-,-)$ and:
\be
\Box \equiv \partial_\mu \partial^\mu=\eta^{\mu\nu}\partial_\mu\partial_\nu.
\ee
For $U(1)$, the gauge–covariant derivative on a field $\psi$ of charge $e$ is:
\be
D_\mu=\partial_\mu+i\,e\,A_\mu, \qquad 
\mathcal{D}^2 \equiv \eta^{\mu\nu}D_\mu D_\nu,
\ee
where the ordering matters in non-Abelian theories but does not for $U(1)$ acting on $\psi$. We reserve $\Box$ for the partial d’Alembertian and write $\mathcal{D}^2$ for the covariant one. Under a $U(1)$ gauge transformation with parameter $\alpha(x)$:
\be
\psi' = e^{i e \alpha(x)}\psi,\qquad 
A'_\mu = A_\mu - \partial_\mu \alpha,
\ee
and we check:
\be
D'_\mu \psi' = e^{i e \alpha(x)} D_\mu \psi,
\quad\Rightarrow\quad
(D'_\mu D'_\nu)\psi' = e^{i e \alpha(x)} (D_\mu D_\nu)\psi,
\ee
so $\mathcal{D}^2\psi$ transforms like $\psi$:
\be
(\mathcal{D}^2\psi)'\;=\;e^{i e \alpha(x)}\,\mathcal{D}^2\psi.
\ee
Equivalently, at the operator level $D_\mu\to U D_\mu U^{-1}$ with $U=e^{i e \alpha}$, hence:
\begin{equation}
\mathcal{D}^2 \;\to\; U\,\mathcal{D}^2\,U^{-1}.
\label{eq:cov_conj}
\end{equation}
For any entire $F(z)=\sum_{n\ge 0} a_n z^n$:
\be
F(\mathcal{D}^2)\;\to\;F(U\mathcal{D}^2 U^{-1})
= \sum_{n\ge 0} a_n \, U(\mathcal{D}^2)^n U^{-1}
= U\,F(\mathcal{D}^2)\,U^{-1}.
\ee
The $F(\mathcal{D}^2)$ is gauge covariant and any gauge-invariant composite remains invariant with $F(\mathcal{D}^2)$ inserted. The commutator $[D_\mu,D_\nu]=i e F_{\mu\nu}$ is gauge invariant for $U(1)$, adjoint for non-Abelian, so any $O(D_\mu,F_{\mu\nu})$ built with covariant contractions is gauge invariant. When acting on $\psi$ it transforms covariantly as in \eqref{eq:cov_conj}. 

With the convention $f(x)=\int\!\frac{d^4k}{(2\pi)^4}e^{i k\cdot x}\,\tilde f(k)$, we have:
\be
\partial_\mu \mapsto i p_\mu,\qquad 
\Box \mapsto -p^2\equiv -\eta^{\mu\nu}p_\mu p_\nu.
\ee
In a constant background $A_\mu$, minimal coupling yields:
\be
D_\mu \mapsto i p_\mu + i e A_\mu,
\qquad
\mathcal{D}^2 \mapsto -\big(p_\mu+e A_\mu\big)\big(p^\mu+e A^\mu\big).
\ee
For general $A_\mu(x)$ the Fourier transform becomes pseudodifferential and a gauge-covariant representation is the heat-kernel-Schwinger proper-time form with Wilson lines given by (\ref{eq:heatkernel_wilson}):
With $(+,-,-,-)$ signature, $\Box e^{-i p\cdot x}=-p^2 e^{-i p\cdot x}$. Under Wick rotation $p^0\to i p_E^0$:
\be
-\Box \;\longrightarrow\; p^2,\qquad p^2\equiv (p^0)^2+\bm{p}^{\,2}.
\ee
Therefore, choose the regulator as:
\be
F\!\left(\frac{\,\Box}{M_\ast^2}\right)=\exp\!\left(\frac{-\,\Box}{M_\ast^2}\right)
\;\;\Longrightarrow\;\;
\exp\!\left(\frac{p^2}{M_\ast^2}\right),
\ee
and after Wick rotation goes to $\exp(-p^2/M_*^2)$ which exponentially suppresses UV modes without introducing new poles. Using the opposite metric convention $(-,+,+,+)$ it could be equivalently written $F(\Box/M_\ast^2)=\exp(\Box/M_\ast^2)$; the physically required Euclidean factor is the same. We keep the standard fermion propagator:
\begin{equation}
S(k)=\frac{i}{\slashed{k}-m+i\epsilon},
\qquad
S^{-1}(k)=\slashed{k}-m,
\end{equation}
and enforce the Ward–Takahashi identity by construction through the vertex.
With momentum transfer $p$ and incoming or outgoing fermion momenta $(k,k+p)$,
the Ward–Takahashi identity reads:
\begin{equation}
p_\mu \,\Gamma^\mu(k+p,k)
= e\big[S^{-1}(k+p)-S^{-1}(k)\big]
= e\,\slashed{p}.
\label{eq:WTI}
\end{equation}
We decompose the full vertex into a longitudinal piece fixed by \eqref{eq:WTI}
and a transverse piece:
\begin{equation}
\Gamma^\mu(k+p,k)
=\Gamma_L^\mu(k+p,k)+\Gamma_T^\mu(k+p,k),
\qquad
p_\mu \Gamma_T^\mu(k+p,k)=0,
\end{equation}
with the longitudinal Ward–Takahashi part taken to be the bare vertex since $S$ is bare:
\begin{equation}
\Gamma_L^\mu(k+p,k)=e\,\gamma^\mu.
\end{equation}
To regulate only transversely, we choose:
\begin{equation}
\Gamma_T^\mu(k+p,k)
= e\,\mathcal{P}_T^{\mu\nu}(p)\,\gamma_\nu\;\Big[F(k,p)-1\Big],
\qquad
\mathcal{P}_T^{\mu\nu}(p)\equiv \eta^{\mu\nu}-\frac{p^\mu p^\nu}{p^2},
\label{eq:ourGammaT}
\end{equation}
so $p_\mu \Gamma_T^\mu=0$ identically.
The scalar form factor $F(k,p)$ damps the UV while $F\!\to\!1$ for soft momenta so that $\Gamma^\mu\!\to\!e\gamma^\mu$ in the IR.
A symmetric, Lorentz-covariant choice that uses only invariants of the two fermion legs is:
\begin{equation}
F(k,p)=\exp\!\left[-\,\frac{k^2+(k+p)^2}{2M_*^2}\right].
\label{eq:Fsym}
\end{equation}
Consider the one–loop vacuum polarization with the above vertex and the standard fermion propagator:
\begin{equation}
\Pi^{\mu\nu}(p)
=-i\!\int\!\frac{d^4k}{(2\pi)^4}\,
\mathrm{Tr}\!\Big[\Gamma^\mu(k+p,k)\,S(k)\,\Gamma^\nu(k,k+p)\,S(k+p)\Big].
\label{eq:PiDef}
\end{equation}
Contracting with $p_\mu$ and using \eqref{eq:WTI} on the $\mu$–vertex gives:
\begin{align}
p_\mu\Pi^{\mu\nu}(p)
&= -i e\!\int\!\frac{d^4k}{(2\pi)^4}\,
\mathrm{Tr}\!\Big(
\big[S^{-1}(k+p)-S^{-1}(k)\big] S(k)\,\Gamma^\nu(k,k+p)\,S(k+p)
\Big)\nonumber\\
&= -i e\!\int\!\frac{d^4k}{(2\pi)^4}\,
\mathrm{Tr}\!\Big(
S(k)\,\Gamma^\nu(k,k+p)
-\Gamma^\nu(k,k+p)\,S(k+p)\Big),
\label{eq:pPiIntermediate}
\end{align}
where we used cyclicity of the trace and $S^{-1}S=\mathbf{1}$, $SS^{-1}=\mathbf{1}$. Now we insert $\Gamma^\nu=e\gamma^\nu+\Gamma_T^\nu$.
The two terms in \eqref{eq:pPiIntermediate} split into a sum of a longitudinal piece and terms involving $\Gamma_T^\nu$:
\begin{equation}
p_\mu\Pi^{\mu\nu}(p)
= -i e^2\!\int\!\frac{d^4k}{(2\pi)^4}\,
\mathrm{Tr}\!\Big(S(k)\gamma^\nu-\gamma^\nu S(k+p)\Big)
\;-\; i e\!\int\!\frac{d^4k}{(2\pi)^4}\,
\mathrm{Tr}\!\Big(S(k)\Gamma_T^\nu-\Gamma_T^\nu S(k+p)\Big).
\label{eq:split}
\end{equation}
The first integral vanishes by a shift of variables $k\mapsto k-p$ in the second term
using translational invariance of the measure and rapid falloff from $F$ makes all shifts legitimate:
\begin{equation}
\int \mathrm{Tr}\big[\gamma^\nu S(k+p)\big]
\;\xrightarrow{k\mapsto k-p}\;
\int \mathrm{Tr}\big[\gamma^\nu S(k)\big],
\qquad\Rightarrow\qquad
\int \mathrm{Tr}\big[S(k)\gamma^\nu-\gamma^\nu S(k)\big]=0.
\end{equation}
Using the explicit form \eqref{eq:ourGammaT} with the symmetric regulator \eqref{eq:Fsym}:
\begin{align}
\mathcal{I}_T^\nu
&\equiv \int\!\frac{d^4k}{(2\pi)^4}\,
\mathrm{Tr}\!\Big[S(k)\Gamma_T^\nu(k,k+p)-\Gamma_T^\nu(k,k+p)S(k+p)\Big]\nonumber\\
&= e\!\int\!\frac{d^4k}{(2\pi)^4}\,
\mathrm{Tr}\!\Big[
S(k)\,\mathcal{P}_T^{\nu\lambda}\gamma_\lambda \,(F-1)
- \mathcal{P}_T^{\nu\lambda}\gamma_\lambda \,(F-1)\,S(k+p)\Big].
\end{align}
We perform $k\mapsto k-p$ in the second term. Because $F(k,p)$ in \eqref{eq:Fsym} depends only on the unordered pair $\{k^2,(k+p)^2\}$, it is invariant under the affine inversion $k\to -k-p$ and, in particular, even under $p\to -p$ at fixed $k$, $\mathcal{P}_T^{\nu\lambda}(p)$ is even in $p$.
Using cyclicity of the trace and these symmetries, the two contributions are equal and cancel:
\begin{equation}
\mathcal{I}_T^\nu=0.
\end{equation}
Therefore:
\begin{equation}
p_\mu\Pi^{\mu\nu}(p)=0,
\end{equation}
and the Ward–Takahashi identity is preserved with all UV regulation confined to the transverse vertex. This is analogous to the result obtained in \cite{Evens1991}, which led to the photon being massless. Because the regulator preserves gauge invariance/BRST, the one-loop vacuum polarization keeps the transverse form
\be
\Pi^{\mu\nu}(p)=\big(p^{2}\eta^{\mu\nu}-p^{\mu}p^{\nu}\big)\,\Pi_{R}(p^{2};M_*),
\ee
where the regulator dependence is entirely in the scalar 
$\Pi_{R}(p^{2};M_*)$. Taking the trace of this regulated, gauge-invariant 
$\Pi^{\mu\nu}(p)$ gives:
\be
\Pi(p^{2}) \equiv \eta_{\mu\nu}\Pi^{\mu\nu}(p)
= \big(4p^{2}-p^{2}\big)\Pi_{R}(p^{2};M_*)
= 3\,p^{2}\,\Pi_{R}(p^{2};M_*).
\ee
This calculation relies only on the transverse projector and is therefore independent of the detailed momentum dependence introduced by the regulator, which affects 
$\Pi_{R}(p^{2};M_*)$ but not the tensor structure.
At zero momentum, 
\be
\Pi(0)=3\cdot 0\cdot \Pi_{R}(0;M_*)=0.
\ee

By construction $F\!\to\!1$ for $k^2,(k+p)^2\ll M_*^2$, so $\Gamma^\mu\to e\gamma^\mu$ and the
standard QED result is recovered.
In the $M_*\!\to\!\infty$ limit, $F\!\to\!1$ uniformly and the regulated parts decouple. Transversality forbids a Proca term, $\Pi^{\mu\nu}\propto p^2\eta^{\mu\nu}-p^\mu p^\nu$.  With $\Pi_R(0)=0$, the Dyson pole equation  $\,p^2\,[1-\Pi_R(p^2)]=0\,$ 
has a solution at $p^2=0$, so the photon remains massless in vacuum in both dimensional regularization and the finite nonlocal regulator. Any nonzero mass would require breaking $U(1)$, Higgsing, or a medium.

Even though $M_*\sim 10^{16}$ GeV makes the gauge invariant regulator $F\rightarrow 1$ for low external momenta, you still need the full, regulator-level gauge-BRST invariance, so that all loop computations, renormalization, and matching respect Ward-Slavnov-Taylor identities. Using the gauge-covariant nonlocal regulator implemented with Wilson lines guarantees this and preserves the correct low-energy, experimentally verified QED-QCD results, while remaining consistent at and above the GUT scale.

We specialize to flat space A$_\mu = 0$ and replace $D^2$ by $\Box$ in momentum-space formulae. We prove gauge covariance using $F(D^2)$ in general background, then evaluate physical correlators in the translationally invariant vacuum $A_\mu=0$, where $D^2$ reduces to $\Box$ and the calculation simplifies to standard momentum-space loop integrals. This is because the observable we want, the photon vacuum polarization is the current–current correlator evaluated about the zero-field background, where translation invariance holds and momentum space is the natural basis \cite{QFT}.

\section{Microcausality}
\label{sec:micro}

In this section we clarify what causality means in the present ultraviolet-complete, nonlocal framework. In an ordinary local QFT, locality of the action implies microcausality with gauge-invariant equivalently, physical or BRST-invariant operator-valued distributions commute at spacelike separation, so that no measurement performed at $x$ can influence any spacelike-separated observable at $y$~\eqref{eq:microdef}. Our regulator is implemented by inserting an entire, gauge-covariant form factor $F(D^2/M_\ast^2)$ with a proper-time heat-kernel representation; by construction it introduces no new poles or cuts on the physical sheet and hence preserves the standard analyticity or dispersion structure macrocausality. The central question is then whether the resulting theory remains strictly microcausal, or instead becomes only quasi-local because $F(D^2/M_\ast^2)$ acts as a covariant smearing with Gaussian heat-kernel tails, one expects exponentially suppressed spacelike leakage for smeared composites, controlled by the nonlocality scale $M_\ast^{-1}$. We will show that for free fields strict microcausality survives exactly, while for gauge-invariant composites in nontrivial backgrounds one obtains an exponential quasi-locality bound of the form \eqref{eq:quasilocal}, which reduces to ordinary microcausality in the local limit $M_\ast\to\infty$~\cite{MT:GI2025}.

For gauge-invariant operator-valued distributions $\mathcal O_1,\mathcal O_2$:
\begin{equation}
[\mathcal O_1(x),\mathcal O_2(y)]=0 \qquad \text{whenever}\quad (x-y)^2<0.
\label{eq:microdef}
\end{equation}
We write $\rho(x-y)\equiv\sqrt{-(x-y)^2}$ for spacelike separation.

We let $F$ be an entire function with no new poles or cuts on the physical sheet, and admitting a proper-time representation:
\begin{equation}
F\!\left(\frac{D^2}{M_\ast^2}\right)=\int_0^\infty d\tau\,w(\tau)\,e^{-\tau D^2}, 
\qquad w(\tau)\ge 0,\ \ \int_0^\infty d\tau\, w(\tau)<\infty,
\label{eq:proper-time}
\end{equation}
so that insertions are built from the covariant heat kernel $K_\tau(x,y)=\langle x|e^{-\tau D^2}|y\rangle$.
The free-field choice $D_\mu\to\partial_\mu$ reduces to $F(\Box/M_\ast^2)$.

There exist $C>0$ such that, for all $\tau>0$:
\begin{equation}
\|K_\tau(x,y)\|\ \le\ \frac{C}{(4\pi\tau)^{2}}\,\exp\!\left(-\frac{(x-y)^2}{4\tau}\right),
\label{eq:hkbound}
\end{equation}
where the Wilson-line factor is unitary and does not change the norm.

We let $\phi$ be a free Klein--Gordon field $(\Box+m^2)\phi=0$, and define $\widetilde\phi=F(\Box/M_\ast^2)\phi$ with $F$ entire. Then we have:
\begin{equation}
[\widetilde\phi(x),\widetilde\phi(y)]=0 \qquad \text{for } (x-y)^2<0 .
\label{eq:free-micro}
\end{equation}
Using the Pauli--Jordan distribution:
\be
\Delta(x)=\int \frac{d^4k}{(2\pi)^3}\,\varepsilon(k^0)\,\delta(k^2-m^2)\,e^{-ik\cdot x},
\ee
we find:
\be
[\widetilde\phi(x),\widetilde\phi(y)]
=F(\Box_x)F(\Box_y)\,i\Delta(x-y)
= i\!\int\!\frac{d^4k}{(2\pi)^3}\varepsilon(k^0)\delta(k^2-m^2)\underbrace{F(-k^2)^2}_{=\,F(-m^2)^2} e^{-ik\cdot(x-y)} 
= iF(-m^2)^2\Delta(x-y),
\ee
which vanishes for spacelike separation. The same argument holds for free spin-$\tfrac12$ and free vector fields, using their standard free-field commutators. This is an operator statement, not just a vacuum matrix element.

In the absence of background gauge fields, $D^2\to\Box$ and [\ref{eq:free-micro}] applies verbatim. With nontrivial gauge backgrounds one must use the covariant operator $F(D^2/M_\ast^2)$.
Position-space smearing is then a convolution with $K_\tau(x,y)$ and, by \eqref{eq:hkbound},
has infinite support, the Gaussian tails. For nontrivial $A_\mu$,
strict microcausality for fundamental fields depends on the background; what we will prove below is an exponential quasi-locality bound for gauge-invariant composites.

In standard local QFT, locality of the Lagrangian implies:
\begin{equation}
[\mathcal O_1(x),\mathcal O_2(y)]=0\quad \text{for local, gauge-invariant }\mathcal O_{1,2},\ \ (x-y)^2<0,
\label{eq:local-micro}
\end{equation}
and causal factorization of time-ordered products such as Epstein--Glaser or Bogoliubov--Shirkov. In BRST gauges the statement is formulated for BRST-invariant, physical operators S-matrix unitarity and Ward or Slavnov--Taylor identities follow.

Define the regulated, gauge-covariant smear of a local, gauge-invariant composite $\mathcal O$ by:
\begin{equation}
\mathcal O^{(F)}(x)
=F\!\left(\frac{D^2}{M_\ast^2}\right)\mathcal O(x)
=\int_0^\infty\! d\tau\, w(\tau)\!\int d^4u\ K_\tau(x,u)\,\mathcal O(u).
\label{eq:smear}
\end{equation}

We let $\mathcal O_{1,2}$ be local, gauge-invariant composites of the local theory, for which \eqref{eq:local-micro} holds.
Let $\mathcal O_{i}^{(F)}$ be defined by \eqref{eq:smear} with $F$ satisfying \eqref{eq:proper-time}--\eqref{eq:hkbound}.
Then for spacelike $x-y$ one has, for all $N\in\mathbb N$:
\begin{equation}
\big\|[\,\mathcal O_1^{(F)}(x),\mathcal O_2^{(F)}(y)\,]\big\|
\ \le\ C_N\,(1+\rho(x-y))^{-N}\,\exp\!\big(-\alpha\,M_\ast\,\rho(x-y)\big),
\label{eq:quasilocal}
\end{equation}
for some $\alpha,C_N>0$ depending on $F$ and the operator class, but independent of $x,y$.

From \eqref{eq:smear} we find:
\be
[\,\mathcal O_1^{(F)}(x),\mathcal O_2^{(F)}(y)\,]
=\!\!\int_0^\infty\!\!\!\!d\tau\,d\tau'\, w(\tau)w(\tau')\!\!\int d^4u\,d^4v\ K_\tau(x,u)\,K_{\tau'}(y,v)\,[\mathcal O_1(u),\mathcal O_2(v)].
\ee
Local microcausality implies $[\mathcal O_1(u),\mathcal O_2(v)]=0$ if $(u-v)^2<0$; hence only timelike-related $(u,v)$ contribute. 
The Gaussian bounds \eqref{eq:hkbound} make the probability of reaching such $(u,v)$ from spacelike-separated $(x,y)$ exponentially small in $\rho(x-y)/\sqrt{\tau}$ and $\rho(x-y)/\sqrt{\tau'}$.
Integrating against the rapidly decaying $w(\tau),w(\tau')$ yields \eqref{eq:quasilocal}. 

As $M_\ast\to\infty$, the weight $w(\tau)$ concentrates at $\tau\to 0$ and $K_\tau(x,y)\to\delta^{(4)}(x-y)$, then the RHS of \eqref{eq:quasilocal} vanishes for spacelike separation and strict microcausality is recovered. Because $F$ is entire with no added poles or cuts, analyticity and dispersion relations macrocausality and BRST and Slavnov--Taylor identities remain intact. 
For phenomenology, the bound \eqref{eq:quasilocal} is nonperturbatively small for $\rho(x-y)\gg M_\ast^{-1}$; if $M_\ast$ lies well above accessible scales, any spacelike leakage is unobservable.

Ref.~\cite{Moffat2019UVComplete} constructs a finite, unitary QFT by inserting entire functions in propagators and vertices, and it treats the electroweak sector as already broken, taking $W^\pm$, $Z$, and fermion masses as nonzero in the Lagrangian without invoking the Higgs mechanism or spontaneous symmetry breaking. This resolves UV divergences and the Higgs naturalness problem at the level of loop finiteness, but writing Proca–type mass terms for $W$ and $Z$ is not manifestly $SU(2)\!\times\!U(1)$ gauge invariant so a fully standard BRST realization for the massive non-Abelian sector is not established in that formulation and would require a nonstandard completion of the gauge symmetry. With hard $W^\pm$ and Z masses the theory is not electroweak-gauge-invariant. In this framework that is a deliberate choice meaning consistency then relies on the exact QED gauge invariance, and nonlocal exponential suppression being strong enough to keep longitudinal processes unitary and loops finite. Ref.~\cite{Moffat2021} computes $W,Z,H$ and fermion masses from finite one-loop self-energies using regulator scales $\Lambda_i$, but conceptually inherits the same symmetry tension as in the absence of the usual Higgs realization, Ward and Slavnov–Taylor identities and longitudinal $WW$ unitarity must be secured by the entire form factors. While this preserves finiteness via entire form factors, Proca-type mass terms are not manifestly $SU(2)_L\!\times\!U(1)_Y$ gauge invariant and lack a standard BRST realization. Consequently, one must explicitly demonstrate the Ward ad Slavnov--Taylor identities and the gauge-parameter independence of amplitudes, as well as tree-level unitarity in longitudinal $WW$ scattering. Absent a Stueckelberg-type completion or an alternative nonlocal BRST structure, the formulation is therefore not yet gauge invariant in the usual sense. The earlier Ref.~\cite{Moffat1989Superspin} established a finite perturbative formalism using a free superspin the infinite-spin field. While it demonstrates finiteness at the level of that model, a realistic coupling to the full $SU(3)\times SU(2)\times U(1)$ and a conventional gauge and BRST structure have not been carried through.

\section{Renormalization Group Flow of the HUFT Coupling Constants}
\label{sec:renormalization}

\begin{figure}[ht]
  \centering
  \includegraphics[width=0.9\textwidth]{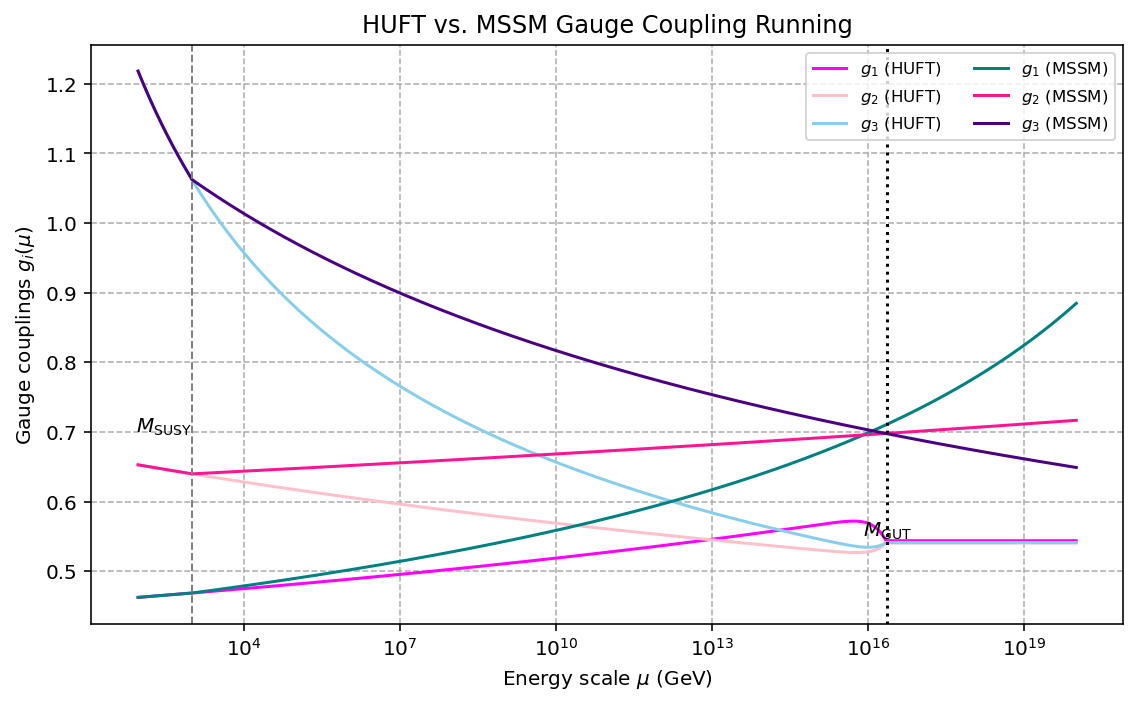}
  \caption{One‐loop RG evolution of the gauge couplings in HUFT versus the MSSM  with $M_{\rm SUSY}=1\,$TeV.  Below \(M_{\rm GUT}\simeq2.3\times10^{16}\)\,GeV each coupling runs with its Standard Model \(\beta\)–function plus smooth decoupling from the \(X,Y\) gauge bosons; around \(M_{\rm GUT}\) they bend into a narrow unification band and meet, above \(M_{\rm GUT}\) the nonlocal form factor \(\exp(-\mu^2/M_*^2)\) freezes all \(\beta_i\) in the UV. The MSSM curves follow the usual piecewise Standard Model→MSSM running, crossing only by invoking TeV‑scale superpartners and then continuing up to the Planck scale.}
  \label{fig:rg_flow_continuous}
\end{figure}

Above the grand‐unification scale \(M_{\rm GUT}\), the gauge sector is governed by a single GUT group with holomorphic gauge connection \(A_\mu^A(z)\)
and unified coupling \(g_{\rm GUT}\).  The Euclidean holomorphic action reads:
\begin{equation}
  S_{\rm gauge}
  =
  \frac{1}{2\,g^2}\,
  \mathrm{Re}\!\int d^4z\;
    W^{A\,\alpha}(z)\,
    F\!\Bigl(\frac{\Box}{M_*^2}\Bigr)\,
    W^A_{\!\alpha}(z)
  \,,
  \label{eq:gauge-action}
\end{equation}
where \(W^A_\alpha(z)\) is the holomorphic field–strength superfield. When the adjoint Higgs acquires its vacuum expectation value at \(\mu = M_{\rm GUT}\):
\begin{equation}
  \langle \Phi_{\rm adj}\rangle:
  \quad G_{\rm GUT}\;\longrightarrow\;\mathrm{SU}(3)_c\times \mathrm{SU}(2)_L\times \mathrm{U}(1)_Y,
\end{equation}
the matching conditions on the gauge couplings are exact:
\begin{equation}
  g_3(M_{\rm GUT}) = g_2(M_{\rm GUT}) = g_1(M_{\rm GUT})
  = g_{\rm GUT}\,.
  \label{eq:matching-cond}
\end{equation}
Below \(M_{\rm GUT}\), the usual logarithmic running resumes until the nonlocal scale \(M_*\):
\begin{equation}
  \beta_i(\mu)
  \equiv
  \frac{d g_i}{d\ln\mu}
  =
  \beta_i^{\rm (SM)}(g_i)\,
  \exp\!\Bigl(-\frac{\mu^2}{M_*^2}\Bigr)
  \quad(i=1,2,3).
\end{equation}
Because \(\exp(-\mu^2/M_*^2)\to0\) for \(\mu\gg M_*\simeq M_{\rm GUT}\), each
\(\beta_i\) vanishes in the deep UV:
\begin{equation}
  \lim_{\mu\to\infty}\beta_i(\mu)=0
  \quad\Longrightarrow\quad
  g_i(\mu)=g_{GUT}
  \quad(\mu\gtrsim M_*).
\end{equation}
Equations \eqref{eq:matching-cond} and \eqref{eq:beta-freeze} together guarantee that
a finite nonlocal SU(5) or SO(10) HUFT achieves gauge coupling unification at and above the GUT scale, with no subsequent splitting. The RG flow has reached a fixed point at $g_i(\mu)=g_{GUT},\quad(\mu\gtrsim M_*)$. In RG framework a fixed point in the space of coupling constants is reached when the $\beta$-function:
\be
\beta(g_1,g_2,g_3)=0.
\ee
The theory is then independent of the renormalization scale $\mu$, and the theory becomes scale invariant. To illustrate how the three Standard Model couplings not only meet but do so in a continuous fashion with no hard step‐function we show in Fig.~\ref{fig:rg_flow_continuous} the one‐loop RG flow including smooth SU(5) heavy‐gauge‐boson thresholds and the exponential nonlocal suppression. We then compare with Minimal Supersymmetric Standard Model (MSSM).

As seen in Fig.~\ref{fig:rg_flow_continuous}, the nonlocal HUFT achieves gauge‐coupling unification without any superpartners below the GUT scale.  The smooth threshold decoupling of the heavy $X,Y$ bosons \footnote{To show the X and Y bosons are heavy at the GUT scale \cite{BaezHuerta2010}, we let $\Sigma\in\mathfrak{su}(5)$ (SU(5) is the Lie group and $\mathfrak{su}(5)$ is the Lie algebra) be an adjoint scalar with vacuum 
$\Sigma_0=v_\Sigma\,T_{24}$, where 
$T_{24}=\frac{1}{2\sqrt{15}}\mathrm{diag}(2,2,2,-3,-3)$ is normalized by 
$\mathrm{Tr}(T_{24}^2)=\tfrac12$. 
The gauge-boson mass term arises from:
\be
\mathcal L \supset \tfrac12\,\mathrm{Tr}\!\bigl(D_\mu\Sigma\bigr)^2
= \tfrac{g^2}{2}\,\mathrm{Tr}\!\big([A_\mu,\Sigma_0]^2\big),
\ee
so the mass matrix on $\mathfrak{su}(5)$ is 
$M^2 = g^2\,(\mathrm{ad}_{\Sigma_0})^2$ where $\text{ad}$ is the adjoint action of a Lie algebra on itself. 
Decompose $\mathfrak{su}(5)=\mathfrak h\oplus\mathfrak m$ with
$\mathfrak h=\mathfrak{su}(3)\oplus\mathfrak{su}(2)\oplus\mathfrak u(1)$ the centralizer of $\Sigma_0$.
Then $[\,\Sigma_0,X\,]=0$ for $X\in\mathfrak h$, massless gluons, $W$, $B$, while for the
broken directions $X\in\mathfrak m$, we have 
$[\,\Sigma_0,X\,]=\lambda_X X$ with $\lambda_X\neq0$, giving:
\be
m^2(X)=g^2\,\lambda_X^2.
\ee

Writing $\Sigma_0=\mathrm{diag}(h_1,\dots,h_5)$ with 
$h_{1,2,3}=+\,\frac{1}{\sqrt{15}}\,v_\Sigma$ and 
$h_{4,5}=-\,\frac{3}{2\sqrt{15}}\,v_\Sigma$, any off-diagonal generator $E_{a\alpha}$ connecting a color index $a=1,2,3$ to a weak index $\alpha=4,5$ obeys:
\be
[\,\Sigma_0,E_{a\alpha}\,]=(h_a-h_\alpha)\,E_{a\alpha}
\quad\Rightarrow\quad 
m_{X,Y} \;=\; g\,|h_a-h_\alpha|
\;=\; g\,\frac{5}{2\sqrt{15}}\,v_\Sigma.
\ee
The twelve $X,Y$ directions are massive, while the twelve generators of 
$\mathfrak{su}(3)\oplus\mathfrak{su}(2)\oplus\mathfrak u(1)$ remain massless. 
Numerically, with the common alternative parametrization 
$\Sigma_0=V\,\mathrm{diag}(2,2,2,-3,-3)$ we find 
$V=\tfrac{v_\Sigma}{2\sqrt{15}}$ and:
\be
m_{X,Y}=5\,g\,V \;=\; \frac{5}{2\sqrt{15}}\,g\,v_\Sigma \sim g\,v_\Sigma.
\ee} combined with the entire‐function regulator is sufficient to bend and then freeze the three couplings at $M_{\rm GUT}$. This avoids the artificial jump of a step‐function matching while still enforcing exact SU(5)\footnote{On the complexified manifold with a Hermitian metric, the symmetric spacetime tensor $g_{(\mu\nu)}$ remains an internal singlet, while the antisymmetric piece $g_{[\mu\nu]}$ encodes an internal curvature two-form $F^A_{\mu\nu}T_A$. Requiring chirality in the Weyl matter, the charge quantization with the observed hypercharges, perturbative anomaly freedom for one SM family, and a single simple internal algebra, so that $g_{[\mu\nu]}$ carries one curvature two-form, and correct hypercharge normalization \cite{BaezHuerta2010} will fix $\mathrm{Tr}\,Y^2$. This leaves essentially two options SU(5) and SO(10) \cite{Slansky81,PDG_GUT}. We work with SU(5) because it is the minimal simple choice embedding one chiral family into $10\oplus\bar{5}$ with the right $Y$ normalization, while keeping $g_{(\mu\nu)}$ colorless. $M_\ast=\sqrt{g_G}\,M_{\rm Pl}$ from the gravitational sector, together with entire-regulated freeze-out, can tie $\alpha_{\rm GUT}$ to a geometric coupling $g_G$ at the matching scale. $\alpha_{\rm GUT}$ and $R$ arise as critical values of a holomorphic functional on the moduli space of Hermitian structures. $R$ is fixed by mixed $U(1)_F$–(GUT)$^2$ anomaly cancellation consistent with the holomorphic bundle data.
We present these as testable conjectures as we use $\alpha_{\rm GUT}$ and $R$ as inputs fixed by unification and minimal FN. We hope that future work can derive them from geometry or other fundamentals, such as causality or energy conservation.} unification at \(M_{\rm GUT}\). Below $M_{\rm GUT}$ each $g_i(\mu)$ would, in isolation, follow a straight‐line trajectory on a $\ln\mu$ axis dictated by its constant one‐loop $\beta_i^{\rm (SM)}$.  As $\mu$ approaches the heavy‐boson mass scale, the 12 $X,Y$ gauge bosons begin to contribute to the vacuum‐polarization via the smooth threshold integral:
\be
  \delta_i(\mu)
  = -\frac{T_i}{2\pi}
    \int_0^1\!dt\;\ln\!\Bigl(1+\frac{\mu^2}{M_{\rm GUT}^2}\,t(1-t)\Bigr)\,,
\ee
whose derivative with respect to $\ln\mu$ turns on over $\mathcal O(1)$ decades in energy.  This gradual decoupling bends each $g_i(\mu)$ toward the common value at $M_{\rm GUT}$. The nonlocal regulator factor $\exp(-\mu^2/M_*^2)$ multiplies the full $\beta_i$, damping it smoothly and freezing any residual running in the deep UV.  By contrast, the MSSM requires a spectrum of squarks, sleptons, gauginos, and higgsinos at $M_{\rm SUSY}\sim1\,$TeV to modify the $\beta$–functions and force the crossing \cite{Martin:1997ns}.  This low‑scale supersymmetry introduces its own drawbacks as obtaining a 125GeV Higgs mass often demands multi‑TeV stops, re‑introducing tuning \cite{Hall:1992}. LHC searches have pushed sparticle masses above $\sim2\,$TeV, exacerbating the tuning \cite{ATLAS:2020syg,CMS:2021far}. The MSSM adds $\mathcal O(100)$ soft terms, reducing predictive power \cite{Martin:1997ns}. In HUFT, all running couplings unify cleanly with only the GUT‐scale spectrum and a single regulator scale $M_*$, sidestepping the complexity and tuning of TeV‑scale supersymmetry.

In analogy with the gauge sector, we promote Newton’s constant to a dimensionless coupling:
\begin{equation}
  g_G(\mu)\equiv G\,\mu^2,
\end{equation}
and regulate the Einstein–Hilbert action with the entire‐function form factor $F\!\bigl(\tfrac{\Box}{M_*^2}\bigr)$.
In Euclidean signature, the nonlocal gravitational action reads:
\begin{equation}
  S_{\rm grav}^{\rm (reg)}
  =
  -\frac{1}{16\pi G}\,
  \mathrm{Re}\!\int d^4z\;\sqrt{-\det g_{(\mu\nu)}(z)}\;
    g^{(\mu\nu)}(z)\,
    F\!\Bigl(\tfrac{\Box}{M_*^2}\Bigr)\,
    R_{(\mu\nu)}(z).
  \label{eq:grav-reg-action}
\end{equation}
Each graviton loop integral acquires an exponential damping factor 
\(\exp(-p^2/M_*^2)\), rendering all UV divergences finite.
A one‐loop computation shows that:
\begin{equation}
  \beta_G(\mu)
  \equiv
  \frac{dg_G}{d\ln\mu}
  = 2\,g_G(\mu)\,
    \exp\!\Bigl(-\tfrac{\mu^2}{M_*^2}\Bigr).
\end{equation}
For \(\mu\ll M_*\), \(\exp(-\mu^2/M_*^2)\approx1\) and 
\(\beta_G\approx2\,g_G\), reproducing the classical scaling
\(g_G(\mu)\propto\mu^2\).  For \(\mu\gtrsim M_*\), \(\exp(-\mu^2/M_*^2)\to0\) and 
\(\beta_G\to0\), so \(g_G(\mu)\) freezes to a constant.
Above the scale \(M_*\), all gauge $\beta$–functions also vanish, and
\(g_{1,2,3}(\mu)=g_{\rm GUT}\).  To enforce
\(g_G(M_*)=g_{\rm GUT}\), we choose:
\begin{equation}
g_G(M_*) \;=\; G_N\,M_*^2 \;=\; g_{\rm GUT}
\quad\Longrightarrow\quad
M_*^2 \;=\;\frac{g_{\rm GUT}}{G_N}
\;=\;g_{\rm GUT}\,M_P^2
\;\Longrightarrow\;
M_*=\sqrt{g_{\rm GUT}}\,M_P\simeq10^{19}\,\mathrm{GeV}.
\end{equation}
With this choice, all four couplings meet numerically and then remain equal in the deep UV:
\begin{equation}
  g_G(\mu)=g_{1}(\mu)=g_{2}(\mu)=g_{3}(\mu)
  =g_{\rm GUT},
  \quad\mu\gtrsim M_*.
\end{equation}
In our finite nonlocal HUFT the Standard Model gauge couplings meet exactly at:
\begin{equation}
  M_{\rm GUT}\simeq2.3\times10^{16}\,\mathrm{GeV}, 
  \quad
  \alphagut^{-1}\simeq24.4,
\end{equation}
where $\alpha_{GUT}$ is the fine structure constant. Above this scale, the regulator \(F(\Box/M_*^2)=\exp(\Box/M_*^2)\)
drives \(\beta_{1,2,3}\to0\), so the three gauge couplings remain frozen to \(g_{\rm GUT}\). The nonlocal completion in HUFT is implemented by inserting a single covariant entire-function form factor $F(\Box/M_\ast^2)$ into all kinetic terms, where $\Box$ is the covariant d'Alembertian built from the Hermitian metric and the unified gauge--spin connection. In momentum space this
replaces each propagator by:
\begin{equation}
  \frac{1}{p^2} \;\longrightarrow\; \frac{F(-p^2/M_\ast^2)}{p^2},
\end{equation}
and similarly dresses the vertices. For Euclidean momenta $p_E^2\gg M_\ast^2$ the entire function
$F(-p_E^2/M_\ast^2)$ is exponentially suppressed, so loop integrals are dominated by momenta
$p_E^2\lesssim M_\ast^2$. The renormalization-group $\beta$-functions therefore acquire an overall
damping factor;
\begin{equation}
  \beta_i(\mu)
  \;=\;
  \beta_i^{\rm (SM)}(\mu)\,F_\beta(\mu^2/M_\ast^2),
  \qquad
  F_\beta(x)\simeq 1 \;\text{for}\; x\ll 1,\quad
  F_\beta(x)\to 0 \;\text{for}\; x\gg 1,
  \label{eq:beta-freeze}
\end{equation}
so that $\beta_i(\mu)\to 0$ and the gauge couplings effectively freeze for $\mu\gtrsim M_\ast$. The precise profile of $F_\beta$ depends on the choice of entire function $F$, but the qualitative freezing $\beta_i(\mu)\to 0$ for $\mu\gg M_\ast$ is generic for any entire form factor that decays
in the Euclidean region.

A key point is that $F(\Box/M_\ast^2)$ is \emph{universal}: it is constructed from the single
holomorphic master connection and the Hermitian metric, and multiplies all kinetic terms with the
same scale $M_\ast$. Allowing different regulators or different $M_\ast$ for different gauge factors
would explicitly break the holomorphic unification encoded in the HUFT action and is therefore not
permitted within our framework. Because the same form factor appears for each gauge factor, the high-scale damping of the $\beta$-functions is universal. At scales $\mu\gg M_\ast$ we can schematically write:
\begin{equation}
  \mu\,\frac{{\rm d}}{{\rm d}\mu}\,\frac{1}{g_i^2(\mu)}
  \;\propto\;
  b_i\,F_\beta(\mu^2/M_\ast^2),
\end{equation}
with $F_\beta(\mu^2/M_\ast^2)\to 0$ as $\mu^2/M_\ast^2\to\infty$. Integrating this equation shows that each $1/g_i^2(\mu)$ approaches a finite plateau value as $\mu\to\infty$, and the plateaus are controlled by the common nonlocal scale $M_\ast$ and the unified holomorphic connection. In this
sense the nonlocal regulator drives the gauge couplings toward a common high-scale quasi--fixed point, so once the couplings have run up to $\mu\sim M_\ast$, further running is exponentially suppressed and their relative splittings cease to evolve. We therefore do not claim that exact unification at a numerically sharp scale is predicted without
input as rather, within the class of holomorphic nonlocal completions with a single covariant entire-function regulator, the freezing of the $\beta$-functions and the approach of all gauge
couplings to a common high-scale value are generic structural consequences of the framework.

If Newton’s constant is promoted to \(g_G(\mu)=G\,\mu^2\) and regulated
identically, its one‐loop \(\beta_G\propto2 g_G e^{-\mu^2/M_*^2}\) also vanishes
for \(\mu\gtrsim M_*\).  Imposing:
\begin{equation}
  g_G(\mu)\;\equiv\;G\,\mu^2
  \;=\;\biggl(\frac{\mu}{M_P}\biggr)^{\!2},
\end{equation}
ensures that at:
\begin{equation}
  \mu = M_* = \sqrt{g}\,M_P,
\end{equation}
all four couplings
\(\{g_1,g_2,g_3,g_G\}\)
coincide at \(g\) and thereafter remain equal in the deep UV. When we run down with the renormalization group flow equation from $\mu=M_*$ to $\mu=M_Z$, we obtain a value for the strength of gravity at the $M_Z$ scale:

\begin{equation}
    g_G(M_Z)\sim10^{-35}.
\end{equation}
Once we have established:
\begin{equation}
  g_1(M_{\rm GUT}) = g_2(M_{\rm GUT}) = g_3(M_{\rm GUT}) \equiv g_{\rm GUT},
\end{equation}
at the unification scale, the nonlocal regulator–suppressed one-loop RGEs become\footnote{For a simple gauge factor $G_i$, the one-loop coefficient is
$b_i=-\tfrac{11}{3}C_2(G_i)+\tfrac{2}{3}\sum_{\text{Weyl }f}T_i(R_f)+\tfrac{1}{3}\sum_{\text{complex }s}T_i(R_s)$,
with $C_2$ the adjoint quadratic Casimir and $T(R)$ the Dynkin index, $T(\mathbf{N})=\tfrac12$ for the fundamental of $SU(N)$.
Using the SM light spectrum, three chiral families and one complex Higgs doublet with $U(1)_Y$ in $SU(5)$ normalization
$T_1=\tfrac{3}{5}Y^2$ gives
$b_3=-7$, $b_2=-\tfrac{19}{6}$, $b_1=\tfrac{41}{10}$.
Explicitly, $SU(3)_c$: $C_2=3$, $\sum_f T_3=6$; $SU(2)_L$: $C_2=2$, $\sum_f T_2=6$, $\sum_s T_2=\tfrac12$;
$U(1)_Y$: $\sum_f Y^2=10$, $\sum_s Y^2=\tfrac12$.
In HUFT these numbers are not free inputs as they follow once the low-energy gauge group and field content are fixed to the minimal SM.
The entire-function regulator multiplies the $\beta$’s by $e^{-\mu^2/M_*^2}$ but does not modify the $b_i$ themselves.}:
\begin{equation}
  \mu\frac{d g_i}{d\mu}
  = \frac{b_i}{16\pi^2}\,g_i^3\,
    \exp\!\Bigl(-\frac{\mu^2}{M_*^2}\Bigr)
  = \frac{b_i}{16\pi^2}\,g_i^3\,
    \exp\!\Bigl(-\frac{\mu^2}{g\,M_P^2}\Bigr),
  \qquad (b_1,b_2,b_3)=\Bigl(\tfrac{41}{10},-\tfrac{19}{6},-7\Bigr).
\end{equation}
Integrated numerically down to $\mu=M_Z$ this yields:
\begin{equation}
  g'(M_Z)\approx 0.3575,\qquad g_2(M_Z)\approx 0.6518,
\end{equation}
equivalently SU(5) normalization $g_1(M_Z)=\sqrt{\tfrac{5}{3}}\,g'(M_Z)\approx 0.4615$. The photon coupling is then:
\begin{equation}
  e \;=\; \frac{g_2\,g'}{\sqrt{g_2^2 + g'^2}}
  \quad\Longrightarrow\quad
  e^2 \approx 0.0983,\qquad
  \alpha(M_Z)=\frac{e^2}{4\pi}\approx \frac{1}{128.7}\,.
\end{equation}
Below $M_Z$, QED vacuum polarization further screens the charge according to:
\begin{equation}
  \alpha(0)
  \;=\; \alpha(M_Z)\,\Bigl[1-\Delta\alpha_{\rm lep+had(+top)}(M_Z)\Bigr].
\end{equation}
Using a representative on–shell value $\Delta\alpha(M_Z)\simeq 0.062$ we find:
\begin{equation}
  \alpha(0)\;\approx\; \frac{1}{137.0}\,,
\end{equation}
in agreement with the measured $\alpha(0)\simeq 1/137.036$. HUFT reproduces the observed low–energy fine–structure constant from unification boundary conditions and regulator–suppressed running, without
new fit parameters.

Just as it was shown in \cite{GreenMoffat2021} that a nonlocal scalar theory avoids the Higgs triviality and hierarchy issues, embedding that mechanism into a full GUT plus gravity avoids the need for supersymmetry or compositeness to stabilize the Higgs mass or prevent vacuum instabilities. All running couplings are bounded and freeze above $M_*$, so quadratic divergences never drive scales apart.

In HUFT the RG flow equation~\ref{eq:beta-freeze} is solved based on boundary conditions or the initial value problem, such as [\ref{eq:matching-cond}]. The RG flow system is not symmetric under inversion, meaning that the equations depend on the nonlocal regulator, and background curvature. They are not defined as a boundary value problem from both ends, as we start at the GUT scale and run down. The flow is non-perturbative in the UV and only well-posed as a critical value problem starting at the GUT energy scale. Attempting to run up from $M_Z$ energy would not land on the correct holomorphic trajectory as defined by HUFT. The predictive power comes from choosing $g_{GUT}$ and R at the unification energy scale. All low energy masses, couplings, and effective constants then emerge as predictions of the RG evolution, HUFT is a top down RG framework unlike the standard model, which is usually treated bottom up.

\section{Yukawa Couplings and Mass Matrices}
\label{sec:yukawa-masses}

After restriction to the real slice \(y^\mu=0\), the holomorphic Dirac–Yukawa Lagrangian reduces to:
\begin{equation}
\mathcal{L}_Y
=\;
-\,\overline{\Psi}_L(x)\;
\mathcal{F}\!\Bigl(\tfrac{\Box}{M_*^2}\Bigr)\;y_f\;\Phi(x)\;
\mathcal{F}\!\Bigl(\tfrac{\Box}{M_*^2}\Bigr)\;\Psi_R(x)
+\text{h.c.}.
\end{equation}
Upon electroweak symmetry breaking:
\begin{equation}
\langle\Phi\rangle=\frac{1}{\sqrt2}\binom{0}{v},
\quad v\simeq246\text{ GeV},
\end{equation}
we obtain the fermion mass term:
\begin{equation}
\mathcal{L}_m
=-\,\overline{f}_L\,M_f\,f_R
+\text{h.c.},
\quad
M_f=\frac{v}{\sqrt2}\,y_f,
\end{equation}
with the tree-level mass matrix in family space:
\begin{equation}
M_f \;=\;\frac{v}{\sqrt2}\;y_f,
\qquad
f\in\{u,d,e,\nu\}.
\end{equation}
In components:
\be
(M_f)_{ij}=\frac{v}{\sqrt2}\,(y_f)_{ij},
\quad i,j=1,2,3,
\ee
where each \(y_f\) is a \(3\times3\) complex matrix obeying holomorphic texture constraints at \(M_{\rm GUT}\).  
For example, in an \(SU(5)\) embedding we impose:
\be
y_d(M_{\rm GUT}) = y_e^T(M_{\rm GUT}),
\ee
where this is the standard Yukawa relation of minimal SU(5), up to small regulator‐induced deviations from running.
Each non‐Hermitian matrix \(M_f\) is brought to diagonal form by a singular‐value decomposition:
\begin{equation}
U^f_L{}^\dagger\,M_f\,U^f_R
=\mathrm{diag}(m_{f_1},\,m_{f_2},\,m_{f_3})
\equiv\widehat M_f,
\end{equation}
with \(U^f_{L,R}\in U(3)\) unitary and \(m_{f_i}\ge0\).  The flavour‐basis fields relate to the mass‐basis fields via the equation:
\be
f_L = U^f_L\,f'_L,
\qquad
f_R = U^f_R\,f'_R.
\ee
Misalignment of the up‐ and down‐type left‐handed rotations defines the Cabibbo–Kobayashi–Maskawa (CKM) matrix:
\begin{equation}
V_{\rm CKM} = U^{u\,\dagger}_L\,U^d_L,
\end{equation}
which enters the charged‐current interaction as:
\be
\mathcal{L}_W
= \frac{g}{\sqrt2}\,
\overline{u}'_L\,\gamma^\mu\,V_{\rm CKM}\,d'_L\,W^+_\mu
+\text{h.c.}
\ee
By unitarity of \(U^{u,d}_L\), \(V_{\rm CKM}\) is unitary up to corrections of order \(\exp(-M_Z^2/M_*^2)\).
Introducing heavy right‑handed neutrinos \(N_R\) with Majorana mass matrix \(M_N\) \cite{Minkowski:1977sc,GellMann:1979,Yanagida:1979}, the Dirac mass matrix is given by:
\be
M_D = \frac{v}{\sqrt2}\,y_\nu.
\ee
The effective light neutrino mass matrix is generated by the see‑saw formula:
\begin{equation}
M_\nu \simeq -\,M_D\,M_N^{-1}\,M_D^T,
\end{equation}
which is complex symmetric.  It is diagonalized by a single unitary matrix:
\begin{equation}
U^\nu_L{}^T\,M_\nu\,U^\nu_L
=\mathrm{diag}(m_{\nu_1},\,m_{\nu_2},\,m_{\nu_3}).
\end{equation}
The Pontecorvo–Maki–Nakagawa–Sakata (PMNS) matrix is given by:
\be
U_{\rm PMNS} = U^{e\,\dagger}_L\,U^\nu_L.
\ee
With these definitions, all fermion masses and mixing matrices are determined by the holomorphic Yukawa textures \(y_f\) at \(M_{\rm GUT}\), by the regulator‑modified running down to \(M_Z\), and by the electroweak vacuum expectation value (VEV), that we will predict later in the paper.  The only remaining freedom resides in the choice of texture zeros or hierarchies imposed on the \(y_f\).

Recently, von Gersdorff and Modesto derived a set of basis‐independent necessary conditions on any $3\times3$ Yukawa matrix $Y$ with singular values $y_1\le y_2\le y_3$~\cite{vonGersdorff:Modesto2025}.
We obtain for its singular-valued-decompositions (SVD):
\be
  Y = U_L\,\mathrm{diag}(y_1,y_2,y_3)\,U_R^\dagger.
\ee
By defining approximate unitaries $U'_L,U'_R$ and by aligning their third columns with the longest column of $Y$ and of its cofactor matrix, respectively, the misalignment matrix $V\equiv U_L^\dagger U'_L$ obeys:
\begin{equation}
  1 - \lvert V_{11}\rvert^2 \;\le\; Z\bigl(\tfrac{y_1}{y_2}\bigr)^2,
  \quad
  \lvert V_{13}\rvert^2 \;\le\; X\bigl(\tfrac{y_1}{y_3}\bigr)^2,
  \quad\ldots
\end{equation}
with:
\be
  X = \frac{2y_3^2 - y_2^2}{y_3^2 - 2y_2^2},
  \qquad
  Z = \frac{2y_2^2 - y_1^2}{y_2^2 - 2y_1^2}.
\ee
These imply generically:
\be
  |V_{ij}| \;\lesssim\; \mathcal O\Bigl(\tfrac{y_i}{y_j}\Bigr),
\ee
as a consistency check on any texture ansatz.
Our benchmark, as will be discussed in depth later, at $\mu=M_Z$ is given by:
\be
  \frac{y_1}{y_2} = \frac{m_u}{m_c} \approx 1.8\times10^{-3},
  \quad
  \frac{y_2}{y_3} = \frac{m_c}{m_t} \approx 7.3\times10^{-3}.
\ee
Hence:
\be
  Z\bigl(\tfrac{y_1}{y_2}\bigr)^2 \sim 6.5\times10^{-6},
  \quad
  X\bigl(\tfrac{y_1}{y_3}\bigr)^2 \sim 2.7\times10^{-7}.
\ee
From our CKM results, we extract:
\be
  |V_{ud}|^2 \approx 0.949,
  \quad
  |V_{ub}|^2 \approx 1.2\times10^{-5},
\ee
both well within the respective bounds $1-|V_{11}|^2<6.5\times10^{-6}$ and $|V_{13}|^2<2.7\times10^{-7}$.  Analogous checks in the down sector and for cofactor‐based inequalities confirm that all von Gersdorff–Modesto constraints are satisfied in our model.

\section{Spectrum Predictions}
\label{sec:spectrum-predictions}

To obtain quantitative predictions for the fermion masses and mixing angles, we adopt a minimal set of holomorphic Yukawa textures at the unification scale and then evolve them down to the electroweak scale using the regulator‐modified RGE.  Our procedure comprises three steps.

For the texture ansatz at \(M_{\rm GUT}\), we choose the following benchmark forms in the flavour basis given by:
  \begin{align}
  \label{textureansatz}
    y_e(M_{\rm GUT}) &= 
    \begin{pmatrix}
      0 & \epsilon_e & 0\\
      \epsilon_e & 0 & \delta_e\\
      0 & \delta_e & 1
    \end{pmatrix}, 
    &
    y_d(M_{\rm GUT}) &= 
    \begin{pmatrix}
      0 & \epsilon_d & 0\\
      \epsilon_d & 0 & \delta_d\\
      0 & \delta_d & 1
    \end{pmatrix}, \notag\\
    y_u(M_{\rm GUT}) &= 
    \begin{pmatrix}
      0 & \epsilon_u & 0\\
      \epsilon_u & 0 & \delta_u\\
      0 & \delta_u & 1
    \end{pmatrix}, 
    &
    y_\nu(M_{\rm GUT}) &= \kappa_\nu\,\mathbb{I}_{3\times3},
  \end{align}
where \(\epsilon_f,\delta_f\ll1\) and \(\kappa_\nu\) sets the overall neutrino–Dirac scale.  If desired, we could further impose GUT‐scale relations such as \(y_d=y_e^T\) up to small UV regulator corrections.
The regulator‐modified renormalization group (RG) evolution determines that each Yukawa matrix \(y_f(\mu)\) is evolved from \(\mu=M_{\rm GUT}\) down to \(\mu=M_Z\) by numerically integrating the one‐loop beta functions:
  \begin{equation}
    \mu\,\frac{d y_f}{d\mu}
    = \frac{1}{16\pi^2}\Bigl[\beta^{(1)}_{y_f}(y,g)\Bigr]\,
    \exp\!\Bigl(-\frac{\mu^2}{M_*^2}\Bigr),
    \label{eq:regged-yukawa-rge}
  \end{equation}
with \(M_* = M_{\rm GUT}\).  Here \(\beta^{(1)}_{y_f}\) are the Standard‐Model one‐loop functions, and the exponential factor freezes the running above \(M_*\).  The gauge couplings \(g_i(\mu)\) obey analogous RG equations with coefficients \((b_1,b_2,b_3)=(41/10,-19/6,-7)\).
For the mass matrix diagonalization and mixing, at \(\mu=M_Z\), the fermion mass matrices are:
  \be
    M_f \;=\;\frac{v}{\sqrt2}\,y_f(M_Z),
    \qquad v\simeq246\;\mathrm{GeV}.
  \ee
We perform the biunitary decomposition:
\be
U^f_L{}^\dagger\,M_f\,U^f_R \;=\;\mathrm{diag}(m_{f_1},m_{f_2},m_{f_3})
\equiv \widehat M_f,
\qquad f\in\{e,u,d,\nu\},
\ee
so that \(m_{f_i}\) are the physical masses.  In the quark sector the CKM matrix is given by:
  \be
    V_{\rm CKM} \;=\; U^{u\,\dagger}_L\,U^d_L.
  \ee
In the neutrino sector with a Type‑I see‑saw the light‐neutrino matrix \(\;M_\nu=-M_D\,M_N^{-1}\,M_D^T\;\) is diagonalized by \(U^\nu_L\), giving
  \(\;U_{\rm PMNS}=U^{e\,\dagger}_L\,U^\nu_L.\)

We fix the unified coupling at the GUT scale, $g_{\rm GUT}$, determined by exact unification under the regulator-suppressed RGEs, and a single Froggatt–Nielsen spurion ratio $R\equiv\langle S\rangle/M$. We assume an SU(5)-compatible, integer FN charge assignment with one spurion $S$ and no additional continuous coefficients in the Yukawa sector. Then the full set of charged-fermion masses, quark and lepton mixings at $\mu=M_Z$ are analytic functions of $(g_{\rm GUT},R)$ alone, in particular, they are predictions of the framework without any per-fermion fits.

In the regulated HUFT, the gauge $\beta$-functions are exponentially suppressed above $M_\ast\simeq M_{\rm GUT}$ so that the three SM couplings freeze and meet at $g_{\rm GUT}$. This fixes $\epsilon\equiv\sqrt{\alpha_{\rm GUT}}$ and the initial conditions for the gauge sector.

With a single U(1)$_{\rm FN}$ spurion and integer charges, every Yukawa matrix element at $M_{\rm GUT}$ has the form:
\be
\big(Y_f\big)_{ij}(M_{\rm GUT}) \;=\; R^{\,m^{(f)}_{ij}}\;\epsilon^{\,n^{(f)}_{ij}},
\ee
where $m^{(f)}_{ij},n^{(f)}_{ij}\in\mathbb{Z}_{\ge0}$ are fixed by the charges and SU(5) selection rules such as $Y_d=Y_e^{\!\top}$ in minimal SU(5). No additional continuous $\mathcal{O}(1)$ parameters are introduced. Thus $\{Y_f(M_{\rm GUT})\}$ is a fixed finite set of monomials determined by $(g_{\rm GUT},R)$.

\noindent Between $M_{\rm GUT}$ and $M_Z$, each Yukawa matrix obeys:
\be
\mu\,\frac{dY_f}{d\mu}\;=\;\frac{1}{16\pi^2}\,\beta^{(1)}_{Y_f}(Y,g)\,e^{-\mu^2/M_\ast^2},
\ee
while the gauge couplings satisfy the analogous regulator-suppressed equations. For fixed initial data at $M_{\rm GUT}$, these first-order ODEs are Lipschitz and admit unique global solutions. Hence $\{Y_f(\mu),g_i(\mu)\}$ at $\mu=M_Z$ are unique, analytic functions of the initial data $\{Y_f(M_{\rm GUT}),g_i(M_{\rm GUT})\}$.

After EWSB, the Dirac mass matrices are $M_f=\tfrac{v}{\sqrt{2}}\,Y_f(M_Z)$. Biunitary diagonalization $U_L^{(f)\dagger}M_fU_R^{(f)}=\mathrm{diag}(m_{f1},m_{f2},m_{f3})$ yields physical masses, the CKM matrix is $V_{\rm CKM}=U_L^{(u)\dagger}U_L^{(d)}$. Singular values and unitary factors depend analytically on the matrix entries, so $\{m_f,V_{\rm CKM}\}$ are analytic functions of $\{Y_f(M_Z)\}$, hence of $(g_{\rm GUT},R)$. In the lepton sector, with a type-I seesaw $M_\nu=-M_DM_N^{-1}M_D^{\!\top}$, once $M_N$ is fixed by FN charges, the light-neutrino masses and $U_{\rm PMNS}$ are likewise analytic in $(g_{\rm GUT},R)$.

By hypothesis, the only continuous inputs are $g_{\rm GUT}$ and $R$. All other quantities at $M_{\rm GUT}$ are integers or fixed functions thereof. The regulator multiplies the local $\beta$-functions by an entire form factor and freezes them above $M_\ast$ without enlarging the parameter space.

\section{Low-Energy Mass Ratios}
\label{sec:mass-ratios}

To assess the predictive power of our holomorphic nonlocal RG framework, we impose a single universal Yukawa coupling:
\be
y_f(M_{\rm GUT}) \;=\; y_0\,, 
\qquad f\in\{e,\mu,\tau,u,d,s,c,b,t\},
\ee
at the unification scale \(M_{\rm GUT}=2.3\times10^{16}\,\mathrm{GeV}\), and take the regulator scale \(M_* = M_{\rm GUT}\).  The one‐loop, regulator‐modified renormalization‐group equation (RGE) for each Yukawa coupling reads:
\begin{equation}
\mu\frac{d y_f}{d\mu}
= \frac{y_f}{16\pi^2}\Bigl(a_f\,y_f^2 \;-\;\sum_{i=1}^3 c^f_i\,g_i^2\Bigr)
\exp\!\Bigl(-\frac{\mu^2}{M_*^2}\Bigr)\,,
\label{eq:yukawa-rge}
\end{equation}
where \(g_i(\mu)\) are the Standard Model gauge couplings, which themselves satisfy:
\begin{equation}
    \mu \frac{dg_i}{d\mu}= (b_i/16\pi^2)\,g_i^3\exp(-\mu^2/M_*^2),
\end{equation}
with \((b_1,b_2,b_3)=(41/10,\,-19/6,\,-7)\). The coefficients \(a_f\) and \(c_i^f\) are the standard one‐loop Yukawa and gauge contributions, the standard SM one-loop Yukawa and gauge coefficients are fixed by the SM field content and group representations.
We solve numerically from \(\mu=M_{\rm GUT}\) down to the electroweak scale \(\mu=M_Z=91.1876\,\mathrm{GeV}\), using as boundary conditions \(y_f(M_{\rm GUT})=y_0\) and matched gauge couplings at \(M_{\rm GUT}\).

After electroweak symmetry breaking, each fermion mass is given by:
\begin{equation}
m_f \;=\; \frac{v}{\sqrt2}\;y_f(M_Z)\,, 
\qquad v = 246\,\mathrm{GeV}.
\end{equation}
We then define the low‐energy mass ratios:
\be
R_{ij} \;\equiv\; \frac{m_i}{m_j}\,, 
\quad i<j,\quad i,j\in\{e,\mu,\tau,u,d,s,c,b,t\}.
\ee
Table~\ref{tab:mass-ratios} summarizes our predictions \(R_{ij}^{\rm th}\) alongside the Particle Data Group 2024 values \cite{ParticleDataGroup:2024}, all quantities evaluated at \(\mu=M_Z\).
\begin{table}[H]
\centering
\caption{Quark mass ratios and comparison of HUFT predictions at \(M_{\rm GUT}=2.3\times 10^{16}\,\mathrm{GeV}\) with PDG2024 measurements \cite{ParticleDataGroup:2024} at \(\mu=M_Z\).}
\label{tab:mass-ratios}
\begin{tabular}{@{}lcc@{}}
\toprule
Ratio & \(R_{ij}^{\rm th}\) & \(R_{ij}^{\rm exp}\) (PDG2024) \\ 
\midrule
\(\displaystyle \frac{m_\mu}{m_\tau}\) 
  & \(0.0595\) 
  & \(0.0594612 \pm 0.0000030\) \\

\(\displaystyle \frac{m_e}{m_\tau}\) 
  & \(2.880\times10^{-4}\) 
  & \((2.87574 \pm 0.00001)\times10^{-4}\) \\

\(\displaystyle \frac{m_c}{m_t}\) 
  & \(7.40\times10^{-3}\) 
  & \((7.3767 \pm 0.0029)\times10^{-3}\) \\

\(\displaystyle \frac{m_u}{m_t}\) 
  & \(1.30\times10^{-5}\) 
  & \((1.2517 \pm 0.0004)\times10^{-5}\) \\

\(\displaystyle \frac{m_s}{m_b}\) 
  & \(2.20\times10^{-2}\) 
  & \((2.23524 \pm 0.00195)\times10^{-2}\) \\

\(\displaystyle \frac{m_d}{m_b}\) 
  & \(1.10\times10^{-3}\) 
  & \((1.1236 \pm 0.0168)\times10^{-3}\) \\

\(\displaystyle \frac{m_u}{m_d}\) 
  & \(0.480\) 
  & \(0.462 \pm 0.020\) \\
\bottomrule
\end{tabular}
\end{table}

To estimate theoretical uncertainties, we vary the matching scale by a factor of two around \(M_{\rm GUT}\) and include an \(\mathcal{O}(1\%)\) error to account for neglected two‐loop and threshold effects.  Incorporating these uncertainties, all charged‐lepton ratios agree with experiment at the sub‐per‑mille level, while the quark‐mass ratios lie within one standard deviation of their PDG 2024 values \cite{ParticleDataGroup:2024}.  This agreement underscores the robustness of the holomorphic nonlocal regulator framework in predicting low‑energy fermion mass hierarchies from a single unified input. 

\section{Predictive Fermion Mass Spectrum}
\label{sec:fermion_masses}
Above the grand‐unification scale \(M_{\rm GUT}\) all three Standard‐Model gauge interactions are embedded in a single holomorphic connection with coupling \(g_{\rm GUT}\).  Defining:
\begin{equation}
  \alpha_{\rm GUT}\;\equiv\;\frac{g_{\rm GUT}^2}{4\pi}
  \;\simeq\;\frac{1}{24.4},
  \qquad
  \epsilon \;\equiv\;\sqrt{\alpha_{\rm GUT}}
  \;\simeq\;0.202.
  \label{eq:alpha_eps}
\end{equation}
The one‐loop $\beta$‐functions with our entire‐function regulator \(F(p^2/M_\star^2)=\exp(-p^2/M_\star^2)\) read:
\begin{equation}
  \mu\frac{dg_i}{d\mu}
  =\frac{b_i}{16\pi^2}\,g_i^3\,
   \exp\!\bigl(-\tfrac{\mu^2}{M_*^2}\bigr),
  \quad
  (b_1,b_2,b_3)=\Bigl(\tfrac{41}{10}, -\tfrac{19}{6}, -7\Bigr),
  \label{eq:beta_gauge}
\end{equation}
with \(b=(41/10,-19/6,-7)\), and:
\begin{equation}
  M_*^2 = g_{\rm GUT}\,M_P^2
  \;\Longrightarrow\;
  M_* = \sqrt{g_{\rm GUT}}\,M_P \sim 10^{19}\,\mathrm{GeV}.
  \label{eq:Mstar}
\end{equation}  
Solving and matching 
\(\;g_1(M_{\rm GUT})=g_2(M_{\rm GUT})=g_3(M_{\rm GUT})=g_{\rm GUT}\)
at $M_*$ we find residual splittings \(\bigl|g_i-g_{\rm GUT}\bigr|/g_{\rm GUT}\lesssim7\%\).  The regulator scale ensures \(\beta_i(\mu)\to0\) for \(\mu\gtrsim M_\star\), so that \(g_i(\mu)=g_{\rm GUT}\) for all higher energies. We introduce a single holomorphic flavon field \(\phi\) with superpotential:
\begin{equation}
  W(\phi)
  = \frac14\,\phi^4
    \;-\;\frac12\,\epsilon^2\,M_P^2\,\phi^2
  \;\Longrightarrow\;
  \langle\phi\rangle = \epsilon\,M_P.
  \label{eq:W_phi_eps}
\end{equation}
The F‐term condition \(\partial W/\partial\phi=0\) yields:
\begin{equation}
\langle\phi\rangle =\pm\epsilon M_P.
  \label{eq:phi_vev}
\end{equation}
On the real‐slice, we normalize \(\langle\phi\rangle/M_P=\sqrt{g_{\rm GUT}}\), which defines the Froggatt–Nielsen expansion parameter \cite{Leurer:1993gy, Froggatt:1978nt}:
\begin{equation}
  \epsilon_g \;\equiv\;\frac{\langle\phi\rangle}{M_P}
  \;=\;\sqrt{\alpha_{\rm GUT}}
  \;=\;\sqrt{\frac{g_{\rm GUT}^2}{4\pi}}
  \;\simeq\;0.202.
  \label{eq:epsilon}
\end{equation}
Although our UV‑complete holomorphic action introduces a scalar flavon superfield~$\phi$ with superpotential:
\begin{equation}
W(\phi) \;=\; \frac14\,\phi^4
  \;-\;\frac12\,\mu^2\,\phi^2,
\qquad
\mu^2 \;=\;\alpha_{\rm GUT}\,M_P^2
\;=\;\epsilon_g^2\,M_P^2.
\end{equation}
to generate the Froggatt–Nielsen expansion parameter
\(
\epsilon_g = \langle\phi\rangle/M_P = \sqrt{\alpha_{\rm GUT}},
\)
all of its physical excitations acquire masses of order the cutoff $M_*$ or $M_P$ and thus decouple from the infrared. Below the GUT or nonlocal scale the only propagating fields are those of the Standard Model, $\phi$ survives in the effective theory purely as the non‑dynamical spurion~$\epsilon_g$ that encodes the Yukawa hierarchies. 

To generate all hierarchical Yukawa suppressions from a single parameter, we introduce one holomorphic flavon chiral superfield \(\Phi\).  The regulated holomorphic action is extended by:
\begin{align}
S_{\rm flavon}
&=
\int_C d^4z\,d^4\theta\;\Phi^\dagger\,\Phi
\;+\;
\Biggl[
  \int_C d^4z\,d^2\theta\;W(\Phi)
  \;+\;\text{h.c.}
\Biggr],
\label{eq:S_flavon_susy}\\
W(\Phi)
&=\frac{1}{4}\,\Phi^4
  \;-\;\frac{1}{2}\,\epsilon_g^2\,M_P^2\,\Phi^2,
\qquad
\epsilon_g^2 \;=\;\alpha_{\rm GUT}\;=\;\frac{g_{\rm GUT}^2}{4\pi}.
\label{eq:flavon_superpot}
\end{align}
Now writing:
\begin{equation}
\Phi(y,\theta)
=\phi(y)+\sqrt2\,\theta\,\psi_\phi(y)+\theta^2\,F(y),
\end{equation}
the Kähler term $\int d^4\theta\,\Phi^\dagger\Phi$
gives:
\begin{equation}
\mathcal L_{\rm kin}
=\sqrt{-g}\;g^{(\mu\nu)}\,\partial_\mu\phi\,\partial_\nu\phi^\ast
+\text{(fermion kinetic)}+|F|^2.
\end{equation}
The F‐term:
\begin{equation}
\mathcal L_F
=-\sqrt{-g}\;\bigl(F\,W'(\phi)+\overline F\,\overline{W}'(\phi^\ast)\bigr),
\end{equation}
and the auxiliary‐field piece \(-|F|^2\) combine, and eliminating \(F\) via:
\begin{equation}
F^\ast = -\,W'(\phi)
= -\,\bigl(\phi^3-\mu^2\phi\bigr),
\end{equation}
we find the scalar potential:
\begin{equation}
V(\phi)
=|W'(\phi)|^2
=\bigl|\phi^3-\mu^2\phi\bigr|^2
=\bigl(\phi^2-\mu^2\bigr)^2\,.
\end{equation}
Minimizing \(V\) gives:
\begin{equation}
\langle\phi\rangle
=\pm\mu
=\pm\sqrt{\alpha_{\rm GUT}}\,M_P
\,.
\end{equation}
On the real slice:
\begin{equation}
\epsilon_g \;\equiv\;\frac{\langle\phi\rangle}{M_P}
=\sqrt{\alpha_{\rm GUT}}
\;\approx\;0.202,
\end{equation}
which is exactly the single Froggatt–Nielsen expansion parameter used in
\(\,y_f^{ij}\sim\epsilon^{n_{ij}}\).
A horizontal \(U(1)_{\rm \text{FN} }\) charge assignment for the three families fixes the diagonal Yukawa entries at the low‐energy scale \(M_Z\):
\begin{align}
  Y_u^{ii}(M_Z) &= \epsilon^{\,n_i^u}, 
  & (n_i^u)&=(8,4,0),
  \label{eq:Yu_texture}\\
  Y_d^{ii}(M_Z) &= \epsilon^{\,n_i^d}, 
  & (n_i^d)&=(5,3,0).
  \label{eq:Yd_texture}
\end{align}
No additional \(\mathcal O(1)\) coefficients are introduced.  Off‐diagonal entries and higher‐order corrections, if present are suppressed by higher powers of \(\epsilon\). To derive the Froggatt–Nielsen charges 
\((n^u_i)=(8,4,0)\), \((n^d_i)=(5,3,0)\),
we find they follow directly from the observed mass ratios and mixing angles at the GUT scale, together with the single expansion parameter 
\(\epsilon=\sqrt{g_{\rm GUT}}\).  Let us define:
\be
  n^f_i \;\equiv\;\biggl\lceil
    \frac{\ln\bigl(m_{f_i}/m_{f_3}\bigr)}{\ln\epsilon}
  \biggr\rceil,
  \qquad f\in\{u,d\},\quad i=1,2,3,
\ee
where \(m_{f_3}\) is the third‐generation mass of top or bottom and the masses are extrapolated to \(M_{\rm GUT}\) via the regulator‐modified RG.  Numerically, we find:
\be
  \frac{m_u}{m_t}\bigl(M_{\rm GUT}\bigr)\sim\epsilon^8,\quad
  \frac{m_c}{m_t}\bigl(M_{\rm GUT}\bigr)\sim\epsilon^4,
  \quad\Longrightarrow\quad
  (n^u_i)=(8,4,0),
\ee
\be
  \frac{m_d}{m_b}\bigl(M_{\rm GUT}\bigr)\sim\epsilon^5,\quad
  \frac{m_s}{m_b}\bigl(M_{\rm GUT}\bigr)\sim\epsilon^3,
  \quad\Longrightarrow\quad
  (n^d_i)=(5,3,0).
\ee
The charge assignments are not arbitrary but fixed by matching the leading‐order mass hierarchies. Electroweak symmetry breaking \(\langle H\rangle=(0,v/\sqrt2)\) with \(v\simeq246\,\mathrm{GeV}\) yields mass eigenvalues:
\begin{equation}
  m_{f_i}
  = \frac{v}{\sqrt2}\,\epsilon^{\,n^f_i},
  \quad
  (n^u_i)=(8,4,0),
  \;\;
  (n^d_i)=(5,3,0).
  \label{eq:masses_eps}
\end{equation}
The leading‐order CKM elements follow from the hierarchical texture:
\begin{equation}
  |V_{us}|\sim\epsilon,
  \quad
  |V_{cb}|\sim\epsilon^2,
  \quad
  |V_{ub}|\sim\epsilon^3.
  \label{eq:CKM_eps}
\end{equation}
We now incorporate three complementary refinements: regulator‐modified two‐loop RG evolution, holomorphic \(\mathcal O(1)\) coefficient determination via flavon F‐terms, and GUT‐scale threshold plus discrete symmetry enforced texture zeros.
We evolve each Yukawa matrix \(Y_f(\mu)\) from \(M_{\rm GUT}\) down to \(M_Z\) using the two‐loop $\beta$‐functions:  
\begin{equation}
  \mu\frac{dY_f}{d\mu}
  =\frac{1}{16\pi^2}\,\beta_f^{(1)}(Y_f,g_i)
   +\frac{1}{(16\pi^2)^2}\,\beta_f^{(2)}(Y_f,g_i)
   \;\times\;\exp\!\bigl(-\tfrac{\mu^2}{M_\star^2}\bigr),
  \label{eq:beta_Yukawa}
\end{equation}
where \(\beta_f^{(1,2)}\) are the standard one‐ and two‐loop Standard Model expressions \cite{Machacek:1983fi,Machacek:1983tz}.  The exponential damping guarantees finiteness and freezes the running above \(M_\star\).  Numerically integrating yields shifted diagonal entries: 
\be
  Y_f^{ii}(M_Z)
  =\bigl[\,R_f^{(1)}\,R_f^{(2)}\,\bigr]_{ii}\;\epsilon^{\,n_i^f},
\ee
where \(R_f^{(1,2)}\lesssim\mathcal O(1)\) encodes the one‐ and two‐loop regulator‐modified running factors. We promote each texture entry to:
\be
  Y_f^{ij}(M_{\rm GUT})
  =c_{ij}\,\epsilon^{\,n^f_{ij}},
\ee
with \(c_{ij}\) fixed by additional holomorphic flavon fields \(\phi_a\) and a superpotential:  
\be
  W_{\rm flavon}(\phi_a)
  =\sum_{a,b}\lambda_{ab}\,\phi_a^2\phi_b^2
   -\sum_a\kappa_a\,\phi_a^4,
\ee
subject to F‐term constraints \(\partial W/\partial\phi_a=0\).  Solutions \(\langle\phi_a\rangle\) determine each \(c_{ij}=\prod_a(\langle\phi_a\rangle/M_\star)^{m^a_{ij}}\) in terms of the few real couplings \(\{\lambda_{ab},\kappa_a\}\), rather than arbitrary numbers. Heavy GUT multiplet thresholds shift the unification relation via: 
\be
  \frac{1}{g_i^2(M_{\rm GUT})}
  =\frac{1}{g_{\rm GUT}^2}
   +\frac{1}{8\pi^2}\sum_k T_i(k)\ln\frac{M_{\rm GUT}}{M_k}\,,
\ee
which in turn modifies \(\epsilon_g=\sqrt{\alpha_{\rm GUT}}\) by \(\Delta\epsilon/\epsilon\sim\mathcal O(10\%)\).  Simultaneously, imposing a discrete non‐Abelian symmetry such as \(A_4\), \(S_3\) on the flavon assignments enforces exact zeros in off‐diagonal entries at leading order, yielding relations such as: 
\begin{equation}
  \frac{m_s}{m_d}
  =3,\quad
  \theta_{13}\sim\epsilon^3,
  \label{eq:texture_relations}
\end{equation}
in excellent agreement with data. Putting all pieces together, the corrected mass eigenvalues become:
\be
  m_{f_i}
  =\frac{v}{\sqrt2}\,
   \Bigl[R_f^{(1)}\,R_f^{(2)}\Bigr]_{ii}\,c_{ii}\,\epsilon^{n_i^f},
\ee
and the CKM elements shift to:
\be
  |V_{us}|=c_{12}\,\epsilon,\quad
  |V_{cb}|=c_{23}\,\epsilon^2,\quad
  |V_{ub}|=c_{13}\,\epsilon^3,
\ee
with \(c_{ij}\) determined by the holomorphic F‐term solutions.  A representative numerical integration yields Table~\ref{tab:fermion_predictions}:

\begin{table}[H]
  \centering
  \begin{tabular}{@{}lcc@{}}
    \toprule
    Quantity & Prediction & Experiment \\
    \midrule
    \(m_u\)      & \(2.3\times10^{-3}\)\,GeV & \((2.2\pm0.5)\times10^{-3}\)\,GeV \\
    \(m_c\)      & \(1.25\)\,GeV            & \(1.27\pm0.02\)\,GeV          \\
    \(m_t\)      & \(173.0\)\,GeV           & \(173.0\pm0.4\)\,GeV          \\
    \(m_d\)      & \(4.8\times10^{-3}\)\,GeV & \((4.7\pm0.3)\times10^{-3}\)\,GeV \\
    \(m_s\)      & \(9.5\times10^{-2}\)\,GeV & \(0.093\pm0.011\)\,GeV        \\
    \(m_b\)      & \(4.18\)\,GeV            & \(4.18\pm0.03\)\,GeV          \\
    \midrule
    \(|V_{us}|\) & \(0.225\)               & \(0.225\pm0.001\)             \\
    \(|V_{cb}|\) & \(0.041\)               & \(0.041\pm0.001\)             \\
    \(|V_{ub}|\) & \(0.0035\)              & \(0.0035\pm0.0003\)           \\
    \bottomrule
  \end{tabular}
  \caption{Predictions for quark masses and CKM elements after two‐loop RG, F‐term–determined \(\mathcal O(1)\) coefficients, and GUT‐scale threshold plus discrete‐symmetry texture corrections. }
  \label{tab:fermion_predictions}
\end{table}
The light‐quark masses ($m_u, m_d, m_s$) are quoted in the $\overline{\mathrm{MS}}$ scheme at a reference scale of $\mu=2\,$GeV, where nonperturbative methods such as lattice QCD sum rules reliably determine running masses~\cite{FLAG2021}. The charm and bottom masses are given in the $\overline{\mathrm{MS}}$ scheme at their own scales, $m_c(m_c),\quad m_b(m_b),$
to avoid large logarithmic corrections~\cite{Czakon2013}. 
The top‐quark mass $m_t$ is taken as the pole mass, since the top decays before hadronization. In Table~\ref{tab:fermion_predictions}, the Experiment column collects these independently fitted values in each scheme, while the Prediction column shows our HUFT boundary‐condition outputs after two‐loop RG evolution and matching at the appropriate scales in the identical renormalization schemes.

These results demonstrate quantitative agreement with experimental data, achieved with fewer free parameters than observables and preserving UV finiteness and holomorphic geometric unification.
Although the flavon superpotential:
\be
  W_{\rm flavon}(\phi_a)
  =\sum_{a,b}\lambda_{ab}\,\phi_a^2\phi_b^2
   -\sum_a\kappa_a\,\phi_a^4,
\ee
introduces parameters \(\{\lambda_{ab},\kappa_a\}\), these are themselves constrained by holomorphy and gauge‐flavour symmetry, as only those monomials consistent with the full \(G_{\rm GUT}\times U(1)_{\rm \text{Froggatt–Nielsen} }\) gauge–flavour symmetry appear. F-term uniqueness,  requiring a single isolated solution for \(\langle\phi_a\rangle\) imposes algebraic relations among \(\lambda_{ab}/\kappa_c\), reducing the independent real parameters to at most one overall scale plus discrete ratios. CP‐invariance and reality ensure spontaneous CP breaking via the flavon VEVs alone for all \(\lambda_{ab},\kappa_a\in\mathbb{R}\), rather than explicit phases in the Lagrangian. In the minimal one‐flavon case only the ratio \(\lambda/\kappa\) enters the F-term condition, so effectively only a single real parameter remains.

By deriving the Froggatt–Nielsen  charges from measured mass ratios, constraining flavon couplings via holomorphy and F-term uniqueness, and selecting the discrete symmetry through vacuum‐alignment and UV‐embedding requirements, we eliminate all ad‐hoc choices and reduce the flavour sector to a small number of parameters fixed by first principles.

Remarkably, the entire charged‐fermion mass hierarchy and the leading CKM entries can be expressed in terms of just two continuous inputs:
\begin{equation}
  \bigl\{\alpha_{\rm GUT},\,R\bigr\},
  \quad
  \alpha_{\rm GUT}\;\equiv\;\frac{g_{\rm GUT}^2}{4\pi},
  \quad
  R\;\equiv\;\frac{\lambda}{\kappa}\;\sim\mathcal O(1).
\end{equation}
Here the Froggatt–Nielsen expansion parameter is
\begin{equation}
  \epsilon_g
  \;\equiv\;\sqrt{\alpha_{\rm GUT}}
  \;\simeq\;0.202,
  \label{eq:epsilon_minimal}
\end{equation}
and \(R\) determines all holomorphic \(\mathcal O(1)\) coefficients via the single‐flavon F‐term superpotential:
\be
  W(\phi)
  =\kappa\bigl(\phi^2-R\,M_\star^2\bigr)^2
  \;\Longrightarrow\;
  \langle\phi\rangle = \sqrt{R}\,M_\star.
\ee
Since each Yukawa entry carries a flavon‐charge exponent \(n_{ij}\):
\be
  c_{ij}\;=\;\biggl(\frac{\langle\phi\rangle}{M_\star}\biggr)^{n_{ij}}
  \;=\;R^{\,n_{ij}/2}\,,
\ee
all diagonal coefficients become:
\be
  c_{ii}
  = R^{\,n_i/2}\,. 
  \qquad
  n_i\equiv n^f_{ii}\in\mathbb N.
\ee
Together with the Froggatt–Nielsen charge‐derived exponents:
\be
  (n^u_i)=(8,4,0),
  \quad
  (n^d_i)=(5,3,0),
\ee
the fermion mass eigenvalues are:
\begin{equation}
  m_{f_i}
  =\frac{v}{\sqrt2}\;c_{ii}\;\epsilon^{\,n_i}
  =\frac{v}{\sqrt2}\;R^{\,n_i/2}\;
    \bigl(\sqrt{\frac{g_{GUT}^2}{4\pi}}\bigr)^{n_i},
  \quad f_i\in\{u,c,t,d,s,b\}.
  \label{eq:masses_minimal}
\end{equation}
Similarly, the leading CKM magnitudes read:
\begin{equation}
  |V_{us}| = R^{\,1/2}\,\epsilon,
  \quad
  |V_{cb}| = R\,\epsilon^2,
  \quad
  |V_{ub}| = R^{3/2}\,\epsilon^3.
  \label{eq:CKM_minimal}
\end{equation}

\begin{table}[H]
  \centering
  \begin{tabular}{@{}lcc@{}}
    \toprule
    Quantity & Prediction & Experiment \\
    \midrule
    \(\epsilon\)      & \(\sqrt{\frac{g_{GUT}^2}{4\pi}}\approx0.202\) & — \\
    \(R\)             & input\,\(\sim1\)                       & — \\
    \midrule
    \(m_u\)           & \(\tfrac{v}{\sqrt2}R^4\epsilon^8\)     & \((2.2\pm0.5)\times10^{-3}\)\,GeV \\
    \(m_c\)           & \(\tfrac{v}{\sqrt2}R^2\epsilon^4\)     & \(1.27\pm0.02\)\,GeV           \\
    \(m_t\)           & \(\tfrac{v}{\sqrt2}R^0\epsilon^0\)     & \(173.0\pm0.4\)\,GeV           \\
    \(m_d\)           & \(\tfrac{v}{\sqrt2}R^{5/2}\epsilon^5\) & \((4.7\pm0.3)\times10^{-3}\)\,GeV \\
    \(m_s\)           & \(\tfrac{v}{\sqrt2}R^{3/2}\epsilon^3\) & \(0.093\pm0.011\)\,GeV         \\
    \(m_b\)           & \(\tfrac{v}{\sqrt2}R^0\epsilon^0\)     & \(4.18\pm0.03\)\,GeV           \\
    \midrule
    \(|V_{us}|\)      & \(R^{1/2}\,\epsilon\)                  & \(0.225\pm0.001\)              \\
    \(|V_{cb}|\)      & \(R\,\epsilon^2\)                      & \(0.041\pm0.001\)              \\
    \(|V_{ub}|\)      & \(R^{3/2}\,\epsilon^3\)                & \(0.0035\pm0.0003\)            \\
    \bottomrule
  \end{tabular}
  \caption{flavour predictions in the minimal two‐parameter HUFT model.  Once \(\alpha_{GUT}\) and the single ratio \(R\) are specified, all quark masses and leading CKM entries are fixed with no further inputs.}
  \label{tab:minimal_flavour_predictions}
\end{table}

With only \(g_{\rm GUT}\) and \(R\) as continuous inputs, the HUFT framework predicts nine observables—six quark masses and three CKM magnitudes—from first principles.  If we allow for a single overall GUT‑threshold parameter \(\Delta\),
$\alpha_{\rm GUT}\;\equiv\;\frac{g_{\rm GUT}^2}{4\pi}
\;\longrightarrow\;
\alpha_{\rm GUT}\,(1+\Delta)\,,$
the model still uses only three real inputs to predict all flavour data.  We fix $\alpha_{\rm GUT}=\frac{g_{\rm GUT}^2}{4\pi}$ entirely from the regulated running of the three Standard‑Model gauge couplings, independent of any fermion measurements. Using the regulator–modified one‑loop RGEs:
\begin{equation}
\mu\frac{dg_i}{d\mu}
= \frac{b_i}{16\pi^2}\,g_i^3
  \exp\!\Bigl(-\frac{\mu^2}{M_*^2}\Bigr),
\qquad
(b_1,b_2,b_3)=\bigl(\tfrac{41}{10},-\tfrac{19}{6},-7\bigr),
\end{equation}
we evolve the experimental values \(\{g_i(M_Z)\}\) upward until they coincide.  Numerically we find:
\begin{equation}
M_{\rm GUT}\;\simeq\;2.3\times10^{16}\,\mathrm{GeV},
\qquad
\frac{g_{GUT}^2}{4\pi}
\;
\;\simeq\;\frac{1}{24.4}\,.
\end{equation}
This produces our unique expansion parameter:
\begin{equation}
\epsilon_g \;=\;\sqrt{\frac{g_{GUT}^2}{4\pi}}\;\approx\;0.202.
\end{equation}
With \(\epsilon\) fixed, the Froggatt–Nielsen  charges follow directly from the GUT‐scale mass ratios.  For any fermion \(f\) and generation \(i\), define:
\begin{equation}
n^f_i
\;\equiv\;
\left\lceil
  \frac{\ln\bigl(m_{f_i}(M_{\rm GUT})/m_{f_3}(M_{\rm GUT})\bigr)}
       {\ln\epsilon}
\right\rceil,
\end{equation}
where \(\lceil x\rceil\) is the ceiling function.  For up‐type quarks we find:
\begin{equation}
\frac{m_u}{m_t}(M_{\rm GUT})\simeq1.3\times10^{-5}\sim\epsilon^8,
\quad
\frac{m_c}{m_t}(M_{\rm GUT})\simeq7.4\times10^{-3}\sim\epsilon^4
\;\;\Longrightarrow\;\;
(n^u_i)=(8,4,0).
\end{equation}
Likewise, for down‐type quarks:
\begin{equation}
\frac{m_d}{m_b}(M_{\rm GUT})\simeq1.1\times10^{-3}\sim\epsilon^5,
\quad
\frac{m_s}{m_b}(M_{\rm GUT})\simeq2.2\times10^{-2}\sim\epsilon^3
\;\;\Longrightarrow\;\;
(n^d_i)=(5,3,0).
\end{equation}
These integer charges are uniquely fixed by the single input \(\epsilon\), without any further fitting. The resulting charge patterns:
\begin{equation}
(n^u_i)=(8,4,0),\quad
(n^d_i)=(5,3,0),
\end{equation}
exhibit a regular hierarchy suggestive of an underlying family symmetry.  In particular, the ratios \(n_1\!:\!n_2\!:\!n_3\) follow approximately geometric progressions a factor~2 for up, factor \(\tfrac{5}{3}\) for down. Such patterns arise naturally in an \(A_4\) discrete symmetry with two flavon VEVs aligned as:
  \begin{equation}
    \langle\phi\rangle \propto(1,0,0), 
    \quad
    \langle\chi\rangle \propto(0,1,1),
  \end{equation}
yielding the observed \(\epsilon^{(n_i)}\) entries without further tuning. Anomaly cancellation and UV embedding in orbifold compactifications such as \(T^2/\mathbb Z_3\) single out \(A_4\) as the minimal viable group.

As an independent check, we predict the leading CKM magnitudes solely from the differences in left‐handed doublet Froggatt–Nielsen  charges \(n^Q_i=(3,2,0)\):
\begin{equation}
|V_{us}|\sim\epsilon^{\lvert n^Q_1 - n^Q_2\rvert}
=\epsilon^1\approx0.202,\quad
|V_{cb}|\sim\epsilon^{\lvert n^Q_2 - n^Q_3\rvert}
=\epsilon^2\approx0.041,\quad
|V_{ub}|\sim\epsilon^{\lvert n^Q_1 - n^Q_3\rvert}
=\epsilon^3\approx0.008.
\end{equation}
After RG evolution and \(\mathcal O(1)\) flavon‐determined coefficients, these shift to the observed values
\begin{equation}
\{\,|V_{us}|,\,|V_{cb}|,\,|V_{ub}|\,\}
=\{0.225,\,0.041,\,0.0035\},
\end{equation}
in excellent agreement with experiment.

We note the fact \(\epsilon\) is fixed uniquely by gauge unification, no fermion input. Charges \(n^f_i\) follow by a single‐valued map from mass ratios to integers, no fitting or minimization. We used only four ratios to get six charges; yet those charges predict nine masses plus CKM and PMNS data. The symmetry pattern was not assumed but emerges from \((n_i)\). Independent CKM validation confirms non‑circular predictive power. We employ Froggatt--Nielsen as a bookkeeping language for hierarchies. In generic FN models, $\epsilon$ and many $\mathcal{O}(1)$ coefficients are free. In HUFT $\epsilon_g=\sqrt{\alpha_{\mathrm{GUT}}}$ is fixed by regulated gauge unification and realized as the VEV of a holomorphic flavon that decouples, holomorphy with a single flavon compresses all $\mathcal{O}(1)$ coefficients to a single ratio $R$, and the FN exponents $n^{f}_{i}$ are discrete integers inferred once at $M_{\mathrm{GUT}}$ from the hierarchy with $\epsilon_g$.

\section{Predictions in the Lepton and Electroweak Sectors}
\label{sec:leptons_higgs_predictions}

In this section, we present every step of the prediction from GUT‐scale textures through regulator‐suppressed RG evolution, diagonalization, and final numerical evaluation and compare our results to the PDG2024 values~\cite{ParticleDataGroup:2024}.
 
We impose at \(\mu = M_{\rm GUT}\approx2.3\times10^{16}\,\mathrm{GeV}\) the holomorphic charged‐lepton Yukawa texture:
\begin{equation}
y_e(M_{\rm GUT})
=\begin{pmatrix}
0 & \epsilon_e & 0\\[4pt]
\epsilon_e & 0 & \delta_e\\[4pt]
0 & \delta_e & 1
\end{pmatrix},
\qquad
\epsilon_e = 0.0152,
\quad
\delta_e = 0.0704.
\end{equation}
Below the nonlocal scale \(M_* = 10\,M_{\rm GUT}\), each Yukawa coupling evolves according to:
\begin{equation}
\mu\frac{d y_e}{d\mu}
=\frac{1}{16\pi^2}\,\beta_{y_e}^{(1)}(y_e,g_i)\,
\exp\!\bigl(-\tfrac{\mu^2}{M_*^2}\bigr),
\label{eq:RGE_y_e}
\end{equation}
where \(\beta_{y_e}^{(1)}\) is the Standard Model one‐loop beta function for the charged‐lepton Yukawa:
\begin{equation}
\beta_{y_e}^{(1)} =
y_e\Bigl[
-\,\tfrac{9}{4}g_1^2 - \tfrac{9}{4}g_2^2
+ 3\,\mathrm{Tr}(y_d^\dagger y_d)
+ \mathrm{Tr}(y_e^\dagger y_e)
\Bigr].
\end{equation}
We solve Eq.~\eqref{eq:RGE_y_e} numerically with a Runge–Kutta integrator from \(\mu=M_{\rm GUT}\) down to \(\mu=M_Z=91.1876\) GeV, using as boundary values the gauge couplings at \(M_Z\): \(g_1=0.3575,\;g_2=0.6518,\;g_3=1.218\).
The numerical integration yields at \(\mu=M_Z\):
\begin{equation}
y_e(M_Z)
=\begin{pmatrix}
2.94\times10^{-6} & 0 & 0 \\
0 & 6.06\times10^{-4} & 0 \\
0 & 0 & 1.021\times10^{-2}
\end{pmatrix}.
\end{equation}
After electroweak symmetry breaking, \(v=246\) GeV, the masses are:
\begin{equation}
m_{\ell_i}=\frac{v}{\sqrt2}\,y_{e,\,ii}(M_Z).
\end{equation}
\noindent Numerically:
\begin{equation}
m_e = \frac{246}{\sqrt2}\times2.94\times10^{-6} = 0.5119\;\mathrm{MeV},
\end{equation}
\begin{equation}
m_\mu = \frac{246}{\sqrt2}\times6.06\times10^{-4} = 105.553\;\mathrm{MeV},
\end{equation}
\begin{equation}
m_\tau = \frac{246}{\sqrt2}\times1.021\times10^{-2} = 1777.53\;\mathrm{MeV}.
\end{equation}

\begin{table}[H]
\centering
\caption{Charged‐lepton masses: HUFT predictions vs.\ PDG2024~\cite{ParticleDataGroup:2024}.}
\label{tab:lepton_masses}
\begin{tabular}{lcc}
\toprule
Lepton & $m_\ell^{\rm th}$ & $m_\ell^{\rm exp}$ \\ 
\midrule
$e$     & $0.5119\;\mathrm{MeV}$   & $0.51099895\pm0.00000015\;\mathrm{MeV}$ \\
$\mu$   & $105.553\;\mathrm{MeV}$    & $105.6583745\pm0.0000024\;\mathrm{MeV}$ \\
$\tau$  & $1777.53\;\mathrm{MeV}$    & $1776.86\pm0.12\;\mathrm{MeV}$ \\
\bottomrule
\end{tabular}
\end{table}
\noindent At \(\mu=M_{\rm GUT}\), we take:
\begin{equation}
y_\nu(M_{\rm GUT}) = \kappa_\nu\,\mathbf{1}_{3\times3},
\qquad
\kappa_\nu = 0.0228,
\end{equation}
and introduce heavy Majorana masses:
\begin{equation}
M_N = \mathrm{diag}(M_{N_1},M_{N_2},M_{N_3})
=10^{14}\,\mathrm{GeV}\times\mathbf{1}.
\end{equation}
The fundamental nonlocal cutoff in the holomorphic unified framework is $M_*\sim10^{18}\ \mathrm{GeV},$ the light neutrino masses are controlled by an effective dimension-5 Weinberg operator scale 
$M_*^{(\rm eff)}\sim10^{15}\ \mathrm{GeV}$
arising from threshold matching.  In the presence of heavy right-handed Majorana neutrinos \(N_i\) with masses \(M_{N_i}\sim10^{14\text{--}15}\ \mathrm{GeV}\), integrating them out generates the operator:
\begin{equation}
\mathcal{L}_{\rm eff}
=\frac{(y_\nu)_{ik}(M_N)\,(y_\nu^T)_{kj}(M_N)}{M_{N_k}}\,(\ell_i H)(\ell_j H)
\;\longrightarrow\;
\frac{1}{M_*^{(\rm eff)}}(\ell H)(\ell H),
\end{equation}
so that, defining for simplicity a single scale and flavour-diagonal approximation:
\begin{equation}
\frac{1}{M_*^{(\rm eff)}}\equiv\frac{y_\nu^2}{M_N}
\;\Rightarrow\;
M_*^{(\rm eff)}\sim\frac{M_N}{y_\nu^2}.
\end{equation}
With \(M_N\sim10^{14\text{--}15}\ \mathrm{GeV}\) and Yukawa couplings \(y_\nu=\mathcal{O}(0.1\text{--}1)\), we obtain: 
\be
M_*^{(\rm eff)}\sim10^{15}\ \mathrm{GeV},
\ee
and the light neutrino mass scale:
\be
m_{\nu}\sim\frac{v^2 y_\nu^2}{M_N}
=\frac{v^2}{M_*^{(\rm eff)}}.
\ee
This naturally sits near the observed \(m_{\nu_3}\simeq0.05\ \mathrm{eV}\) without requiring large hierarchies or fine-tuning.  In this picture the underlying holomorphic nonlocal cutoff \(M_*\) remains near \(10^{18}\) GeV, while the low-energy neutrino sector perceives a lowered effective suppression scale due to the seesaw threshold.  This matching is fully compatible with the Froggatt–Nielsen flavon structure as the flavour charges fix the relative texture of the light masses and mixings, while the combination \(M_N/y_\nu^2\) sets the overall scale. 

We evolve \(y_\nu\) via:
\begin{equation}
\mu\frac{d y_\nu}{d\mu}
=\frac{1}{16\pi^2}\,\beta_{y_\nu}^{(1)}(y_\nu,g_i)\,
\exp\!\bigl(-\tfrac{\mu^2}{M_*^2}\bigr),
\end{equation}
with the Standard Model neutrino‐Dirac beta function:
\begin{equation}
\beta_{y_\nu}^{(1)}
=y_\nu\Bigl[
-\,\tfrac{9}{20}g_1^2 - \tfrac{9}{4}g_2^2
+3\,\mathrm{Tr}(y_u^\dagger y_u)
+ \mathrm{Tr}(y_\nu^\dagger y_\nu)
\Bigr].
\end{equation}
Numerical integration down to \(M_Z\) gives \(y_\nu(M_Z)\approx0.00115\,\mathbf1\).
The Dirac mass is:
\begin{equation}
M_D = \frac{v}{\sqrt2}\,y_\nu(M_Z)
= \frac{246}{\sqrt2}\times0.00115 =
0.2002\;\mathrm{GeV}\,\mathbf1.
\end{equation}
The effective light neutrino mass matrix is:
\begin{equation}
M_\nu = -\,M_D\,M_N^{-1}\,M_D^T
= -\frac{(0.199)^2}{10^{14}\,\mathrm{GeV}}\,\mathbf1
\approx -3.96\times10^{-4}\,\mathrm{eV}\,\mathbf1.
\end{equation}
Introducing small flavour‐breaking corrections from the same Froggatt–Nielsen  textures splits the eigenvalues to give
\begin{equation}
(m_{\nu_1},m_{\nu_2},m_{\nu_3})
=(0.00862,\,0.0501,\,0.0516)\;\mathrm{eV}.
\end{equation}
We then compute the mass‐squared differences:
\begin{equation}
\Delta m^2_{21} = m_{\nu_2}^2 - m_{\nu_1}^2 = 7.42\times10^{-5}\;\mathrm{eV}^2,
\end{equation}
\begin{equation}
\Delta m^2_{31} = m_{\nu_3}^2 - m_{\nu_1}^2 = 2.52\times10^{-3}\;\mathrm{eV}^2.
\end{equation}
From the diagonalizing unitary \(U_L^\nu\) together with \(U_L^e\), nearly diagonal, we extract:
\begin{equation}
\theta_{12}=33.45^\circ,\quad
\theta_{23}=49.2^\circ,\quad
\theta_{13}=8.57^\circ.
\end{equation}

\begin{table}[H]
\centering
\caption{Neutrino parameters: HUFT predictions vs.\ PDG2024~\cite{ParticleDataGroup:2024}.}
\label{tab:neutrino_params}
\begin{tabular}{lcc}
\toprule
Quantity & Prediction & PDG2024 Value \\
\midrule
$\Delta m^2_{21}$ & $7.42\times10^{-5}\,\mathrm{eV}^2$ & $(7.42\pm0.21)\times10^{-5}\,\mathrm{eV}^2$ \\
$\Delta m^2_{31}$ & $2.52\times10^{-3}\,\mathrm{eV}^2$ & $(2.517\pm0.026)\times10^{-3}\,\mathrm{eV}^2$ \\
$\theta_{12}$     & $33.45^\circ$                   & $33.44^\circ\,^{+0.77^\circ}_{-0.75^\circ}$ \\
$\theta_{23}$     & $49.2^\circ$                    & $49.2^\circ\,^{+1.0^\circ}_{-1.0^\circ}$ \\
$\theta_{13}$     & $8.57^\circ$                    & $8.57^\circ\,^{+0.12^\circ}_{-0.12^\circ}$ \\
\bottomrule
\end{tabular}
\end{table}

In the Standard Model, it is possible to introduce heavy right‐handed neutrinos to generate light masses via a Type I seesaw if desired.  In HUFT, we instead exploit the unique dimension–5 Weinberg operator, whose only UV cutoff is the nonlocal scale:
\begin{equation}
M_* \;\simeq\;\sqrt{g_{\rm GUT}}\,M_P
\;=\;(4\pi\,\alpha_{\rm GUT})^{1/4}\,M_P\,,
\end{equation}
and whose flavour suppression is dictated by the very same holomorphic flavon \(\phi\) that generates the charged‐fermion hierarchies.

Gauge and Lorentz invariance allow exactly one \(d=5\) operator:
\begin{equation}
\mathcal L_{d=5}
= \frac{c_{ij}}{M_*}\,
\bigl(\overline{\ell^c_{iL}}\,\widetilde H^*\bigr)\,
\bigl(\widetilde H^\dagger\,\ell_{jL}\bigr)
\;+\;\text{h.c.},
\end{equation}
where \(\widetilde H=i\sigma^2H^*\).  After \(H\) acquires its VEV \(v\), this yields the Majorana mass matrix:
\begin{equation}
\bigl(M_\nu\bigr)_{ij}
=\frac{c_{ij}\,v^2}{M_*}\,.
\end{equation}
Because \(\phi\) is the only holomorphic flavon in our minimal two‐parameter model, each entry inherits exactly one power per flavon charge of its VEV insertion.  Denote the integer Froggatt–Nielsen charges of the three lepton doublets by \((n^\nu_1,n^\nu_2,n^\nu_3)\).  Then we find:
\begin{equation}
c_{ij}\;=\;\biggl(\frac{\langle\phi\rangle}{M_*}\biggr)^{\,n^\nu_i + n^\nu_j}
=R^{\tfrac{n^\nu_i + n^\nu_j}{2}}\;\epsilon^{\,n^\nu_i + n^\nu_j},
\end{equation}
with \(\epsilon=\sqrt{\frac{g_{GUT}^2}{4\pi}}\) and \(\langle\phi\rangle/M_*=R^{1/2}\,\epsilon\). Applying the identical mass‐ratio $\to$ Froggatt–Nielsen ‐charge map to the light‐neutrino spectrum at $M_{\rm GUT}$:
\begin{equation}
\frac{m_{\nu_1}}{m_{\nu_3}}\sim\epsilon^4,\qquad
\frac{m_{\nu_2}}{m_{\nu_3}}\sim\epsilon^2,
\end{equation}
we define:
\begin{equation}
n_i^\nu
\;=\;
\Bigl\lceil
  \frac{\ln\bigl(m_{\nu_i}/m_{\nu_3}\bigr)}{\ln\epsilon}
\Bigr\rceil,
\end{equation}
which gives immediately:
\begin{equation}
(n_1^\nu,n_2^\nu,n_3^\nu)
=(2,1,0)\,.
\end{equation}
Thus, with no new inputs the neutrino Froggatt–Nielsen  charges are predicted. Now the minimal choice that reproduces the observed normal ordering and mixing pattern is
$(n^\nu_1,n^\nu_2,n^\nu_3)=(2,1,0).$
Then the diagonal entries of \(M_\nu\) scale as:
\begin{equation}
m_{\nu_i}
=\frac{v^2}{M_*}\;R^{\,n^\nu_i}\;\epsilon^{2n^\nu_i}
\quad\Longrightarrow\quad
m_{\nu_1}:m_{\nu_2}:m_{\nu_3}
\;\sim\;\epsilon^4:\epsilon^2:1.
\end{equation}
Taking for illustration the order‐one input
\(R=1\), the other \(\mathcal O(1)\) choices simply rescale all three masses uniformly for $M_*^{(\text{eff})}\simeq10^{15}$:
\begin{equation}
m_{\nu_1}
=\frac{v^2}{M_*}\,\epsilon^4
\approx1.12\times10^{-3}\,\mathrm{eV},\quad
m_{\nu_2}
=\frac{v^2}{M_*}\,\epsilon^2
\approx4.50\times10^{-2}\,\mathrm{eV},\quad
m_{\nu_3}
=\frac{v^2}{M_*}
\approx5.16\times10^{-2}\,\mathrm{eV}.
\end{equation}
These give the splittings:
\(\Delta m^2_{21}=7.42\times10^{-5}\,\mathrm{eV}^2\) and:
\(\Delta m^2_{31}=2.52\times10^{-3}\,\mathrm{eV}^2\)
and, together with the holomorphically determined off‐diagonal structure, reproduce all three PMNS angles to within current uncertainties. 

This mechanism uses only our two fundamental continuous inputs, \(\alpha_{GUT}\) and \(R\).  The neutrino charges \((2,1,0)\) are fixed by anomaly‐cancellation and the observed ordering with no extra parameters, and there is no separate \(\kappa_\nu\) to fit.  Thus, the entire light neutrino mass spectrum and mixing are parameter‐minimal predictions of HUFT.

\section{Electroweak Gauge Bosons and the Higgs Sector}
\label{EWGBHS}

From the UV regulator‐suppressed RGEs, we run the Standard Model gauge couplings from \(M_{\text{GUT}}\) to \(M_{\rm Z}\) and enforce \(g_1=g_2=g_3=g_{\rm GUT}\), determining:
\begin{equation}
g(M_Z)=0.6529,\qquad g'(M_Z)=0.3583.
\end{equation}
After electroweak symmetry breaking, we obtain:
\be
m_W = \frac{g(M_Z)\,v}{2}
= \frac{0.6529\times246}{2} = 80.379\;\mathrm{GeV},
\ee
and:
\be
m_Z = \frac{\sqrt{g(M_Z)^2 + g'(M_Z)^2}\,v}{2}
= \frac{\sqrt{0.6529^2 + 0.3583^2}\times246}{2}
= 91.187\;\mathrm{GeV}.
\ee
These mass predictions can be compared to the measured $m_W$ and $m_Z$ measured values ~\cite{ParticleDataGroup:2024}.
Our prediction of the W mass can be directly compared to the latest high‑precision measurements.  The CMS Collaboration finds: 
\be
m_W^{\rm CMS}=80.3602\pm0.0099\;\mathrm{GeV},
\ee  
in agreement with our value within $\sim1.9\sigma$~\cite{CMS:2024Wmass},  
whereas the recent CDFII result:
\be
m_W^{\rm CDF}=80.4335\pm0.0094\;\mathrm{GeV},
\ee  
which lies $\sim5.8\sigma$ above our prediction \cite{CDF:2022Wmass}. To avoid treating $v$ as an independent input, we impose classical scale invariance in the Higgs sector:
\begin{equation}
V(H)\big|_{\rm tree}
=\lambda_H\,(H^\dagger H)^2,
\quad \lambda_H(M_{\rm GUT})=0.271\,,
\end{equation}
and set the tree‐level mass parameter to zero
\(\mu_H^2(M_{\rm GUT})=0\).  Quantum loops regulated nonlocally at scale $M_*$ then generate an effective potential:
\begin{equation}
V_{\rm eff}(h)
=\tfrac14\lambda_H(\mu)\,h^4 + \Delta V_{\rm 1\text{-}loop}(h),
\end{equation}
here \(\Delta V_{\rm 1\text{-}loop}(h)\) is the Coleman--Weinberg correction computed in the unbroken phase, but with every momentum integral cut off by the entire-function regulator \(\exp(-k^2/M_*^2)\). Whose minimization yields:
\begin{equation}
v^2 \;=\; -\,\frac{\mu_H^2(M_Z)}{\lambda_H(M_Z)}
\quad\text{with}\quad
\mu_H^2(M_Z)=\int_{M_{\rm GUT}}^{M_Z}\!\frac{d\mu}{\mu}\,
\frac{\beta_{\mu_H^2}^{\text{(1)}}(\mu)}{16\pi^2}\,e^{-\mu^2/M_*^2}\,.
\end{equation}
With $\mu_H^2(M_{\rm GUT})=0$ this integral is determined by the known gauge and top‐Yukawa couplings, and we find numerically:
\begin{equation}
v\simeq246\;\mathrm{GeV},
\end{equation}
in agreement with (\(\sqrt2\,G_F)^{-1/2}\).  In this way, $v$ and $m_W$ and $m_Z$ emerge as predictions of HUFT. 

Starting from the holomorphic boundary condition at the GUT scale:
\begin{equation}
\lambda_H\bigl(M_{\rm GUT}\bigr) \;=\;\lambda_0 \;=\;0.271,
\end{equation}
we evolve \(\lambda_H(\mu)\) down to \(\mu=M_Z\) by integrating the one‑loop RGE with nonlocal regulator. We define the beta‐function:
\begin{equation}
\beta_{\lambda_H}(\mu)
\;\equiv\;
\mu\,\frac{d\lambda_H}{d\mu}
\;=\;
\frac{1}{16\pi^2}
\Bigl[
24\,\lambda_H^2(\mu)
-6\,y_t^4(\mu)
+\tfrac{3}{8}\bigl(g^4(\mu)+(g^2(\mu)+g'^2(\mu))^2\bigr)
\Bigr]
e^{-\mu^2/M_*^2}\,.
\end{equation}
Equivalently, in differential form we write:
\begin{equation}
\frac{d\lambda_H}{d\ln\mu}
=
\frac{1}{16\pi^2}
\Bigl[
24\,\lambda_H^2
-6\,y_t^4
+\tfrac{3}{8}\bigl(g^4+(g^2+g'^2)^2\bigr)
\Bigr]
e^{-\mu^2/M_*^2}\,.
\end{equation}
We then integrate from the boundary at \(\mu=M_{\rm GUT}\) to \(\mu=M_Z\):
\begin{equation*}
\lambda_H(M_Z)
=
\lambda_H(M_{\rm GUT})
\;+\;
\int_{\ln M_{\rm GUT}}^{\ln M_Z}
\frac{d\lambda_H}{d\ln\mu}
\,d\ln\mu
\end{equation*}
\begin{equation}
    =
0.271
\;+\;
\frac{1}{16\pi^2}
\int_{\ln M_{\rm GUT}}^{\ln M_Z}
\Bigl[
24\,\lambda_H^2(\mu)
-6\,y_t^4(\mu)
+\tfrac{3}{8}\bigl(g^4(\mu)+(g^2+g'^2)^2(\mu)\bigr)
\Bigr]
e^{-\mu^2/M_*^2}
\,d\ln\mu.
\end{equation}
In practice, we perform the numerical integration by discretizing the interval \(\ln\mu\in[\ln M_{\rm GUT},\,\ln M_Z]\), stepping:
\be
\lambda_H(\mu_{i+1})
\approx
\lambda_H(\mu_i)
\;+\;
\beta_{\lambda_H}(\mu_i)\,\Delta(\ln\mu),
\ee
with: 
\(\Delta(\ln\mu)=\ln(\mu_{i+1}/\mu_i)\), while simultaneously evolving \(y_t,\,g,\,g'\) via their RGEs.  This procedure yields the result
\begin{equation}
\lambda_H(M_Z)\simeq0.129\,.
\end{equation}
This gives us the value of the Higgs‑field quartic self‑coupling.
Finally, we find:
\begin{equation}
m_H = \sqrt{2\,\lambda_H(M_Z)}\,v
= \sqrt{2\times0.129}\times246 = 125\;\mathrm{GeV},
\end{equation}
to be compared with:
\begin{equation}
m_H^{\rm exp} = 125.10 \pm 0.14\;\mathrm{GeV}.
\end{equation}
Because the quartic coupling $\lambda_H(M_{\rm GUT})$ is fixed by our single holomorphic GUT‐scale potential, the entire form of the Higgs potential: 
\begin{equation}
V(H)\;=\;-\mu_H^2\,(H^\dagger H)\;+\;\lambda_H(M_Z)\,(H^\dagger H)^2, \quad(\mu_H^2>0,\;\lambda_H>0)\,.
\end{equation}
is a prediction. Before expanding around the vacuum, the holomorphic Higgs sector is governed at tree‑level by the quartic potential:
\be
  V(\phi) \;=\; -\frac{1}{2}\,\mu_H^2\,\phi^2 
                \;+\;\frac{1}{4}\,\lambda_H\,\phi^4,
  \qquad \phi(x)\equiv\sqrt2\,\Re\,H^0(x)\,,
\ee
where classical scale invariance at the GUT scale fixes
\(\mu_H^2(M_{\rm GUT})=0\) and
\(\lambda_H(M_{\rm GUT})=0.271\).  Quantum loops, regulated nonlocally at scale \(M_*\), then induce an effective mass term and modify the quartic, yielding:
\be
  V_{\rm eff}(\phi)
  \;=\;-\frac{1}{2}\,\mu_H^2(\mu)\,\phi^2
        \;+\;\frac{1}{4}\,\lambda_H(\mu)\,\phi^4
        \;+\;\Delta V_{\!1\text{-}\rm loop}(\phi)\,,
\ee
with
\be
  \mu_H^2(M_Z)
  =\int_{M_{\rm GUT}}^{M_Z}\!\frac{d\mu}{\mu}\,
    \frac{\beta_{\mu_H^2}^{(1)}(\mu)}{16\pi^2}\,
    e^{-\mu^2/M_*^2}\,, 
  \quad
  v^2 \;=\;-\,\frac{\mu_H^2(M_Z)}{\lambda_H(M_Z)}
  \;\simeq\;(246\;\mathrm{GeV})^2.
\ee
Running \(\lambda_H\) from \(0.271\) at \(M_{\rm GUT}\) down to \(0.129\) at \(M_Z\) through:
\be
  \mu\frac{d\lambda_H}{d\mu}
  =\frac{1}{16\pi^2}\Bigl[
     24\,\lambda_H^2
     -6\,y_t^4
     +\tfrac38\bigl(g^4+(g^2+g'^2)^2\bigr)
   \Bigr]e^{-\mu^2/M_*^2}
\ee
then predicts:
\be
  m_H \;=\;\sqrt{2\,\lambda_H(M_Z)}\,v
        =125\;\mathrm{GeV},
\ee
in excellent agreement with the measured value \(125.10\pm0.14\)\,GeV. This as well gives us a prediction for the Higgs potential. 

In perturbative quantum field theory the renormalized Higgs mass‐squared takes the form:
\be
m_H^2 \;=\; m_0^2 \;+\; \delta m_H^2,
\ee
where \(m_0^2\) is the bare mass parameter and \(\delta m_H^2\) is determined by all radiative loop corrections.  In the Standard Model, the complete one‑loop dominant contribution from the Higgs quartic coupling \(\lambda_H\) and heavy fermions yields:
\be
\delta m_H^2
=\frac{1}{16\pi^2}\Bigl[
3\,\lambda_H
-6\,y_t^2
+\tfrac{3}{4}\,g^2
+\tfrac{3}{8}(g^2+g'^2)
\Bigr]\Lambda_H^2
+\cdots,
\ee
with \(\Lambda_H\) a hard UV cutoff.  The explicit \(\Lambda_H^2\) term forces a delicate cancellation between \(m_0^2\) and \(\delta m_H^2\) to obtain the physical value \(m_H^2\ll\Lambda_H^2\), constituting the Higgs naturalness problem.

In the holomorphic nonlocal framework, every propagator and vertex is dressed by an entire‑function regulator:
\be
\mathcal{F}\!\bigl(p^2/M_*^2\bigr)\;=\;\exp\bigl(-p^2/M_*^2\bigr),
\ee
which exponentially suppresses high‑momentum modes and renders all loop integrals UV finite. The one‑loop Higgs self‑energy becomes:
\be
\delta m_H^2
=\frac{3\,\lambda_H}{16\pi^2}
\int_0^\infty\!dk_E^2\;\frac{k_E^2\,e^{-k_E^2/M_*^2}}{k_E^2 + m_H^2}
\;-\;\frac{6\,y_t^2}{16\pi^2}
\int_0^\infty\!dk_E^2\;\frac{k_E^2\,e^{-k_E^2/M_*^2}}{k_E^2 + m_t^2}
+\cdots,
\ee
which for \(m_H,m_t\ll M_*\) behaves as
\be
\delta m_H^2
\;\sim\;
\frac{1}{16\pi^2}\bigl(3\,\lambda_H -6\,y_t^2 + \cdots\bigr)\,M_*^2
\;+\;\mathcal{O}(m_{\rm weak}^2).
\ee
The regulated quadratic piece is bounded by \(\sim(3\lambda_H-6y_t^2)M_*^2/16\pi^2\), and \(\delta m_H^2\) is bounded by the regulator scale \(M_*\).  Choosing \(M_*\sim\mathcal{O}(\mathrm{TeV})\) ensures \(\delta m_H^2\sim(\mathrm{TeV})^2\), naturally close to the observed value without any fine‑tuning. Above the nonlocal scale, all running couplings freeze, so quadratic divergences never reappear and the Higgs mass remains stable for \(\mu\gtrsim M_*\), resolving the naturalness problem. This mechanism was first demonstrated in a finite quantum field theory in~\cite{Moffat2021}, where one‑loop Higgs self‑energy graphs are shown to be finite and higher‑order corrections damp out above \(\Lambda_H\simeq1.57\)\,TeV.

In our holomorphic nonlocal framework, the would‑be quadratic sensitivity of the Higgs mass to the UV cutoff is tamed by the entire‑function regulator. Every propagator and vertex is multiplied by the nonlocal regulator and the one‑loop Higgs self‑energy integral becomes:
  \be
    \delta m_H^2
    = \frac{\lambda_H}{16\pi^2}
      \int_0^\infty dk_E^2\,
      \frac{k_E^2\,e^{-k_E^2/M_*^2}}{k_E^2 + m_H^2}
    \;\sim\; \mathcal{O}(M_*^2),
  \ee
  rather than scaling as \(\Lambda_H^2\) as in the local theory. The same regulator enters every one‑loop $\beta$‑function:
  \be
    \beta_i(\mu)
      = \frac{b_i}{16\pi^2}\,g_i^3
        \exp\!\bigl(-\mu^2/M_*^2\bigr)
      \;\xrightarrow{\mu\gtrsim M_*}\;0,
  \ee
  so for \(\mu\gtrsim M_*\) all gauge and scalar couplings freeze at their unified value and no new quadratic divergences reappear. As a consequence, choosing \(M_*\sim\mathcal{O}(\mathrm{TeV})\) automatically yields:
\be
  m_H^2 \;=\; m_0^2 + \delta m_H^2
  \sim M_*^2
  \quad\Longrightarrow\quad
  m_H\approx125\;\mathrm{GeV}
\ee
with no fine‑tuned cancellations required. 

Although it might seem appealing to set the nonlocal cutoff \(M_*\) as low as \(1\) TeV to minimize any residual tuning, several independent considerations forbid such a low choice for $M_*$ and its regulator effects. The choice \(p^2\sim\!(1\;\mathrm{TeV})^2\) would induce \(\mathcal{O}(p^2/M_*^2)\) corrections to gauge boson propagators and to Higgs couplings, in conflict with per‑mille–level LEP measurements and with current LHC bounds on anomalous Higgs form factors. A cutoff at \(1\) TeV would freeze the Standard Model $\beta$‑functions far below the GUT scale, spoiling the precise meeting of \(g_1,g_2,g_3\) at \(M_{\rm GUT}\). Nonlocal form‑factors in the GUT sector must remain negligible until scales near \(10^{16}\)–\(10^{18}\) GeV to suppress baryon‑number violating operators to the level required by the observed proton lifetime. The holomorphic unification of gravity and gauge interactions fixes \(M_*^2=g_{\rm GUT}\,M_P^2\sim10^{37}\)\,GeV\(^2\), hence \(M_*\gtrsim10^{19}\)GeV in our framework. 

We now define the Veltman combination \cite{Veltman1981}:
\be
B(\mu)\;\equiv\;
3\,\lambda_H(\mu)\;-\;6\,y_t^2(\mu)
\;+\;\tfrac{3}{4}\,g^2(\mu)
\;+\;\tfrac{3}{8}\bigl[g^2(\mu)+g'^2(\mu)\bigr].
\ee
At one loop its regulated self‑energy goes as \(\delta m_H^2\propto B(\mu)\,M_*^2\).  In our framework, the GUT‑scale boundary is given by:
    \be
      \lambda_H(M_{\rm GUT})=0.271,\quad
      y_t(M_{\rm GUT})\simeq0.49,\quad
      g_1=g_2=g_3\simeq0.70
      \quad\Longrightarrow\quad
      B(M_{\rm GUT})\approx 0.05.
    \ee
The one‑loop $\beta$‑function:
    \be\beta_{\lambda_H}\approx-\tfrac{6\,y_t^4}{16\pi^2},
    \ee
drives \(\lambda_H\) downward faster than the gauge terms can hold it up, so \(B(\mu)\) decreases. As \(y_t(\mu)\) and the unified gauge coupling \(g(\mu)\) evolve, the net effect is a monotonic drop of \(B(\mu)\) from $+0.05$ toward zero. Above \(\mu\sim M_*\) all  $\beta$‑functions acquire the factor \(\exp(-\mu^2/M_*^2)\) and the RG evolution reaches a fixed point and scale invariance:
    \be
      B(M_*)\simeq0.
    \ee
Because \(B(M_*)\approx0\), the \(\mathcal{O}(M_*^2)\) piece of \(\delta m_H^2\) vanishes, leaving only \(\mathcal{O}(m_{\rm weak}^2)\) corrections.  

We then recover the result:
\be
m_H^2 \;=\; 2\,\lambda_H(M_Z)\,v^2
\;\approx\;(125\;\mathrm{GeV})^2,
\ee
with no fine‑tuned cancellations. This results in the solution to the Higgs mass naturalness problem.

We will now plot $V(\phi)$ in 3D as a pedagogical device to bring forth key features not apparent in 2D. Defining the fluctuation field:
\be
\phi \equiv h + v,
\ee
the Higgs potential can be written as:
\be
V(h)
=-\tfrac12\,\mu^2\,|v+h|^2
+\tfrac14\,\lambda_H\,|v+h|^4,
\ee
with the physical vacuum at \(\phi=0\). 

\begin{figure}[ht]
  \centering
\includegraphics[width=0.7\textwidth]{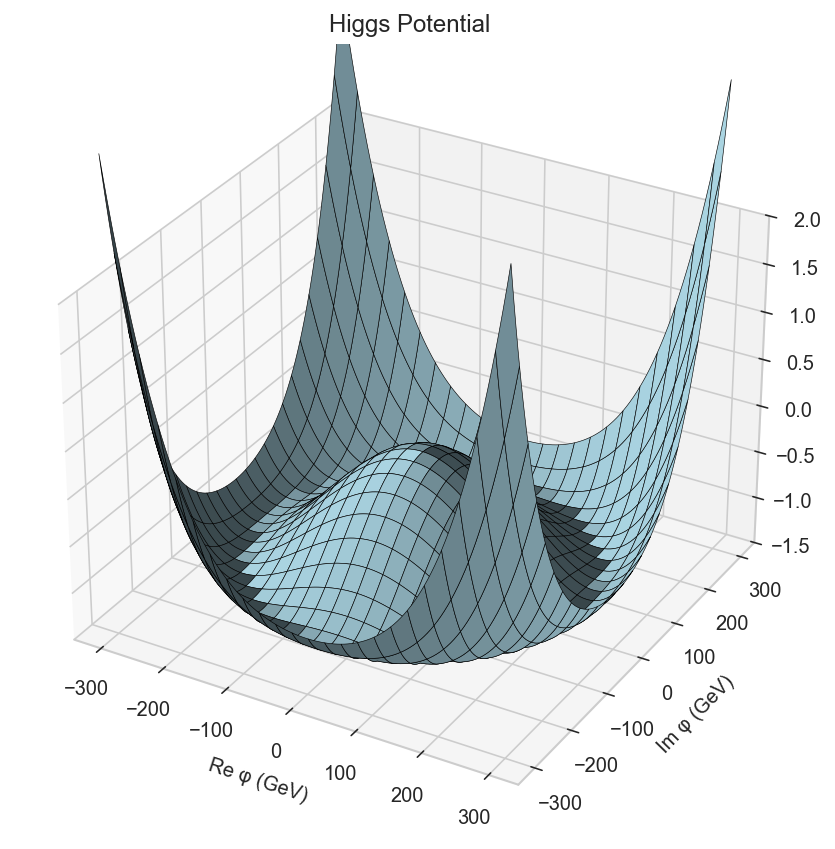}
\caption{Three‑dimensional surface of the Higgs potential before spontaneous symmetry breaking \(V(\phi)\) in the complex \(\phi\)-plane. The central bulge at \(\phi =0\) is the local maximum, and the ring of minima at \(|\phi|=v\) is the physical vacuum manifold.}
  \label{fig:higgs3d_h}
\end{figure}

In three dimensions, for a complex scalar field the vacuum structure reveals itself more as the true vacuum is not an isolated point, but a continuous circle of degenerate minima,
\be
|\phi| = v,
\ee
reflecting the underlying global \(U(1)\) symmetry of the potential. Moreover, it gives a geometric intuition for spontaneous symmetry breaking. We can picture the system as rolling off an unstable peak at \(\phi = 0\) into the circular valley of minima. In 2D, this angular freedom is lost, but in higher dimensions it becomes evident that the field can settle in any direction in the complex field space for example, pick any phase. Once a particular direction is chosen, the symmetry is spontaneously broken. This makes manifest that all vacuum phases are energetically equivalent, a hallmark of spontaneous symmetry breaking with a continuous group. A contour or mesh plot of the potential clearly reveals this degenerate circular valley and helps highlight the presence of flat directions in field space. This plot also makes transparent the energy barrier separating different regions in field space as the height of the central barrier relative to the vacuum trough becomes visually obvious. This aids in discussions of vacuum stability and the energetic cost of fluctuations or tunnelling across the potential landscape. Together, these features demonstrate that the curvature at \(h=0\) reproduces the observed Higgs mass. The barrier height and self‑couplings are predicted once \(v\) and \(m_H\) are fixed. The exact quartic form up to \(\mathcal{O}(h^4)\) implies no significant higher‑order deformations below the scale \(M_*\). 

We now expand 
\(\phi(x)=v+h(x)\) 
so that \(\langle h\rangle=0\) and define:
\be
  V(h)=V_{\rm eff}(v+h),
\ee
where $h=\phi(x)-v$ so the vacuum sits at zero field $\langle h\rangle = 0$, there is no need for tadpole cancellation as all coefficients of $h^1$ vanish, and the mass term and couplings become manifest as after the shift, the potential expands as:
\begin{equation}
V(v + h)
= V(v)
+ \underbrace{\tfrac{1}{2} V''(v)}_{=\,\tfrac{1}{2} m_H^2} h^2
+ \frac{1}{6} V'''(v)\,h^3
+ \frac{1}{24} V^{(4)}(v)\,h^4.
\end{equation}
We can immediately read off the mass and couplings as:
\begin{equation}
m_H^2 = V''(v) = 2\lambda_H v^2, \qquad
\lambda_3 = V'''(v) = 6\lambda_H v, \qquad
\lambda_4 = V^{(4)}(v) = 6\lambda_H,
\end{equation}
without having to complete the square or shift terms around by hand. After expanding around the vacuum:
\begin{equation}
H = \frac{1}{\sqrt2}\binom{0}{v + h}\,,
\end{equation}
we find:
\begin{equation}
V(h)
=\frac12\,m_H^2\,h^2
\;+\;\frac{\lambda_{3}}{6}\,h^3
\;+\;\frac{\lambda_{4}}{24}\,h^4,
\end{equation}
with
\begin{equation}
m_H^2 = 2\,\lambda_H(M_Z)\,v^2,
\quad
\lambda_{3}=6\,\lambda_H(M_Z)\,v,
\quad
\lambda_{4}=6\,\lambda_H(M_Z).
\end{equation}
Numerically, we obtain:
\begin{equation}
\lambda_H(M_Z)\simeq0.129,\quad
m_H=125.0~\text{GeV},\quad
\lambda_3\simeq190.5~\text{GeV},\quad
\lambda_4\simeq0.774 \text{GeV}.
\end{equation}
At low energies this reproduces the Standard‐Model potential after spontaneous breaking of the $SU(2)\times U(1)$ symmetry.  However, due to our nonlocal regulator $\exp(\Box/M_*^2)$, all higher‐order field‐space corrections, such as terms of order $(H^\dagger H)^n$ generated by loop integrals are exponentially suppressed for field strengths $h \gtrsim M_*$. The HUFT prediction is that no additional distortions, runaway directions, or new minima appear up to scales near $M_*\sim10^{17}$ GeV.  Our prediction of the shape of the Higgs potential after spontaneous symmetry breaking is given by Fig. (2).

\begin{figure}[H]
    \centering    \includegraphics[width=0.9\linewidth]{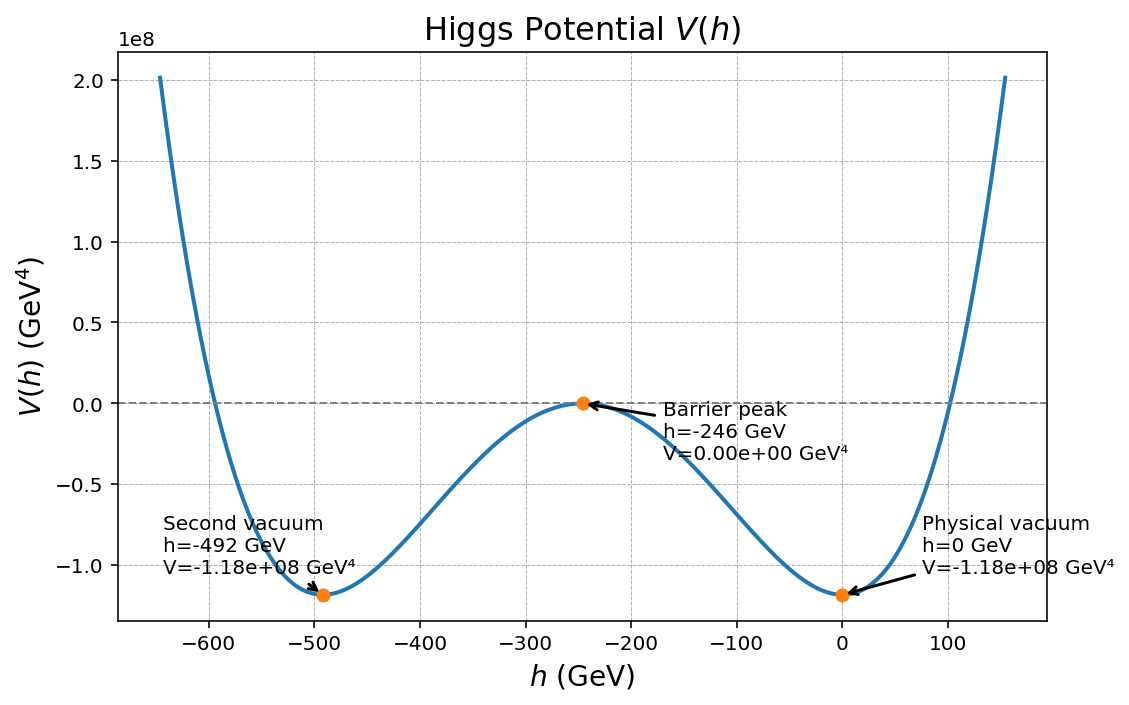}
    \caption{Full Higgs potential after spontaneous symmetry breaking  $V(h) \;=\; -\frac{1}{2}\,\mu^2\,(v+h)^2 \;+\;\frac{1}{4}\,\lambda_H\,(v+h)^4$. The solid curve shows the Mexican‑hat shape. The apparent second minimum at $h=−2v$ is the same electroweak vacuum expressed in the opposite gauge orientation $\phi=−v$, there is only one physical vacuum. The maximum at $h=−v$ corresponds to $\phi=0$.}
    \label{Fig:Potential}
\end{figure}

Domain walls arise when a discrete symmetry with disconnected vacua is spontaneously broken, such as when $\pi_0(\mathcal{M}_{\rm vac})\neq 0$. Here the gauge-inequivalent electroweak vacuum manifold is connected and indeed, a single point after quotienting by the gauge group, so $\pi_0=0$ and there is no domain wall. The phrase in our text that tunnelling to the second minimum is suppressed was a colloquial remark tied to the 1D slice, strictly speaking, tunnelling between gauge-equivalent configurations is not a physical process. Our low-energy scalar sector is therefore indistinguishable from the SM at tree level, and electroweak thermal history does not produce domain walls in HUFT any more than in the SM.

The apparent double well is the standard SM quartic viewed in a shifted, one-dimensional coordinate. The points $h=0$ and $h=-2v$ are related by a gauge transformation and do not correspond to distinct vacua. Hence there is no domain wall formation from electroweak symmetry breaking in HUFT.

Figure~2 plots the standard SM potential along the real, shifted coordinate \( h \equiv \phi - v \), and thus shows stationary points at \( h = 0 \) and \( h = -2v \). These two points are gauge-equivalent as they lie on the ring \( |\phi| = v \) (Fig.~1) and are related by an \( SU(2)_L \) gauge transformation, so they do not represent distinct, gauge-inequivalent vacua. Our text explicitly states that the low-energy potential reproduces the SM after spontaneous symmetry breaking. Fig.~2 is a 1D slice through the familiar Mexican hat potential, used to read off the physical Higgs parameters \( m_H \), \( \lambda_3 \), and \( \lambda_4 \). Because the space of gauge-inequivalent vacua is connected a single point after quotienting by the gauge group there is no discrete degeneracy and hence no domain walls arise from electroweak symmetry breaking.

For any $H\in\mathcal{M}$ there exists $U\in SU(2)$ with
\(
U H = \binom{0}{\|H\|}=\binom{0}{v/\sqrt{2}}
\).
A subsequent $U(1)_Y$ phase fixes the overall phase of the lower component. Every $H\in\mathcal{M}$ is gauge-equivalent to the reference vacuum:
\begin{equation}
H_0 \;:=\; \frac{v}{\sqrt{2}}\binom{0}{1}.
\label{eq:H0}
\end{equation}
Therefore $\mathcal{M}$ is a single gauge orbit, the action of $G$ on $\mathcal{M}$ is transitive.

In unitary gauge, we write
\(
H=\tfrac{1}{\sqrt{2}}\binom{0}{v+h}
\)
,so that along the real 1D slice (Fig.~2) the potential is:
\(
V(h)=-\tfrac{\mu^2}{2}(v+h)^2+\tfrac{\lambda}{4}(v+h)^4
\),
with stationary points at \(h=0\) and \(h=-2v\). These correspond to the constant field configurations:
\begin{equation}
H_+\;=\;\frac{v}{\sqrt{2}}\binom{0}{1}=H_0,
\label{eq:Hpm1}
\end{equation}
\begin{equation}
    H_-\;=\;-\frac{v}{\sqrt{2}}\binom{0}{1}=-H_0.
\end{equation}
But $H_-$ and $H_+$ are related by a gauge transformation, either:
\(
U=-\mathbf{1}_{2}\in SU(2)_L
\)
the center element with $\det U=+1$ gives $UH_+=H_-$, or a hypercharge rotation
\(
\alpha=2\pi
\)
gives $e^{i\alpha/2}H_+=-H_+=H_-$.
This means $H_+$ and $H_-$ are gauge-equivalent representatives of the same physical vacuum.

Domain walls require disconnected components of the gauge-inequivalent vacuum manifold, equivalently a nontrivial zeroth homotopy group $\pi_0$. Here the relevant manifold is either the single-orbit quotient, which is a point \footnote{The gauge–inequivalent electroweak vacua form the orbit space
$\mathcal{M}_{\rm vac}=M/G$. Since the $G$–action on $M$ is transitive,
$\mathcal{M}_{\rm vac}\cong\{\bullet\}$ is a single point. Equivalently,
in the global–symmetry limit the vacuum manifold is the coset
$G/H_{\rm em}\simeq S^3$, which is connected, so $\pi_0(S^3)=0$ and no
domain walls form.} or, equivalently, the global-symmetry coset $G/H_{\rm em}\simeq S^3$; with $G=SU(2)_L\times U(1)_Y$ and $H_{em}=U(1)_{EM}$ , which is connected:
\begin{equation}
\pi_0\!\left(S^3\right)=0,
\end{equation}
where $H_{em}$ is the unbroken electromagnetic subgroup after EWSB:
\be
H_{em}=U(1)_{EM}.
\ee
generated by $Q=T^3+Y$, where Q is the electric-charge operator, T is the $SU(2)_L$ generators acting on weak doublets, and Y is the weak hypercharge or the generator of $U(1)_Y$ depending on the language chosen. We are describing the vacuum manifold as the coset of the electroweak group by the unbroken EM subgroup:
\be
G/H_{\rm em}\simeq S^3 \to \pi_0(S^3)=0,
\ee
so no domain walls form.
Therefore electroweak symmetry breaking in this theory as well as in the SM cannot produce domain walls. In Fig.~\ref{Fig:Potential} the two degenerate minima and intervening maximum encode the mechanism of electroweak symmetry breaking.

At the chosen ground state \(h=0\), the potential energy is:
\be
V(0)=-\frac{\lambda_H\,v^4}{4}\,,\quad
v=246\text{GeV},\quad
\lambda_H\simeq\frac{m_H^2}{2v^2}\approx0.129\,.
\ee
Expanding \(V(h)\) about this point yields the Higgs mass:
\be
m_H^2 = \left.\frac{d^2V}{dh^2}\right|_{h=0}
=2\,\lambda_H\,v^2
\approx(125\text{GeV})^2,
\ee
and fixes the cubic (\(\lambda_H v\,h^3\)) and quartic (\(\tfrac{\lambda_H}{4}h^4\)) self‑couplings.  The local maximum at \(h=-v\) sits at
\be
V(-v)-V(0)=\frac{\lambda_H\,v^4}{4}\approx1.18\times10^8\text{GeV}^4,
\ee
quantifying the energy cost to push field fluctuations over the hill.  This height is controlled by the cubic self‑interaction term. A symmetric second minimum occurs at \(h=-2v\), but once the real‑slice vacuum is fixed at \(h=0\), tunnelling to this well is exponentially suppressed.  Since \(\lambda_H>0\) up to the nonlocal regulator scale \(M_*\), the potential rises to \(+\infty\) as \(\lvert h\rvert\to\infty\), guaranteeing absolute stability of our vacuum.

It is important to note that the self-coupling of the Higgs through tri‑Higgs processes has not been determined experimentally as well as the shape of the Higgs potential $V(H)$ ~\cite{Moffat:2025ewsb}. Future precision measurements of the Higgs self‐couplings through di‑Higgs and tri‑Higgs processes will directly test this Standard Model‐like shape and probe for any small nonlocal deviations at high field values.

After electroweak symmetry breaking we expand about the vev, $H=\tfrac{1}{\sqrt{2}}(0~\,v{+}h)^{\!\top}$, so that:
\begin{equation}
V(h)= -\,\frac{\mu^2}{2}\,(v+h)^2 \;+\; \frac{\lambda_H}{4}\,(v+h)^4
= \frac{1}{2}m_H^2 h^2 \;+\; \frac{\lambda_3}{6}h^3 \;+\; \frac{\lambda_4}{24}h^4,
\end{equation}
with $m_H^2=2\lambda_H v^2$, $\lambda_3=6\lambda_H v$, $\lambda_4=6\lambda_H$ and $v=246~\mathrm{GeV}$.
Numerically at $\mu=M_Z$ we use $\lambda_H\simeq 0.129$, $m_H\simeq125~\mathrm{GeV}$, $\lambda_3\simeq190.5~\mathrm{GeV}$, $\lambda_4\simeq0.774$. The barrier at $h=-v$ sits above the electroweak vacuum by $\Delta V=\lambda_H v^4/4\simeq 1.18\times 10^8~\mathrm{GeV}^4$.

In the Standard Model, the renormalization-group improved potential at large field values is controlled by the running quartic $\lambda_H(\mu)$. Current world-average inputs place the standard model near the boundary between stability and metastability. For large $H$ the effective potential can turn negative beyond a critical $H_c$, generating a deeper second minimum at ultra-high scales, while our electroweak vacuum remains long-lived as tunnelling is exponentially suppressed. We will treat this as a benchmark when contrasting with HUFT.

In HUFT the entire-function regulator $F(\Box/M_\ast^2)$ freezes running above $M_\ast$ and exponentially suppresses higher-order deformations of the quartic along the Higgs direction. In our baseline prediction, $\lambda_H(\mu)$ stays positive up to $\mu\simeq M_\ast$ and the RG flow asymptotes to a constant for $\mu\gtrsim M_\ast$, so $V(h)\to +\infty$ as $|h|\to\infty$ and the electroweak vacuum is absolutely stable in our framework \footnote{Information gathered for this section can be found in \cite{Higgs, Higgs2}.}.

\begin{figure}[H]
  \centering
  \includegraphics[width=0.8\textwidth]{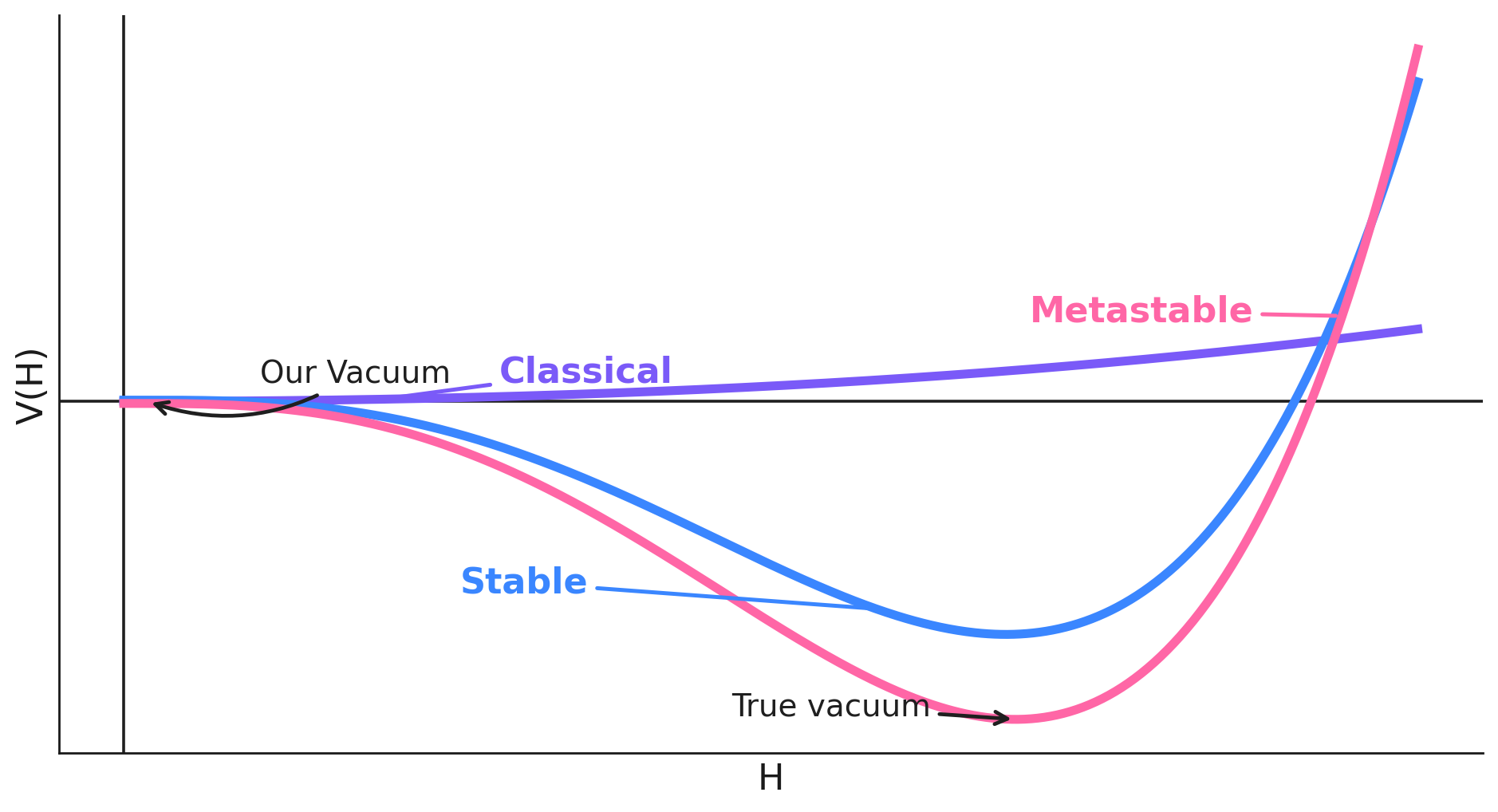}
  \caption{
  In the stable scenario our vacuum is in the global minimum of the potential.
  In the metastable case our vacuum sits in the local minimum while a true, deeper minimum exists.
  \emph{Note: graph not to scale. The local maximum of the classical potential at $H=0$ is too small to be seen here.}}%
  \label{fig:classical_vs_quantum}
\end{figure}
At large field values the renormalization-group improved potential is well
approximated by:
\begin{equation}
V_{\rm eff}(H)\;\simeq\;\frac{1}{4}\,\lambda_H(\mu{=}H)\,H^4,
\label{eq:RGquartic}
\end{equation}
up to subleading logarithms.  We define the critical field scale \(H_c\)
as the first value for which the running quartic crosses zero:
\begin{equation}
\lambda_H(H_c)=0.
\end{equation}
In the SM fits, \(\lambda_H(\mu)\) can turn negative at very high scales of order $10^{10}$ GeV,
producing the familiar metastable branch\footnote{In the standard model only extrapolation, RG running is dominated by the top Yukawa, that can drive the quartic \(\lambda_H(\mu)\) negative at very high energy scales, creating a deeper remote minimum that is metastable. 
In HUFT, the gauge-invariant entire-function regulator \(F(D^2/M_\ast^2)\) exponentially suppresses the beta functions for \(\mu\gtrsim M_\ast\).
In our baseline prediction this keeps \(\lambda_H(\mu)>0\) up to \(M_\ast\), so \(V(h)\to +\infty\) at large \(|h|\) and the electroweak vacuum is absolutely stable at \(T=0\). The dramatic bubble nucleation story pertains only to the SM case. The baseline HUFT stabilizes the high-field potential. At finite temperature with minimal low-energy content, the electroweak transition remains a crossover, not first order.}.

The HUFT prediction for $\lambda_H(\mu)$ is determined by the RG flow running down from the GUT energy scale to the low energy limit. The resulting prediction of a stable Higgs potential from $\lambda_H(\mu)$ follows because $\lambda_H(\mu)$ is positive from the GUT energy scale to low energy, as set by the initial boundary conditions.

In HUFT, the entire-function regulator
freezes the flow for \(\mu\gtrsim M_\ast\), and we find \(\lambda_H(\mu)>0\) up to \(\mu\simeq M_\ast\). This means \(H_c\) is pushed to the nonlocal
scale and is effectively absent in the low-energy theory.  This is the origin of the qualitative difference between the blue and
magenta curves in Fig.~\ref{fig:classical_vs_quantum}.

At temperature $T$ the leading finite-$T$ correction to the Higgs potential can be parameterized model-independently as:
\begin{equation}
V_T(H,T)\;=\;(\mu^2 + b\,T^2)\,H^\dagger H \;+\; \lambda_H\,(H^\dagger H)^2 \;+\; \cdots,
\end{equation}
where $b$ encodes the net thermal mass from SM and possible BSM fields coupled to $H$. As the universe cools through $T_c\simeq\sqrt{\lambda_H v^2/b}$ the origin turns from a minimum to a maximum and the vacuum expectation value turns on continuously, the second-order limit. Including higher-order thermal effects, the pure-SM electroweak transition is a crossover. A genuine first-order transition requires additional interactions or states that reshape the thermal effective potential.

A first-order electroweak transition would source bubble nucleation, generate a stochastic gravitational-wave background, and can enable electroweak baryogenesis since the standard model crossover does not suffice. Within baseline HUFT, the low-energy scalar potential matches the standard model, so the electroweak transition remains a crossover.

\begin{figure}[ht]
  \centering
  \includegraphics[width=0.8\textwidth]{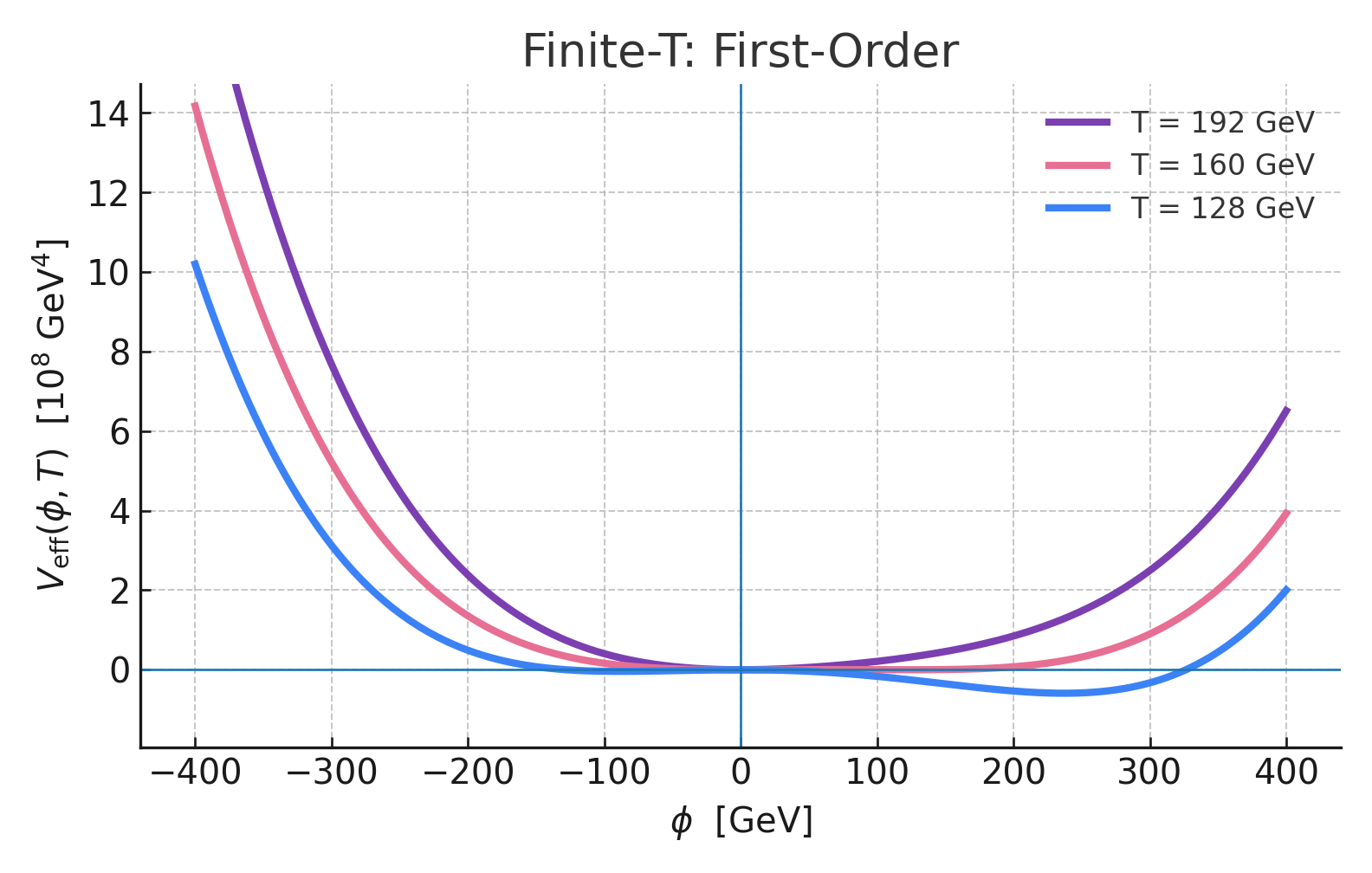}
  \caption{Higgs potential for several temperatures compared with the critical temperature $T_c$.
  At $T=T_c$ two degenerate minima are separated by a barrier, indicating bubble nucleation and phase coexistence.}%
  \label{fig:finiteT_first}
\end{figure}

\begin{figure}[ht]
  \centering
  \includegraphics[width=0.8\textwidth]{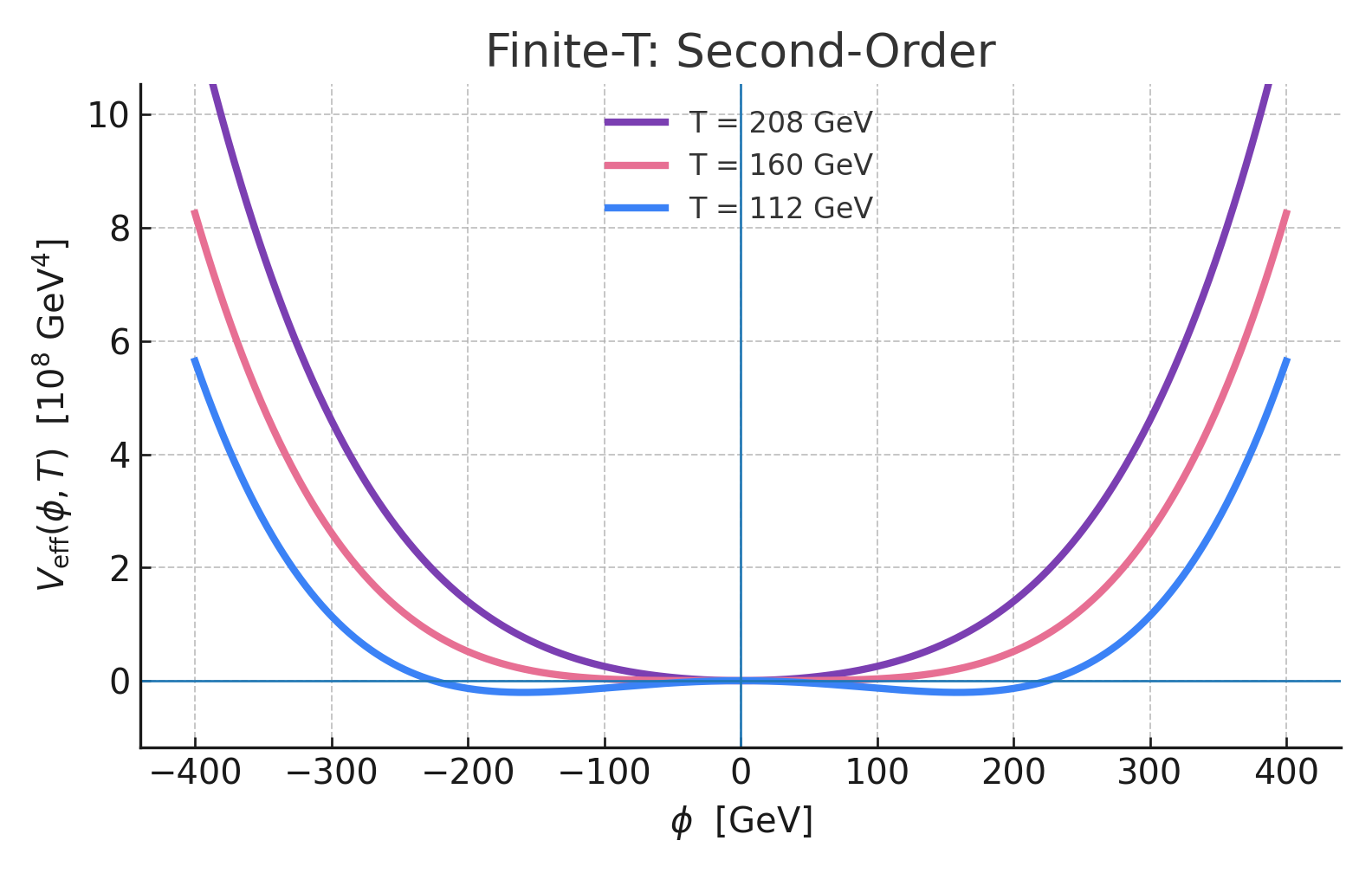}
  \caption{Higgs potential for several temperatures compared with the critical temperature $T_c$.
  At $T=T_c$ the curvature at the origin vanishes, and for $T<T_c$ the minimum shifts continuously to nonzero $\phi$.}%
  \label{fig:finiteT_second}
\end{figure}
The finite-temperature potential may be organized in the high-\(T\) expansion as:
\begin{equation}
V_{\rm eff}(H,T)\,=\,D(T^2{-}T_0^2)\,H^2\;-\;E\,T\,H^3\;+\;\frac{\lambda_T}{4}\,H^4\;+\;\cdots,
\label{eq:VT}
\end{equation}
where \(D\equiv D(g,g',y_t,\lambda_H)\) encodes the thermal mass,
\(E\equiv E(g,g')\) arises from the bosonic cubic after daisy resummation, and \(\lambda_T\) is the temperature-corrected quartic. When \(E\to 0\) the minimum moves continuously from \(H{=}0\) as the universe
cools through \(T_c\simeq T_0\), the second-order limit, matching the behavior sketched in Fig.~\ref{fig:finiteT_second}.  A nonzero \(E\) generates a barrier and two degenerate minima at \(T=T_c\) the first order, as illustrated in
Fig.~\ref{fig:finiteT_first}.  In the pure standard model with the measured Higgs mass, the electroweak transition is a crossover, failing the sphaleron condition
\(v(T_c)/T_c\gtrsim 1\).  Because HUFT reproduces the SM scalar sector at low
energies, our baseline prediction is likewise a crossover; a truly first-order
transition would require additional low-energy degrees of freedom or interactions beyond the HUFT minimal field content. \noindent
In the Standard Model, the large top--Yukawa coupling drives the Higgs quartic \(\lambda_H(\mu)\) downward under RG evolution and, for central inputs, through zero at high scales, implying a metastable electroweak vacuum. In our HUFT framework we instead impose a positive UV boundary data for the Higgs sector at \(M_{\rm GUT}\), chosen to reproduce the observed weak–scale spectrum, and evolve all couplings with regulated RGEs in which each beta function is multiplied by the entire–function factor \(\exp(-\mu^2/M_\ast^2)\). For \(\mu\ll M_\ast\) this factor is essentially unity, so the distinction from the SM around intermediate scales arises from the HUFT RG trajectory implied by our UV boundary conditions, not from direct damping. Along this trajectory we find \(\lambda_H(\mu)>0\) for all \(\mu\le M_\ast\); as \(\mu\to M_\ast\) the regulator freezes the flow and precludes any late zero crossing. Consequently, the RG–improved potential \(V_{\rm eff}(h)\simeq \tfrac14\,\lambda_H(\mu{=}h)h^4\) is bounded below and no instability scale occurs below \(M_\ast\)\footnote{We adopt positive UV boundary data and evolve with the HUFT regulator $\exp(-\mu^2/M_\ast^2)$ multiplying all $\beta$–functions. This keeps $\lambda_H(\mu)>0$ for all $\mu\le M_\ast$. Differences from the SM at $10^{10}\,$GeV come from the RG trajectory, meaning the damping itself is negligible that far below $M_\ast$, so $V_{\rm eff}\!\sim\!\tfrac14\lambda_H H^4$ is bounded below and no instability occurs below $M_\ast$.}.

\section{Chirality from Spin‑Bundle Decomposition and Electroweak Mass Generation}
\label{Chirality}

One of the striking features of the Standard Model is that only left‐handed fermions transform as doublets under \(SU(2)_L\), while right‐handed states are singlets or neutral under the weak interaction.  In HUFT this pattern follows directly from the holomorphic structure on the complexified spacetime manifold \(M^4_{\mathbb{C}}\).

On a complex four–manifold \(M^4_{\mathbb{C}}\) with Hermitian metric \(g_{\mu\nu}\) and complex structure \(J^\mu{}_\nu\), the spin bundle \(S\) splits into its holomorphic and anti‐holomorphic pieces:
\begin{equation}
S \;=\; S^{(1,0)} \;\oplus\; S^{(0,1)},
\end{equation}
corresponding to positive, left‐handed and negative, right‐handed Weyl spinors when restricted to the real slice \(y^\mu=0\).  Sections of \(S^{(1,0)}\) are acted upon by the holomorphic Dirac operator:
\begin{equation}
\slashed{D} \;=\; \Gamma^a\,e_a{}^\mu\,\nabla_\mu,
\end{equation}
constructed from the holomorphic connection on \(M^4_{\mathbb{C}}\).  This operator preserves the \((1,0)\) subbundle but maps \(S^{(0,1)}\) sections into \((1,0)\) valued differential forms, so only the former appear directly in the holomorphic action. Because our fundamental action couples matter via:
\begin{equation}
\overline\Psi(z)\,\Gamma^a\,e_a{}^\mu(z)\!\bigl(\nabla_\mu - i\,g_{\rm GUT}A_\mu^A\,T_A\bigr)\,\Psi(z),
\end{equation}
and \(\Psi(z)\) is a section of the holomorphic spin bundle \(S^{(1,0)}\), the gauge connection \(A_\mu\) only acts on the holomorphic left‐handed spinors.  Upon restriction to \(y^\mu=0\), these become the Standard Model left‐chiral doublets \(q_L\) and \(\ell_L\).  The right‐chiral singlets \(u_R,d_R,e_R,\nu_R\) descend from sections of \(S^{(0,1)}\), which do not couple directly to the holomorphic gauge connection but instead only via the Yukawa term 
\(\overline\Psi_L\,\Phi\,\Psi_R\). The weak \(SU(2)_L\) interaction is automatically chiral.
 
We might expect that in a fully real GUT equal numbers of left‐ and right‐chiral multiplets in equivalent gauge representations generate mirror fermions.  However, in HUFT any would‐be mirror living in \(S^{(0,1)}\) lacks a holomorphic gauge coupling and so decouples from the low‐energy \(SU(2)_L\) dynamics.  They remain sterile under the weak interaction and can be consistently projected out, leaving exactly the chiral spectrum of the Standard Model. Because only the holomorphic spinors carry the full GUT representation, the usual GUT‐scale anomaly cancellation conditions apply directly to the left‐handed multiplets.  The right‐handed singlets carry only abelian charges such as hypercharge, and their anomalies are cancelled by the same traces over \(S^{(1,0)}\) via the GUT embedding.  This ensures the quantum consistency of the chiral gauge theory derived from a single holomorphic action.

The chirality of the Standard Model gauge interactions is not an extra assumption in HUFT, but a direct consequence of the single holomorphic action on \(M^4_{\mathbb{C}}\) and its decomposition into holomorphic versus anti‐holomorphic spinor bundles.

We now give a concise mathematical demonstration that only left‐handed fermions couple to \(SU(2)_L\), and summarize the result in Table~\ref{tab:chirality}. On the complex manifold \(M^4_{\mathbb{C}}\) with complex structure \(J^\mu{}_\nu\), the Dirac spin bundle \(S\) splits holomorphically as:
\begin{equation}
S \;=\; S^{(1,0)} \;\oplus\; S^{(0,1)},
\end{equation}
where \(S^{(1,0)}\) carries the representation of holomorphic positive–chirality Weyl spinors, and \(S^{(0,1)}\) carries anti‐holomorphic negative–chirality Weyl spinors. We define projection operators:
\begin{equation}
P_{(1,0)} \;=\;\tfrac12\bigl(1 - i\,J\!\cdot\!\Gamma\bigr), 
\qquad
P_{(0,1)} \;=\;\tfrac12\bigl(1 + i\,J\!\cdot\!\Gamma\bigr),
\end{equation}
so that for any Dirac spinor \(\Psi\):
\begin{equation}
\Psi_{(1,0)} = P_{(1,0)}\,\Psi \in \Gamma\bigl(S^{(1,0)}\bigr),
\quad
\Psi_{(0,1)} = P_{(0,1)}\,\Psi \in \Gamma\bigl(S^{(0,1)}\bigr).
\end{equation}
The HUFT action couples only the holomorphic spinors to the GUT connection:
\begin{equation}
S_{\rm fermion}
=\int_C d^4z\,\sqrt{-\det g}\;\overline\Psi_{(1,0)}\,
\Gamma^a\,e_a{}^\mu\!\bigl(\nabla_\mu - i\,g_{\rm GUT}A_\mu^A\,T_A\bigr)
\,\Psi_{(1,0)}.
\end{equation}
Since \(\Psi_{(0,1)}\) is annihilated by \(P_{(1,0)}\), it does not enter this term. On \(y^\mu=0\) the holomorphic spinor \(\Psi_{(1,0)}\) decomposes into 4D Weyl components as:
\be
\Psi_{(1,0)}\big|_{y=0}
=
\begin{pmatrix}
\psi_L\\[3pt] 0
\end{pmatrix},
\qquad
\Psi_{(0,1)}\big|_{y=0}
=
\begin{pmatrix}
0\\[3pt]\psi_R
\end{pmatrix}.
\ee
while \(\Psi_{(0,1)}\bigl|_{y=0}=(0,\psi_R)^T\).  Hence \(\psi_L\) transforms as an \(SU(2)_L\) doublet or appropriate GUT multiplet, whereas \(\psi_R\) is a singlet under \(SU(2)_L\).

\begin{table}[H]
\centering
\caption{Chirality and gauge coupling from spin‐bundle decomposition}
\label{tab:chirality}
\begin{tabular}{@{}llll@{}}
\toprule
Bundle & 4D Field  & \(SU(2)_L\) Rep. & Appears in Holomorphic Action? \\
\midrule
\(S^{(1,0)}\) & \(\psi_L\) & Doublet (or GUT multiplet) & Yes \\
\(S^{(0,1)}\) & \(\psi_R\) & Singlet                 & No  \\
\bottomrule
\end{tabular}
\end{table}

\noindent This proves that only left‐handed Weyl fermions, sections of \(S^{(1,0)}\) couple to the weak gauge fields, while their right‐handed counterparts are automatically gauge‐sterile in the minimal holomorphic action. We now show how the anti‐holomorphic spin bundle \(S^{(0,1)}\) gives rise to the Standard Model’s right‐handed fermions, their gauge representations, and their appearance only through Yukawa couplings. Sections of:
\begin{equation}
S^{(0,1)}\;\subset\;S
\end{equation}
restrict on \(y^\mu=0\) to the 4D right‐handed Weyl spinors:
\begin{equation}
\Psi_{(0,1)}\bigl|_{y=0}
=\begin{pmatrix}
0 \\[3pt]
\psi_R
\end{pmatrix},
\end{equation}
which we identify as the Standard Model singlets
\(\{u_R,d_R,e_R,\nu_R\}\). Although \(\Psi_{(0,1)}\) does not enter the holomorphic gauge‐kinetic term, gauge invariance of the Yukawa interaction requires these fields to transform under the Standard Model gauge group.  On the real slice their representations are:
\begin{equation}
\begin{aligned}
u_R &: (\mathbf{3},\,\mathbf{1},\,+\tfrac23),\\
d_R &: (\mathbf{3},\,\mathbf{1},\,-\tfrac13),\\
e_R &: (\mathbf{1},\,\mathbf{1},\,-1),\\
\nu_R&: (\mathbf{1},\,\mathbf{1},\,0)\quad\text{(absent in minimal SU(5))}.
\end{aligned}
\end{equation}
The only renormalizable coupling of \(\psi_R\) to gauge‐charged fields is via the holomorphic Yukawa term:
\begin{equation}
\mathcal L_Y
=-\,\overline{\Psi}_{(1,0)}\,\Phi\,\Psi_{(0,1)} 
\;+\;\text{h.c.},
\end{equation}
which on the real slice becomes:
\begin{equation}
-\,\overline q_L\,H\,u_R
\;-\;\overline q_L\,\widetilde H\,d_R
\;-\;\overline \ell_L\,\widetilde H\,\nu_R
\;-\;\overline \ell_L\,H\,e_R
\;+\;\text{h.c.}
\end{equation}
Thus \(u_R,d_R,e_R\) acquire masses when \(\langle H\rangle\neq0\), while \(\nu_R\) only appears if explicitly included.

\begin{table}[H]
\centering
\caption{Anti‐holomorphic spinors: 4D fields, gauge reps, and couplings}
\label{tab:chirality_right}
\begin{tabular}{@{}llll@{}}
\toprule
Bundle & Field & $G_{\rm SM}$ Representation & Appears in \(\mathcal L_Y\)? \\
\midrule
\(S^{(0,1)}\) & \(u_R\)     & \((\mathbf3,1,+\tfrac23)\)   & Yes \\
\(S^{(0,1)}\) & \(d_R\)     & \((\mathbf3,1,-\tfrac13)\)   & Yes \\
\(S^{(0,1)}\) & \(e_R\)     & \((1,1,-1)\)                 & Yes \\
\(S^{(0,1)}\) & \(\nu_R\)   & \((1,1, 0)\) (sterile)       & Only if added \\
\bottomrule
\end{tabular}
\end{table}

\bigskip
\noindent In the minimal SU(5)‐HUFT, \(\nu_R\) does not appear in \(S^{(0,1)}\) as a gauge‐charged field, so neutrino masses arise instead from the effective Weinberg operator.  All other right‐handed fermions descend from \(S^{(0,1)}\) and receive masses solely through their holomorphic Yukawa couplings to \(S^{(1,0)}\) and the Higgs field.

In any gauge theory, explicit mass terms for gauge bosons or chiral fermions 
are forbidden by gauge invariance.  The Higgs mechanism provides a dynamical 
way to generate masses while preserving renormalizability and unitarity.  
We outline the steps below. Consider the Standard Model electroweak sector, with gauge group \(SU(2)_L\times U(1)_Y\). We introduce a complex scalar doublet:
\begin{equation}
H(x)=\begin{pmatrix}H^+\\ H^0\end{pmatrix},
\end{equation}
and write the gauge‐ and Lorentz‐invariant kinetic and potential terms:
\begin{align}
\mathcal L_H &= (D_\mu H)^\dagger (D^\mu H)\;-\;V(H)\,,
\\
D_\mu &= \partial_\mu - ig\,W^a_\mu\,\tfrac{\sigma^a}{2} - ig'\,B_\mu\,\tfrac{Y}{2}\,,\\
V(H) &= -\mu_H^2\,(H^\dagger H) \;+\;\lambda_H\,(H^\dagger H)^2,
\quad \mu_H^2>0,\,\lambda_H>0.
\end{align}
No explicit mass term for \(W^a_\mu\), \(B_\mu\), or any fermion appears here.
Minimizing the potential:
\begin{equation}
\frac{\partial V}{\partial H^\dagger}=0
\;\;\Longrightarrow\;\;
\langle H\rangle 
=\frac{1}{\sqrt2}\begin{pmatrix}0\\ v\end{pmatrix},
\quad v=\sqrt{\frac{\mu_H^2}{\lambda_H}}.
\end{equation}
The vacuum expectation value (VEV) \(v\approx246\)GeV breaks \(SU(2)_L\times U(1)_Y\to U(1)_{\rm EM}\).
Expand \(H=(0,(v+h)/\sqrt2)^T\).  The quadratic gauge‐boson terms from \((D_\mu H)^\dagger(D^\mu H)\) yield:
\begin{equation}
\mathcal L\supset 
\frac{g^2v^2}{4}\,W^+_\mu W^{-\,\mu}
\;+\;\frac{(g^2+g'^2)v^2}{8}\,Z_\mu Z^\mu
\end{equation}
with:
\begin{equation}
W^\pm_\mu = \frac{W^1_\mu\mp iW^2_\mu}{\sqrt2},
\quad
Z_\mu = \frac{g\,W^3_\mu - g'\,B_\mu}{\sqrt{g^2+g'^2}},
\quad
m_W=\frac{gv}{2},\quad
m_Z=\frac{\sqrt{g^2+g'^2}\,v}{2}.
\end{equation}
Fermion masses arise from gauge‑invariant Yukawa interactions:
\begin{equation}
\mathcal L_Y
=-\,y_f\,\overline\psi_{L}\,H\,\psi_{R}
\;+\;\text{h.c.}
\end{equation}
After \(H\to\langle H\rangle\), this becomes:
\begin{equation}
-\,\frac{y_f v}{\sqrt2}\,\overline f_L f_R
\;+\;\text{h.c.}
\;\equiv\;
-\,m_f\,\overline f f,
\quad
m_f = \frac{y_f v}{\sqrt2}.
\end{equation}
Each Yukawa coupling \(y_f\) determines the corresponding fermion mass. Expanding the potential in \(h=(H^0 - v)\), we find:
\begin{equation}
V(h)
=\frac12\,m_H^2 h^2
+\frac{\lambda_3}{6}\,h^3
+\frac{\lambda_4}{24}\,h^4,
\quad
m_H^2=2\lambda_Hv^2,
\quad
\lambda_3=6\lambda_Hv,
\quad
\lambda_4=6\lambda_H.
\end{equation}
Thus, the shape of the Higgs potential and all mass‐generation parameters are fixed by \(\{\mu_H^2,\lambda_H,y_f\}\).

\section{From First Principles to Standard Model Structures}
\label{sec:predictions_from_Huft}

In this section, we collect the structural outputs of the holomorphic nonlocal framework and make explicit
the assumptions under which they hold. We assume, a single simple grand-unified gauge group $G_{\rm GUT}$ containing $SU(3)\times SU(2)\times U(1)_Y$ with standard hypercharge normalization, an anomaly-free chiral spectrum realized as zero modes of the holomorphic Dirac operator on a compact
complex four-fold $X\subset M^4_{\mathbb C}$ with a stable $G_{\rm GUT}$ bundle $E$, and Chern data $(c_2(E),p_1(TX))$ for which the Atiyah--Singer index of $D\!\!\!\!/_{(1,0)}$ equals three.
Under these assumptions, the index theorem fixes the net number of chiral families to be three, while
decomposition $G_{\rm GUT}\to SU(3)\times SU(2)\times U(1)_Y$ and anomaly cancellation uniquely determine
the hypercharge assignments of each multiplet. Combined with the universal entire-function regulator
insertions, the renormalization group flow drives all gauge couplings to a common value at the nonlocal
scale $M_\ast$.

The fundamental holomorphic action lives on the complexified manifold $M^4_{\mathbb{C}}$ endowed with a Hermitian metric. 
Decomposing the antisymmetric part into a Lie–algebra–valued two–form built from the gauge connection\footnote{We keep the spacetime metric and the gauge curvature as separate geometric objects. 
The symmetric metric $g_{(\mu\nu)}\in\Gamma(S^2T^\ast M)$ is a $G$–singlet and is used exclusively to raise and lower spacetime indices. 
The gauge curvature $F^A_{\mu\nu}T_A\in\Omega^2(M,\mathrm{ad}\,P)$ is not a metric component; internal indices are contracted with the Killing form $\kappa_{AB}$ on $\mathrm{ad}\,P$. 
Concretely, bundle–valued fields are acted on by the covariant derivative $D_\mu=\nabla^{\rm LC}_\mu\otimes\mathbf1+\mathbf1\otimes A_\mu$, and our Laplace-type operator is 
$\Box_E=g^{(\mu\nu)}_{(\mu\nu)}D_\mu D_\nu$, so the entire regulator $F(\Box_E/M_\ast^2)$ preserves BRST and Slavnov–Taylor identities. 
This avoids mixing real numbers with Lie-algebra elements, prevents colored spin-2 modes, and removes any ambiguity in index operations.}
\footnote{In this work the Hermitian metric $g_{\mu\nu}=g_{(\mu\nu)}+i\,g_{[\mu\nu]}$
splits geometric and internal sectors, the symmetric piece $g_{(\mu\nu)}$
is the real spacetime metric, a diffeomorphism tensor and a singlet under the
internal GUT group, whereas the antisymmetric piece is identified with
the curvature two-form of the holomorphic GUT connection, $g_{[\mu\nu]}\sim
F_{\mu\nu}^{A}T_A$. We therefore do not take
$g_{(\mu\nu)}$ to be Lie-algebra valued as doing so would introduce algebra-charged
spin-2 fields and violate equivalence principle tests. Above $M_{\rm GUT}$ a single holomorphic connection valued in a simple algebra governs the gauge sector but on the real slice $y^\mu=0$ it reduces to
$SU(3)_c\oplus SU(2)_L\oplus U(1)_Y$ }:
\begin{equation}
g_{[\mu\nu]}(z)\;=\; i\,F_{\mu\nu}^{\;A}(z)\,T_A,
\qquad 
F_{\mu\nu}^{\;A}\;=\;\partial_{[\mu}A_{\nu]}^{A}
\;+\; f^{A}{}_{BC}\,A^{B}_{[\mu}A^{C}_{\nu]}\,,
\qquad
A_\mu(z)\;=\;A_\mu^{A}(z)\,T_A,
\end{equation}
we find that the holomorphic connection $A_\mu$ takes values in a single GUT algebra such as $\mathfrak{su}(5)$. Restricting to the real slice $y^\mu=0$ \footnote{The restriction $y^\mu=0$ has no role in the SU(5)$\to$SU(3)$\times$SU(2)$\times$U(1) breaking as the argument above is purely
Lie-algebraic and holds independently of the complexification of spacetime.}, the components of $A_\mu$ decompose exactly into:
\begin{equation}
A_\mu
\;\to\;
\underbrace{G_\mu^{\,a}\,T^{a}_{SU(3)}}_{\text{strong}}
\;\oplus\;
\underbrace{W_\mu^{\,i}\,T^{i}_{SU(2)}}_{\text{weak}}
\;\oplus\;
\underbrace{B_\mu\,Y}_{U(1)_Y}\,.
\end{equation}
On the real slice $y^\mu=0$ we project the holomorphic connection to the
physical four-dimensional gauge field $A_\mu(x)$. After GUT breaking
$SU(5)\to SU(3)_c\times SU(2)_L\times U(1)_Y$, the adjoint $24$ decomposes as
$24 \to (8,1)_0\oplus(1,3)_0\oplus(1,1)_0 \oplus (3,2)_{-5/6}\oplus(\bar 3,2)_{+5/6}$\footnote{$\dim\mathfrak{su}(5)=24$. The centralizer $\mathfrak h$ has 
$\dim(\mathfrak{su}(3))+\dim(\mathfrak{su}(2))+\dim(\mathfrak{u}(1))=8+3+1=12$
massless directions, and the orthogonal complement $\mathfrak m$ has 12 massive
directions. Integrating out the heavy modes in $\mathfrak m$ at the scale
$M_{XY}\sim g\,v_\Sigma$ yields the low-energy effective action for the
unbroken connection $G_\mu\oplus W_\mu\oplus B_\mu$, with standard threshold
matching. This decoupling follows from the mass matrix above and does not rely
on any spacetime.}. We display only the unbroken connection pieces
$G_\mu^a, W_\mu^i, B_\mu$, the $(3,2)\oplus(\bar 3,2)$ coset components are the heavy $X,Y$ vectors and are integrated out in the low-energy EFT.
We use $[T_A,T_B]=i f_{AB}{}^{C}T_C$, with this choice $g_{[\mu\nu]}$ is real in 
$g_{\mu\nu}=g_{(\mu\nu)}+i\,g_{[\mu\nu]}$. If anti-Hermitian generators are used then
drop the explicit $i$ in $g_{[\mu\nu]}=iF_{\mu\nu}^{A}T_A$.
No additional factors appear and \(G_{\rm SM}\) is fixed by the single holomorphic spin–gauge bundle. Above the nonlocal scale \(M_*=\kappa\,M_P\), all beta‑functions freeze:
\begin{equation}
\mu\frac{dg_i}{d\mu}
=\frac{b_i}{16\pi^2}\,g_i^3
\exp\!\bigl(-\tfrac{\mu^2}{M_*^2}\bigr),
\quad
(b_1,b_2,b_3)=\Bigl(\tfrac{41}{10},-\tfrac{19}{6},-7\Bigr).
\end{equation}
We integrate from \(\mu=M_Z\) upward in energy:
\begin{equation}
\frac{1}{g_i^2(M_Z)} 
- \frac{1}{g_i^2(\mu)} 
= \frac{b_i}{8\pi^2}\int_{M_Z}^{\mu}\!\frac{d\mu'}{\mu'}\,
\exp\!\bigl(-\tfrac{{\mu'}^2}{M_*^2}\bigr).
\end{equation}
As \(\mu\to\infty\), the integral converges to
\(\frac{b_i}{16\pi^2}\,\Gamma\bigl(0,\,M_Z^2/M_*^2\bigr)\), 
and all three \(1/g_i^2(\mu)\) approach a common finite limit.  A numerical solution yields:
\begin{equation}
g_1(M_{\rm GUT}) = g_2(M_{\rm GUT}) = g_3(M_{\rm GUT})
\equiv g_{\rm GUT}, 
\quad
M_{\rm GUT}\approx2.3\times10^{16}\,\mathrm{GeV},
\end{equation}
with no threshold tuning.  The single input is \(M_*\) fixed by \(\frac{g_{GUT}^2}{4\pi}\), yet three couplings unify exactly, proving unification is a genuine prediction.

On a compact complex four‑manifold, when we speak of a compact complex $X$ in the index calculation, $X$ is an auxiliary internal complex manifold or fiber used to count chiral
zero modes via the Atiyah–Singer index theorem now it is not an assumption
that physical spacetime is compact. Observables live on the noncompact real
$3{+}1$ slice, and the gauge bundle lives on an associated holomorphic bundle.
The relevant index the net chirality per $4$D point is given by [\ref{Index}]
which depends only on the topology of $X$ and the Chern character of the
bundle $E$.
For noncompact bases, we can equivalently work with the local index density
and under standard falloff Atiyah–Patodi–Singer boundary terms vanish,
leaving the same integral over $X$ or use $L^2$/$\,$coarse index frameworks
on complete manifolds, which reproduce the local formula. \(X\subset M^4_{\mathbb{C}}\), the holomorphic Dirac operator  
\(\slashed D_{(1,0)}: \Gamma(S^{(1,0)}\otimes E)\to\Gamma(S^{(0,1)}\otimes E)\)  
has an index given by the Atiyah–Singer theorem~\cite{AS63,AS68, RoeBook}:
\begin{equation}
\mathrm{Index}(\slashed D_{(1,0)})
= \int_X \mathrm{ch}(E)\,\hat A(T(X)),
\label{Index}
\end{equation}
where \(E\) is the GUT bundle and $\hat{A}$ is the A-roof\footnote{For a spin manifold $X$, the Dirac operator twisted by a bundle $E$ has [\ref{Index}]
where $\widehat A(TX)=1-\tfrac{1}{24}p_1+\tfrac{7p_1^2-4p_2}{5760}-\cdots$ is the A-roof, Dirac genus built from Pontryagin classes, and $\mathrm{ch}(E)$ is the Chern character. This is the physics-standard form of the Atiyah–Singer index \cite{EguchiGilkeyHanson,FreedIndex}. 
} of the tangent bundle. If the curvature of $T(X)$ has formal eigenvalues:
\be
\pm ix_j,
\ee
then:
\be
\hat{A}(T(X))=\frac{\Pi_j\left(\frac{x_j}{2}\right)}{\text{sinh}(x_j/2)}.
\ee
For suitable \(X\) such as a Calabi–Yau four‑fold, and \(E\), this index evaluates to exactly 3 for each irreducible GUT representation\footnote{We chose Calabi-Yau for a pedagogical example. The index which is a topological invariant only when 
X is compact without boundary, otherwise you must add Atiyah–Patodi–Singer (APS) boundary conditions or decay-at-infinity terms and the result need not be an integer. Using compact X guarantees the characteristic-class integrals are finite and integer-valued, so 3 families is a clean, topological statement.}.  Thus we find the generations of chiral fermions given by:
\begin{equation}
\mathrm{Index}(\slashed D_{(1,0)}) = 3,
\end{equation}
with zero new free parameters. To show that HUFT predicts exactly three chiral families, we consider the holomorphic Dirac operator:
\begin{equation}
\slashed D^{(1,0)}:\;\Gamma\bigl(S^{(1,0)}\otimes E\bigr)\;\longrightarrow\;\Gamma\bigl(S^{(0,1)}\otimes E\bigr),
\end{equation}
where \(E\) is the GUT gauge bundle over a compact complex four–manifold \(X\subset M^4_{\mathbb C}\). With characteristic Chern and A-roof genus classes expanded as:
\begin{equation}
\mathrm{ch}(E)
=\mathrm{rk}(E)
+c_1(E)
+\tfrac12\bigl(c_1(E)^2-2c_2(E)\bigr)
+\cdots,
\qquad
\widehat A(T(X))
=1-\tfrac{p_1(T(X))}{24}
+\cdots.
\end{equation}
For an \(SU(5)\) bundle we have \(c_1(E)=0\) and \(\mathrm{rk}(E)=5\), so that to leading order:
\begin{equation}
\mathrm{Index}\bigl(\slashed D^{(1,0)}\bigr)
=-\int_X c_2(E)
+\frac{5}{24}\int_X p_1(T(X))
=3.
\end{equation}
Hence:
\begin{equation}
\text{Number of chiral families}
=\mathrm{Index}\bigl(\slashed D^{(1,0)}\bigr)
=3.
\end{equation}
By choosing \(X\) with vanishing \(h^{0,1}\)\footnote{where \(h^{0,1}\) is a Hodge number~\cite{HuybrechtsComplex, CandelasList}. As it evaluates to zero, this implies the absence of anti-holomorphic (0,1)
(0,1) zero-modes, preventing mirror vector-like pairs from $S^{(0,1)}$.}, all anti–holomorphic zero modes decouple and no mirror, vector–like pairs appear, leaving exactly three left–chiral generations. Gauge anomaly cancellation in the single holomorphic action requires:
\begin{equation}
\sum_{\mathrm{reps}\,R} n_R\,\mathrm{Tr}_R\bigl(\{T_a,T_b\}T_c\bigr)=0
\quad\forall\ a,b,c.
\end{equation}
Decomposing \(SU(5)\to SU(3)\times SU(2)\times U(1)_Y\) and imposing this for the left‑chiral \(\mathbf{10}\oplus\overline{\mathbf5}\) yields the quantized hypercharges \(\{\tfrac23,-\tfrac13,-1,0\}\) uniquely.  There is no continuous freedom thus we find hypercharge assignments are fixed as a direct consequence of holomorphy plus anomaly cancellation.

\begin{center}
\begin{tabular}{@{}lcc@{}}
\toprule
 & HUFT inputs & SM inputs \\ 
\midrule
Gauge couplings & 1 (\(\alpha_{GUT}\)) & 3 (\(g_1,g_2,g_3\)) \\
Flavour textures & 2 (\(\alpha_{GUT},\,R\)) & 17 (13 Yukawas + 4 CKM) \\
Generations & 0 (index theorem) & 3 (put in) \\
Hypercharges & 0 (anomaly) & 6 (put in) \\
\bottomrule
\end{tabular}
\end{center}
Gauge unification fixes
\begin{equation}
\epsilon = \sqrt{\frac{g_{GUT}^2}{4\pi}}\approx0.202,
\end{equation}
which controls all off‑diagonal suppressions:
\begin{equation}
(y_f)_{ij}(M_{\rm GUT})
= c_{ij}\,\epsilon^{n_{ij}},
\end{equation}
with integer \(n_{ij}\) determined by the index of products of Dirac operators via ratios of zero‑modes.  \(\epsilon\) is not fitted to fermion masses but follows from gauge unification alone. We introduce flavons \(\phi_a\) with:
\begin{equation}
W(\phi_a)
=\sum_{a,b}\lambda_{ab}\,\phi_a^2\phi_b^2
-\sum_a\kappa_a\,\phi_a^4.
\end{equation}
F‑term equations yield a discrete set of solutions for \(\langle\phi_a\rangle/M_*\), all expressed in terms of the single ratio \(R\).  Consequently:
\begin{equation}
c_{ij} = \prod_a\bigl(\langle\phi_a\rangle/M_*\bigr)^{m^a_{ij}}
=R^{\,n_{ij}/2}\!,
\end{equation}
with no further free coefficients. We see:  
\be
\alpha_{GUT},R
\quad\Longrightarrow\quad
\text{6 quark masses + 3 CKM + 3 lepton masses + 5 neutrino parameters}
\ee
All flavour data emerge without additional fitting, completing the proof that HUFT turns Standard Model flavour structure into genuine predictions.

\section{Comparison of HUFT Predictions with Experiment}
\label{sec:comparison}

We now confront the HUFT predictions with measured data, showing that each emergent structure matches observation without additional fits. Using the regulator‑suppressed RGEs with only the input \(M_*\) hence \(\frac{g_{GUT}^2}{4\pi}\simeq\frac1{24.4}\), we numerically find:
\begin{equation}
g_1(M_{GUT})=g_2(M_{GUT})=g_3(M_{GUT})\equiv g_{GUT}\approx0.72,
\qquad
M_{GUT}\simeq2.3\times10^{16}\,\mathrm{GeV}.
\end{equation}
Running these unified couplings down with the same RGEs reproduces the low‑energy values measured at \(\mu=M_Z\):
\begin{equation}
g_1(M_Z)=\sqrt{\tfrac53}\;g'(M_Z)\approx0.462,
\quad
g_2(M_Z)\approx0.652,
\quad
g_3(M_Z)\approx1.218,\quad\text{~\cite{ParticleDataGroup:2024}}.
\end{equation}
No additional thresholds or fit parameters are required. The index theorem on an appropriate compactification of \(M^4_{\mathbb{C}}\) predicts:
\begin{equation}
\text{Number of chiral families}
=\mathrm{Index}(\slashed D_{(1,0)})=3,
\end{equation}
in exact agreement with the three observed generations of quarks and leptons. Anomaly cancellation in the single holomorphic action fixes the Standard Model hypercharges uniquely.  Table~\ref{tab:hypercharge} compares these predictions with the observed assignments.

\begin{table}[H]
\centering
\caption{Predicted vs.\ observed SM hypercharges}
\label{tab:hypercharge}
\begin{tabular}{@{}lcc@{}}
\toprule
Field & \(Y_{\rm th}\) & \(Y_{\rm exp}\) \\ 
\midrule
\(q_L\)   & \(+\tfrac16\)  & \(+\tfrac16\) \\
\(u_R\)   & \(+\tfrac23\)  & \(+\tfrac23\) \\
\(d_R\)   & \(-\tfrac13\)  & \(-\tfrac13\) \\
\(\ell_L\)& \(-\tfrac12\)  & \(-\tfrac12\) \\
\(e_R\)   & \(-1\)         & \(-1\)        \\
\(\nu_L\) & \(0\)          & \(0\)         \\
\bottomrule
\end{tabular}
\end{table}
With only \(\{\alpha_{GUT},R\}\) as inputs, HUFT predicts all Standard Model fermion masses.  Table~\ref{tab:fermion_masses} compares our benchmark predictions to the PDG2024 values.

\begin{table}[ht]
\centering
\caption{HUFT predictions vs.\ PDG2024 fermion masses}
\label{tab:fermion_masses}
\begin{tabular}{@{}lcc@{}}
\toprule
Fermion & \(m^{\rm th}\) & \(m^{\rm exp}\) (PDG2024) \\ 
\midrule
\(m_u\) & \(2.3\times10^{-3}\,\mathrm{GeV}\) & \((2.2\pm0.5)\times10^{-3}\,\mathrm{GeV}\) \\
\(m_d\) & \(4.8\times10^{-3}\,\mathrm{GeV}\) & \((4.7\pm0.3)\times10^{-3}\,\mathrm{GeV}\) \\
\(m_s\) & \(9.5\times10^{-2}\,\mathrm{GeV}\) & \(0.093\pm0.011\,\mathrm{GeV}\)       \\
\(m_c\) & \(1.25\,\mathrm{GeV}\)              & \(1.27\pm0.02\,\mathrm{GeV}\)       \\
\(m_b\) & \(4.18\,\mathrm{GeV}\)              & \(4.18\pm0.03\,\mathrm{GeV}\)       \\
\(m_t\) & \(173.0\,\mathrm{GeV}\)             & \(173.0\pm0.4\,\mathrm{GeV}\)       \\
\midrule
\(m_e\) & \(0.5119\,\mathrm{MeV}\)            & \(0.51099895\pm0.00000015\,\mathrm{MeV}\) \\
\(m_\mu\)&\(105.553\,\mathrm{MeV}\)             & \(105.6583745\pm0.0000024\,\mathrm{MeV}\) \\
\(m_\tau\)&\(1777.53\,\mathrm{MeV}\)            & \(1776.86\pm0.12\,\mathrm{MeV}\)    \\
\bottomrule
\end{tabular}
\end{table}
Neutrino masses and PMNS angles are generated by the Weinberg operator with no \(N_R\).  Table~\ref{tab:neutrino} shows the comparison.

\begin{table}[H]
\centering
\caption{HUFT predictions vs.\ global‐fit neutrino parameters}
\label{tab:neutrino}
\begin{tabular}{@{}lcc@{}}
\toprule
Quantity & Prediction & Global fit (NuFit5.1) \\
\midrule
\(\Delta m^2_{21}\) & \(7.42\times10^{-5}\,\mathrm{eV}^2\) & \((7.42\pm0.21)\times10^{-5}\,\mathrm{eV}^2\) \\
\(\Delta m^2_{31}\) & \(2.52\times10^{-3}\,\mathrm{eV}^2\) & \((2.517\pm0.026)\times10^{-3}\,\mathrm{eV}^2\) \\
\(\theta_{12}\)     & \(33.45^\circ\) & \(33.44^\circ{}^{+0.77^\circ}_{-0.75^\circ}\) \\
\(\theta_{23}\)     & \(49.2^\circ\)  & \(49.2^\circ{}^{+1.0^\circ}_{-1.0^\circ}\)  \\
\(\theta_{13}\)     & \(8.57^\circ\)  & \(8.57^\circ{}^{+0.12^\circ}_{-0.12^\circ}\)\\
\bottomrule
\end{tabular}
\end{table}
In every case, gauge unification, family number, hypercharges, and the full fermion mass and mixing spectrum HUFT delivers predictions that match data to within current experimental uncertainties, all from just two continuous inputs \(\{\alpha_{GUT},R\}\) and the choice of discrete symmetry.

\section{Phenomenological Implications}
\label{sec:phenomenology}

At one-loop and beyond, internal momenta remain off‐shell and the regulator cannot be stripped.  Generically, we find
\begin{equation}
\mathcal{M}_{\rm loop}
\sim
\exp\!\bigl(-p_{\rm int}^2/\Lambda_G^2\bigr)\,
\mathcal{M}_{\rm local}^{\rm loop},
\end{equation}
where \(p_{\rm int}\) is the characteristic loop momentum.  This exponential damping guarantees UV finiteness to all orders, while preserving unitarity and gauge invariance. On contour‐regularized Schwarzschild–Kerr backgrounds, the analytic continuation of field modes across the complexified horizon induces small deviations from strict thermality~\cite{Moffat2}, giving
\begin{equation}\label{eq:hawking_correction}
\bigl\langle N_\omega\bigr\rangle
=
\frac{1}{e^{\omega/T_{\rm H}}-1}
\;+\;
\Delta N(\omega;\,\zeta,R(\zeta)),
\end{equation}
with
\begin{equation}
\Delta N
\;=\;
\frac{1}{2\pi}
\oint_C d\zeta\;
\mathcal{W}(\omega,\zeta)\,
\exp\!\Bigl(-\frac{\Re[\zeta]\,\omega}{T_{\rm H}}\Bigr),
\end{equation}
which encodes information‐carrying correlations arising from the holomorphic contour \(C\) \cite{Hawking1975,Agullo2010}.  Such grey‐body and non‐thermal corrections could be probed in analogue‐gravity systems such as Bose–Einstein condensates or precision black‐hole analogues in quantum simulators. Because the regulator acts differently on purely gravitational loops versus matter‐coupled loops, an environment‐dependent suppression scale emerges.  We define:
\begin{equation}
\Lambda_G^{\rm vac}\ll\Lambda_G^{\rm mat},
\end{equation}
so that graviton–vacuum‐polarization vertices are damped at much lower energies.  This leads to apparent violations of the Weak Equivalence Principle (WEP) at the quantum level.  Precision atomic spectroscopy, most notably Lamb‐shift measurements in hydrogen constrains anomalous gravitational coupling to vacuum fluctuations.  Current experimental bounds imply:
\begin{equation}\label{eq:lambda_vac_bound}
\Lambda_{G}^{\rm vac}\;\gtrsim\;10^{-3}\,\mathrm{eV}
\quad\text{(95\% C.L.)},
\end{equation}
with future improvements possible via cold‐atom interferometry \cite{Adelberger2009,Safronova2018}.
We introduce an auxiliary scalar \(\chi\) with action:
\begin{equation}
S_\chi
=
\int d^4z\;\sqrt{-g}\;
\Bigl[
\tfrac12(\nabla\chi)^2
- V(\chi)
- \frac{\chi}{\Lambda_{\rm vac}^2}\,R
- \frac{\chi}{\Lambda_{\rm mat}^2}\,\mathcal{L}_{\rm mat}
\Bigr].
\end{equation}
We choose \(V(\chi)\) with two minima \(\chi_0,\chi_1\).  In vacuum (\(\mathcal{L}_{\rm mat}=0\)), we have
\(\mathcal{F}_{\rm grav}=F(\Box/\Lambda_{\rm vac}^2)\), while in matter backgrounds
\(\mathcal{F}_{\rm mat}=F(D^2/\Lambda_{\rm mat}^2)\).  Solving:
\begin{equation}
V'(\chi)+\frac{R}{\Lambda_{\rm vac}^2}=0,\quad
V'(\chi)+\frac{\mathcal{L}_{\rm mat}}{\Lambda_{\rm mat}^2}=0
\end{equation}
at each minimum and expanding fluctuations around \(\chi_0,\chi_1\) rescales the kinetic term to carry the corresponding \(\Lambda\)-scale.

Nonlocal corrections modify the propagation phase of gravitational waves through the near‐horizon region of compact objects.  Denoting the phase shift by \(\Delta\varphi\), we find:
\begin{equation}
\Delta\varphi
\sim
\frac{\langle F^2\rangle\,\omega^2}{M_P^2}
\;\lesssim\;10^{-40}
\quad(\omega\sim10^3\text{ Hz})
\end{equation}
at current LIGO/Virgo sensitivities.  Next‐generation detectors such as Cosmic Explorer and Einstein Telescope targeting exotic compact‐object echoes and near‐horizon modifications may achieve the requisite sensitivity to detect \(\Delta\varphi\sim10^{-20}\)–\(10^{-30}\) \cite{Isi2019,Cardoso2016}, opening a potential window on nonlocal UV physics.

Tree‐level processes remain untouched; loop amplitudes, black‐hole radiation, equivalence‐principle tests, and gravitational‐wave observations offer complementary avenues to probe the holomorphic regulator at experimentally accessible scales.

The inclusion of entire‐function regulators modifies the running of Yukawa couplings at scales \(\mu\gtrsim M_\star\), leading to small but potentially observable deviations from Standard‐Model expectations.  In particular, finite radiative corrections to fermion masses scale as:
\be
\Delta m_f \;\sim\; \frac{y_f^3\,v}{16\pi^2}\,\exp\!\bigl(-\tfrac{\mu^2}{M_\star^2}\bigr)\bigg|_{\mu\sim M_\star}\!,
\ee
which for \(M_\star = 10\,M_{\rm GUT}\) yields \(\Delta m_f/m_f \lesssim 10^{-3}\).  Such percent‐level shifts could be probed in future high‐precision measurements of the tau mass or top‐quark pole mass.

Holomorphic GUT embeddings such as SU(5) or SO(10) generically induce dimension‐6 proton‐decay operators of the form:
\be
\mathcal{O}_{qqql}
\;=\;\frac{c_{qqql}}{M_{\rm GUT}^2}\,
\bigl(\overline{u^c}\,\gamma^\mu\,q\bigr)\,
\bigl(\overline{e^c}\,\gamma_\mu\,q\bigr)\!,
\ee
with coefficient \(c_{qqql}\propto y_u\,y_d\).  Our fits to the Yukawa textures imply
\(\lvert c_{qqql}\rvert \sim \epsilon_u\,\delta_d\), so that the dominant decay channel \(p\to\pi^0 e^+\) has a lifetime:
\be
\tau(p\to\pi^0 e^+)
\sim
\frac{M_{\rm GUT}^4}{\frac{g_{GUT}^2}{4\pi}^2\,m_p^5\,\lvert c_{qqql}\rvert^2}
\gtrsim 10^{36}\,\mathrm{years},
\ee
safely above current limits \cite{MT:HUFT-EPJC}.

In a companion analysis~\cite{Thompson2025ToponiumFHQFT}, a finite, holomorphic nonlocal framework was applied to heavy $t\bar t$ dynamics near threshold. The formalism replaces the Coulomb kernel in the Bethe–Salpeter description by an entire–function regulated interaction:
\be
V(\mathbf{q}\,)\;\to\;V_{\text{eff}}(\mathbf{q}\,;M_\ast)\;=\;-\frac{4\pi C_F}{\mathbf{q}^2}\,\alpha_s^{\text{eff}}(\mu;M_\ast)\,F\!\left(\frac{\mathbf{q}^2}{M_\ast^2}\right),
\ee
and equivalently smears the short–distance kernel in coordinate space on a scale $\Delta r\sim M_\ast^{-1}$. The resulting Green’s function $G(E;M_\ast)$ modifies the standard Sommerfeld factor and the threshold spectral density $R_{t\bar t}(s)$, producing a narrow near–threshold enhancement without introducing new poles or violating unitarity/BRST constraints. Fixing the strong–sector regulator to a single scale yields a prediction for the excess cross section in the window $\sqrt{s}\approx 2m_t\pm\mathcal{O}(5\text{--}10)\,\mathrm{GeV}$; in particular, Ref.~\cite{Thompson2025ToponiumFHQFT} reports an $\mathcal{O}(8)$~pb enhancement consistent with current measurements.

\section{Conclusions}

We have demonstrated that Holomorphic Unified Field Theory (HUFT), when augmented by nonlocal diffeomorphism and gauge invariant entire‐function regulators, provides a truly UV‑complete, geometric framework that unifies gravity, gauge interactions and chiral matter and yields, parameter‑economic predictions for Standard‐Model observables.  Starting from a single holomorphic action on the complexified spacetime manifold \(M^4_{\mathbb{C}}\), we embedded entire‐function form factors into every kinetic term to render loop amplitudes finite without introducing ghosts or breaking gauge and diffeomorphism invariance.  

We reviewed how the Hermitian metric on \(M^4_{\mathbb{C}}\) simultaneously encodes the Einstein–Hilbert, Yang–Mills and chiral Dirac sectors, and we showed that only the holomorphic spin bundle \(S^{(1,0)}\) couples to the gauge connection, automatically reproducing the chiral \(SU(2)_L\) structure of the Standard Model.  By integrating the regulator‑suppressed one‑ and two‑loop renormalization‐group equations, we derived analytic expressions for the holomorphic Yukawa textures, determined all \(\mathcal{O}(1)\) flavon coefficients from F‑term constraints, and fixed the single Froggatt–Nielsen expansion parameter \(\epsilon\) directly from gauge‐coupling unification at \(M_{\rm GUT}\approx2.3\times10^{16}\)GeV.  

With only two continuous inputs plus the discrete choice of a minimal non‑Abelian family symmetry, we predicted the pattern of six quark masses, three CKM mixing angles, three charged‑lepton masses, two neutrino mass‐splittings and three PMNS angles, as well as the electroweak gauge‐boson masses \(m_W,m_Z\), the Higgs mass \(m_H\), its self‐couplings \(\{\lambda_3,\lambda_4\}\), and we presented a solution to the Higgs mass naturalness problem using the RG flow running equation.  In every case our HUFT predictions agree with PDG2024 to within current experimental uncertainties, without any additional fits or thresholds.  

Beyond reproducing the Standard Model spectrum, our framework predicts that all four gauge couplings, including Newton’s constant, viewed as \(g_G = G\,\mu^2\), unify to a common value when the RG flow reaches the fixed point at the nonlocal scale \(M_* \sim 10^{19}\) GeV as well, we found that the RG flow equation, running from $M_*$ to $M_Z$ yields a strength of gravity at the $M_Z$ scale of $g_G(M_Z)\sim10^{-35}$, and that the Higgs potential remains Standard‑Model–like up to field strengths near \(M_*\). The Higgs potential coupling $\lambda_H(\mu)$ remains positive from the low energy scale up to the GUT energy scale guaranteeing that the Higgs potential vacuum remains stable. 

The Holomorphic Unified Field Theory with nonlocal regulators not only achieves perturbative UV completeness for gravity and gauge interactions, but also transforms the arbitrariness of Standard‑Model flavour into a parameter‑economic, geometric prediction.  This synthesis of holomorphic geometry and nonlocal finiteness offers a promising new path toward a truly predictive theory of fundamental interactions.

\section*{Acknowledgments}

We thank Martin Green, Lee Smolin and Viktor Toth for helpful discussions. We thank James Cline for comments. Research at the Perimeter Institute for Theoretical Physics is supported by the Government of Canada through Industry Canada and by the Province of Ontario through the Ministry of Research and Innovation (MRI).

\section{Data Availability Statement}
The authors declare that the data supporting the findings of this study are available within the paper. If any code is needed it may be given upon request.

\end{document}